\documentstyle[12pt,epsf]{article}
\newcommand {\m}{\mu}
\newcommand {\n}{\nu}
\newcommand {\pl}{\partial}
\newcommand {\p} {\phi}

\newcommand {\al}{\alpha}
\newcommand {\be}{\beta}
\newcommand {\ga}{\gamma}
\newcommand {\Ga}{\Gamma}

\newcommand {\la}{\lambda}

\newcommand {\si}{\sigma}
\newcommand {\Si}{\Sigma}

\newcommand {\Th}{\Theta}
\newcommand {\om}{\omega}
\newcommand {\Om}{\Omega}

\newcommand {\na}{\nabla}
\newcommand {\del}  {\delta}
\newcommand {\Del}  {\Delta}
\newcommand {\mn}{{\mu\nu}}
\newcommand {\ls}   {{\lambda\sigma}}
\newcommand {\ab}   {{\alpha\beta}}
\newcommand {\half}{ {\frac{1}{2}} }
\newcommand {\fourth} {\frac{1}{4} }
\newcommand {\sqg} {\sqrt{g}}

\newcommand {\Lcal}{{\cal L}}

\newcommand {\Dcal}{{\cal D}}

\newcommand {\ra} {\rightarrow}
\newcommand {\pr}   {{\quad .}}
\newcommand {\com}  {{\quad ,}}
\newcommand {\q}    {\quad}
\newcommand {\qq}   {\quad\quad}
\newcommand {\qqq}   {\quad\quad\quad}

\newcommand {\nl}    {\newline}
\newcommand {\nn}    {\nonumber}
\newcommand {\ul}    {\underline}
\newcommand {\vs}[1]  { \vspace*{#1 cm} }

\newcounter{eq}
\newcounter{sc}

\newcommand {\NP}   {Nucl.Phys.}
\newcommand {\PL}   {Phys.Lett.}
\newcommand {\PR}   {Phys.Rev.}

\newcommand {\CQG}  {Class.Quantum Grav.}

\newcommand {\JMP}   {J.Math.Phys.}
\def\overleftrightarrow#1{\vbox{\ialign{##\crcr
 $\leftrightarrow$\crcr\noalign{\kern-1pt\nointerlineskip}
 $\hfil\displaystyle{#1}\hfil$\crcr}}}

\begin{document}

%%%%%%%%%%%%%%%%%%%%%%%%%%%%%%%%%%%%%%%%%%%%%%%%%%%%%%%%%%%%%%%%%%
%%%%%%%%%%%%%%%%%%%%%%%% Title %%%%%%%%%%%%%%%%%%%%%%%%%%%%%%%%%%%
%%%%%%%%%%%%%%%%%%%%%%%%%%%%%%%%%%%%%%%%%%%%%%%%%%%%%%%%%%%%%%%%%%
\begin{flushright}
DAMTP/97-3\\
Jan.,1997\\
hep-th/9702003
\end{flushright}
\vspace{24pt}

%\magnification=\magstep1
\pagestyle{empty}
\baselineskip15pt
%\font\cmssB=cmss17
%\font\cmssS=cmss10

\begin{center}
{\large\bf 
           Graphical
           Classification of Global SO(n) Invariants\\
           and\\
           Independent General Invariants
\vskip 1mm
}

\vspace{10mm}

Shoichi ICHINOSE
%          \footnote{
%          On leave of absence, until Jan. 31,1997,  from
%          Department of Physics, University of Shizuoka,
%          Yada 52-1, Shizuoka 422, Japan.
%          E-mail address:\ s.ichinose@damtp.cam.ac.uk\ ;\
%          ichinose@momo1.u-shizuoka-ken.ac.jp
%                  }
and Noriaki IKEDA${}^{\dag}$ 
%\footnote
%{ E-mail address:\ nori@kurims.kyoto-u.ac.jp }
\\
\vspace{5mm}
DAMTP, University of Cambridge,       \\
Silver Street, Cambridge CB3 9EW, UK\cite{email1}                  \\
${}^{\dag}$Research Institute for Mathematical Sciences \\
Kyoto University, Kyoto 606-01, Japan\cite{email2}
\end{center}

%\maketitle

\vspace{15mm}
\begin{abstract}
This paper treats some basic points in  general relativity
and in its perturbative analysis. Firstly
a systematic classification of global SO(n) invariants,
which appear in the weak-field expansion of 
n-dimensional gravitational theories,  is presented.
Through the analysis, we explain the following points:\  
a)\ a graphical representation is introduced to express invariants
clearly;\  
b)\ every graph of invariants is specified by a set of indices;\  
c)\ a number, called {\it weight}, is assigned to each invariant.
%'degeneracy' in the contraction of suffixes.
It expresses the symmetry with respect to the suffix-permutation
within an invariant.
Interesting relations among the weights of invariants are given. 
Those relations show the consistency and the completeness of the present
classification;\ 
d)\ some reduction procedures  
%and a graph relation due to the gauge fixing condition 
are introduced in graphs for the purpose of classifying them.
Secondly
the above result is applied to the proof of the independence of 
general invariants
with the mass-dimension $M^6$ for the general geometry
in a general space dimension. We take a
graphical representation for general invariants too.
Finally all relations depending on
each space-dimension are systematically
obtained for 2, 4 and 6 dimensions.

\vspace{15mm}

PACS NO:\ 02.70.-c,\ 04.20.-q,\ 04.60.-m,\ 02.40.Pc\nl
%Keywords:\ Anomaly,\ Heat-kernel,\ Non-abelian anomaly,\ Fujikawa's
%method\nl
\end{abstract}

\newpage
\pagestyle{plain}
\pagenumbering{arabic}
%\setcounter{page}{1}

%%%%%%%%%%%%%%%%%%%%%%%%%%%%%%%%%%%%%%%%%%%%%%%%%%%%%%%%%%%%%%%%%%
%%%%%%%%%%%%%%%%%%%%%%%% Article %%%%%%%%%%%%%%%%%%%%%%%%%%%%%%%%%
%%%%%%%%%%%%%%%%%%%%%%%%%%%%%%%%%%%%%%%%%%%%%%%%%%%%%%%%%%%%%%%%%%

\rm
%%%%%%%%%%%%%%%%%%%%%%%%%%%%%%%%%%%%%%%%%%
%%%%%%%%%%%%%%%%%%%%%%%%%%%%%%%%%%%%%%%%%%%%%%%%%%%%%%%%%%%%%%%%%%%%%
%%%%%%%%%%%%%%%%%%%%%%%%%%%%%%   SEC  1    %%%%%%%%%%%%%%%%%%%%%%%%%%
%%%%%%%%%%%%%%%%%%%%%%%%%%%%%%%%%%%%%%%%%%%%%%%%%%%%%%%%%%%%%%%%%%%%%
\section{Introduction}
In  classical and quantum gravity, the most important elements
are invariants under the general coordinate transformation 
(referred to as general
invariants) because they are independent of a chosen coordinate.
Physical quantities can be expressed as functions of them.
The main problem we address in this paper is how to find
all independent general invariants for each space-dimension.
It is highly non-trivial because of the high symmetry
of Riemann tensors and their products.\cite{FKWC,SI}
As far as general invariants with lower mass-dimensions\cite{foot1}
%\footnote{
%When we say ``dimension'', there are two different meanings in this
%paper. One is the dimension of space, which we call space-dimension
%when clear separation is necessary. The other is the mass dimension
%of an operator. For example, Riemann tensor has $M^2$ dimension
%in any space-dimension. We call it mass-dimension when necessary.
%}
are concerned, it is
practically no problem because we have much experience in the past.
However we encounter general invariants with higher 
mass-dimensions in some cases such as
when we consider gravitational theories 
in the higher space-dimensions (ex. Weyl anomaly in a higher
dimensional gravity-matter theory) 
or when we consider higher-order quantum
corrections there (ex. Counter-terms at higher-order
or higher-order effective action).
As the mass-dimension of general invariants increases, 
the above problem becomes serious. 
At present, there seems to be no general way
of fixing complete and
independent general invariants.  

With such a direction in mind,
an approach to treat general invariants is given in \cite{SI},
where a graphical representation is introduced. The problem of 
listing  all general invariants is transformed to that of
listing  all closed graphs. 
It works for a general geometry in general space-dimension.
Some graph relations are introduced to express some relations between
Riemann tensors such as Bianchi identity and the cyclic identity.
It is a powerful technique to find relations between general
invariants. 
However, as noted in the discussion of \cite{SI}, the approach
does not guarantee the independence between finally listed ones.
It gives only the sufficient terms as the list of complete and
independent general invariants. The final list of terms
could still involves linearly dependent terms.
%It gives only the candidates of independent ones.
%In other words, the approach of \cite{SI} gives only the sufficient
%condition for the list of complete and independent general invariants.
In this paper, we provide 
another approach to prove the independence of general invariants
, as local functions, in the final list. 
%It gives the sufficient condition.
%The proof is for the general geometry in general space-dimension. 
%(We do not use any specific geometry.)

As far as local properties are concerned, it is sufficient to 
consider them in the weak-field perturbation around  flat space.
%****(intro.1)%%%%%%%%%%%%%%%%%%%%
\begin{eqnarray}
g_\mn=\del_\mn+h_\mn\com\q |h_\mn|\ll 1\com
                                                      \label{intro.1}
\end{eqnarray}
%%%%%%%%%%%%%%%%%%%%%%%%%%%%%%
where $\m,\n=1,\cdots,n$\ and $\del_\mn$ is the flat space metric.
The advantages of this ``weak-field''(or ``linear'') 
representation, compared with the use of the full metric $g_\mn$ and 
its inverse $g^\mn$, are 
a) there are no 'inverse' fields and every general invariant is
expressed by $h_\mn$ and its derivatives, and
b) If we express general tensors in terms of ``weak-fields''
representation, some non-linear relations\cite{foot2}
%\footnote{
%The relations come from
%``Type 2'' symmetry in Sec.2 of \cite{SI}.
%Note that the statement does not apply to those relations which
%come from ``Type 3'' symmetry. It
%is the symmetry depending on each dimension.
%         }
, such as the Bianchi identity and the cyclic identity,  
are automatically satisfied at each order of $h$.
Each general invariant is expanded
as an infinite power series in $h_\mn$. Among many expanded terms, we
focus mainly on the 'products' of $\pl_\m\pl_\n h_\ab$, 
because they turn out to give sufficient information to determine
important quantities.
As for general terms, we will make comments in Sec.VI and Sec.X.
In \cite{II2} ( we call this 'paper (I)' ), we  introduced
a graphical representation for the 'products' of $\pl_\m\pl_\n h_\ab$, 
and examined
some basic definitions and lemmas, some features of the graphs.
Paper (I) deals mainly with the case of $\pl\pl h$- and $(\pl\pl h)^2$-tensors.
In this paper we study $(\pl\pl h)^3$-tensors, where we can see a more
general structure valid for general 
invariants with higher mass-dimensions.
We classify $(\pl\pl h)^3$-invariants 
completely. The result is applied 
to the proof of independence of general invariants with dimension $M^6$.
We prove it for a general geometry in a general space-dimension.

After listing all independent general invariants in a
general dimension, we examine them in each space-dimension
in order to find additional relations depending on
the space-dimension.
%Finally we present all relations, between general invariants,
%which depend on dimension. 
The approach of \cite{SI} is
applied and  2, 4 and 6 space-dimensions are considered.

Many graphs are presented to show their usefulness. We can easily
identify a tensor or an invariant with many suffixes involved.
One of its important advantages is we can utilize the graph
topology in explicit tensor calculation ( in computer ).
We introduce some {\it indices} to represent the graph topology.
The explicit calculational result of weak field expansion
of general invariants, presented in App.E, shows the power
of the present approach.

In Sec.II, we review paper (I) and explain the basic ingredients
necessary for the present classification. 
Every SO(n)-invariant is represented by a graph.
Classification is done in a two-fold
way: one by the 'bondless diagram', which is explained in Sec.III,  
and the other by 'reduced graphs', which is explained in Sec.IV. 
Every graph is named respecting both classification 
schemes. In Sec.V,
we introduce some indices in order to specify every graph by a set of
topological numbers. 
The set of indices distinguishes each graph.
Every graph has another number called the 'weight', which shows the
``degree of symmetry'' with respect to suffix-contraction. Various
identities between weights are presented in Sec.VI. 
They show the consistency and completeness of the present 
classification. Disconnected
graphs are treated in Sec.VII. 
We devote ourselves to the  classification of SO(n)-invariants
from Sec.II to Sec.VII.
In Sec.VIII we apply the results
to general relativity and show the independence of general invariants.
All special relations, between general invariants, which depend on
space-dimension are explicitly obtained for 2, 4 and 6 dimensions
in Sec.IX.
The discussion and conclusion are made in Sec.X.
Some appendices are provided in order to show the content of the text
more concretely. 
App.A shows the full list of 
$(\pl\pl h)^3$-invariants
with their graphs and their graph names.  
App.B lists the indices and the weights of all $(\pl\pl h)^3$-invariants.
App.C deals with general invariants of a type
$\na\na R\times R$\, where a graph for
$\pl_\m\pl_\n\pl_\la\pl_\si h_\ab$ is introduced.
App.D deals with general invariants of another type
$\na R\times \na R$\, where a graph for
$\pl_\m\pl_\n\pl_\la h_\ab$ is introduced.
App.E lists the contribution to $(\pl\pl h)^3$-terms of 
some general invariants
with mass-dimension $M^6$.
App.F shows all graphs of general invariants with $M^6$-dimension.
Some anti-symmetrized quantities, 
which are used in Sec.IX, are defined
graphically in App.G.

%%%%%%%%%%%%%%%%%%%%%%%%%%%%%%%%%%%%%%%%%%%%%%%%%%%%%%%%%%%%%%%%%%%%%
%%%%%%%%%%%%%%%%%%%%%%%%%%%%%%   SEC  2    %%%%%%%%%%%%%%%%%%%%%%%%%%
%%%%%%%%%%%%%%%%%%%%%%%%%%%%%%%%%%%%%%%%%%%%%%%%%%%%%%%%%%%%%%%%%%%%%
\section{Graphical Representation of SO(n)-Invariants}
We briefly explain some basic terminology and an important lemma, 
introduced in paper (I), which are necessary for the present paper.

\begin{figure}
   \centerline{
{\epsfxsize=4.80cm  \epsfysize=2.68cm \epsfbox{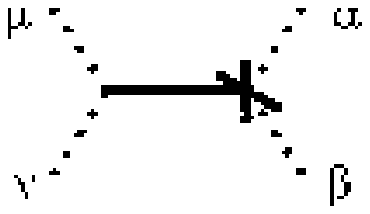}}
               }
\caption{
%*fig1*Fig.1\ 
4-tensor $\pl_\m\pl_\n h_\ab$
        }
\label{fig1}
\end{figure}

\begin{figure}
   \centerline{
{\epsfxsize=2.05cm  \epsfysize=1.83cm \epsfbox{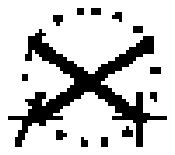}}
               }
\caption{
%*fig2*Fig.2\ 
Graphical representation of 
$A1=\pl_\si\pl_\la h_\mn\cdot\pl_\si\pl_\n h_{\m\la}$. 
        }
\label{fig2}
\end{figure}

The 4-th rank global SO(n) tensor(4-tensor), 
$\pl_\m\pl_\n h_\ab$\ is graphically
represented  in Fig.\ref{fig1}.
%%%%%%%%%%%%%%%%%%%%%%%  Fig.1  %%%%%%%%%%%%%%%%%%%%%%%%%%%%%%%%%%%%
%%%%%%%%%%%%%%%%%%%%%%%%%%%%%%%%%%%%%%%%%%%%%%%%%%%%%%%%%%%%%%%%%%%%%
Dotted lines, a rigid line, a vertex with  and without a crossing mark
are called {\it suffix-lines}, a {\it bond}, a {\it h-vertex} and 
a {\it dd-vertex} respectively. We graphically represent suffix contraction 
by gluing two corresponding suffix-lines. As an example, 
$A1=\pl_\si\pl_\la h_\mn\cdot\pl_\si\pl_\n h_{\m\la}$ is represented
in Fig.\ref{fig2}.
%%%%%%%%%%%%%%%%%%%%%%%  Fig.2  %%%%%%%%%%%%%%%%%%%%%%%%%%%%%%%%%%%%
%%%%%%%%%%%%%%%%%%%%%%%%%%%%%%%%%%%%%%%%%%%%%%%%%%%%%%%%%%%%%%%%%%%%%

Generally suffix-lines in a SO(n)-invariant are closed. We call these 
{\it suffix-loops}. 
Let us state a useful lemma 
on a general SO(n)-invariant made of $s$\ $\pl\pl h$-tensors.
It will be used in Sec.III to classify graphs in terms of 
the vertex\ (h or dd)-distribution in suffix-loops. .
\begin{description}
%%%%%%%%%%%%%%%%%%%%%%%%   Lemma    %%%%%%%%%%%%%%%%%%%%%%%%%%%%%%%%%%%%%%%
\item[Lemma]\ 
Let a general $(\pl\pl h)^s$-invariant ($s=1,2,\cdots$) have $\ul{l}$ 
suffix-loops.
Let
each loop have $v_i$ h-vertices and $w_i$ dd-vertices 
($i=1,2,\cdots, \ul{l}-1,\ul{l}$).
We have the following {\it necessary} conditions for 
$s,\ul{l},v_i \ \mbox{and } w_i$.
%****(rev.0)%%%%%%%%%%%%%%%%%%%%
\begin{eqnarray}
\sum_{i=1}^{\ul{l}}v_i=s\com\q \sum_{i=1}^{\ul{l}}w_i=s\com\nn\\
v_i\geq 0\com\q w_i\geq 0\com\q v_i+w_i\geq 1\com\label{rev.0}\\
v_i\ ,\ w_i\ =0,1,2,\cdots\q ,\q \ul{l}=1,2,3,\cdots,2s-1,2s\pr\nn
\end{eqnarray}
%%%%%%%%%%%%%%%%%%%%%%%%%%%%%
\end{description}
%%%%%%%%%%%%%%%%%%%%%%%%%%%%%%%%%%%%%%%%%%%%%%%%%%%%%%%%%%%%%%%%%%%%%%%%%%%%%

\begin{figure}
   \centerline{
{\epsfxsize=2.19cm  \epsfysize=1.91cm \epsfbox{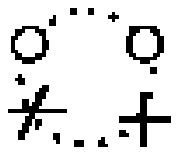}}
               }
\caption{
%*fig3*Fig.3\ 
Bondless diagram for $A1$ of Fig.\protect\ref{fig2}.
dd-vertices are explicitly represented by small circles.
        }
\label{fig3}
\end{figure}

It is useful, for classifying graphs, to introduce a {\it bondless diagram}
which is obtained by deleting all bonds within a graph. For $A1$ of 
Fig.\ref{fig2},
the corresponding bondless diagram is shown in Fig.\ref{fig3}
%%%%%%%%%%%%%%%%%%%%%%%  Fig.3  %%%%%%%%%%%%%%%%%%%%%%%%%%%%%%%%%%%%
%%%%%%%%%%%%%%%%%%%%%%%%%%%%%%%%%%%%%%%%%%%%%%%%%%%%%%%%%%%%%%%%%%%%%
, where a small circle is used to represent a dd-vertex explicitly.

Generally an SO(n)-invariant is composed of some suffix-loops. For each loop,
we define two  indices, the {\it bond changing number} (\ul{bcn}) and
the {\it vertex changing number} (\ul{vcn}), in the following way.
%%%%%%%%%%%%%%%%%%%%%%%%   Def    %%%%%%%%%%%%%%%%%%%%%%%%%%%%%%%%%%%%%%%
\begin{description}
\item[Def ]\ 
\ul{bcn}[\ ] and \ul{vcn}[\ ] are defined for each suffix-loop as
follows\cite{foot3}.
%\footnote{
%\ul{bcn}[\ ] imply an general element of an
%array:\ul{bcn}[0],$\cdots$,\ul{bcn}[$\ul{l}$]. 
%The same thing is for \ul{vcn}[\ ].
%All quantities of indices
%are underlined in the following.
%}
When we trace the suffix-line of a suffix-loop, starting from a vertex 
in a certain direction, we generally pass some vertices,
and  finally come back to the starting vertex. 
When we move, in the tracing, from one vertex to the next vertex, 
we compare the bonds to which the two vertices belong, and their vertex types.
If the bonds are different, we set $\Del\mbox{\ul{bcn}}=1$, otherwise
$\Del\mbox{\ul{bcn}}=0$, 
If the vertex-types are different, we set $\Del\mbox{\ul{vcn}}=1$, otherwise
$\Del\mbox{\ul{vcn}}=0$. 
For the i-th loop, we sum the number $\Del\mbox{\ul{bcn}}$ and
$\Del\mbox{\ul{vcn}}$ while tracing the loop and
assign as
$\sum_{\mbox{along i-loop}}\Delta \mbox{\ul{bcn}}\equiv$ \ul{bcn}[i],
$\sum_{\mbox{along i-loop}}\Delta \mbox{\ul{vcn}}\equiv$ \ul{vcn}[i].
\end{description}
%%%%%%%%%%%%%%%%%%%%%%%%%%%%%%%%%%%%%%%%%%%%%%%%%%%%%%%%%%%%%%%%%%%%%%%%%%
\ul{bcn}[\ ] and \ul{vcn}[\ ] will be used, in Sec.IV and Sec.III
respectively, for classifying graphs.

\begin{figure}
   \centerline{
{\epsfxsize=5.47cm  \epsfysize=1.80cm \epsfbox{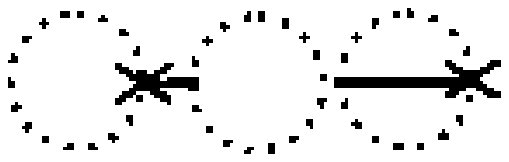}}
               }
\caption{
%*fig4*Fig.4\ 
Graphical representation of 
$PQ
=\pl^2 h_{\la\la}\cdot\pl_\m\pl_\n h_{\mn}$.
        }
\label{fig4}
\end{figure}

In paper (I), we have shown, using the graphical representation, that all
independent invariants are
%****(rev.1)%%%%%%%%%%%%%%%%%%%%
\begin{eqnarray}
P\equiv \pl_\m\pl_\m h_{\al\al}\com\q Q\equiv \pl_\al\pl_\be h_{\ab}
\com\q \label{rev.1}
\end{eqnarray}
%%%%%%%%%%%%%%%%%%%%%%%%%%%%%
for $\pl\pl h$-invariants and
%****(rev.2)%%%%%%%%%%%%%%%%%%%%
\begin{eqnarray}
A1
=\pl_\si\pl_\la h_\mn\cdot\pl_\si\pl_\n h_{\m\la}\ ,\ 
A2
=\pl_\si\pl_\la h_{\la\m}\cdot\pl_\si\pl_\n h_{\mn}\ ,\ 
A3
=\pl_\si\pl_\la h_{\la\m}\cdot\pl_\m\pl_\n h_{\n\si}\ ,\nn\\ 
B1
=\pl_\n\pl_\la h_{\si\si}\cdot\pl_\la\pl_\m h_{\mn}\ ,\ 
B2
=\pl^2 h_{\la\n}\cdot\pl_\la\pl_\m h_{\mn}\ ,\ 
B3
=\pl_\m\pl_\n h_{\la\si}\cdot\pl_\m\pl_\n h_{\ls}\ ,\nn\\
B4
=\pl_\m\pl_\n h_{\la\si}\cdot\pl_\la\pl_\si h_{\mn}\ ,\ 
Q^2
=(\pl_\m\pl_\n h_{\mn})^2\ ,\nn\\
C1
=\pl_\m\pl_\n h_{\la\la}\cdot\pl_\m\pl_\n h_{\si\si}\ ,\ 
C2
=\pl^2 h_{\mn}\cdot\pl^2 h_{\mn}\ ,\ 
C3
=\pl_\m\pl_\n h_{\la\la}\cdot\pl^2 h_{\mn}\ ,\nn\\
PQ
=\pl^2 h_{\la\la}\cdot\pl_\m\pl_\n h_{\mn}\ ,\ 
P^2
=(\pl^2 h_{\la\la})^2\ ,\label{rev.2}
\end{eqnarray}
%%%%%%%%%%%%%%%%%%%%%%%%%%%%%
for $(\pl\pl h)^2$-invariants (totally 13 invariants). 
In Fig.\ref{fig4}, an invariant $PQ$\ in (\ref{rev.2}) is
graphically shown.
%%%%%%%%%%%%%%%%%%%%%%%  Fig.4  %%%%%%%%%%%%%%%%%%%%%%%%%%%%%%%%%%%%
%%%%%%%%%%%%%%%%%%%%%%%%%%%%%%%%%%%%%%%%%%%%%%%%%%%%%%%%%%%%%%%%%%%%%
When a diagram is composed of some parts which are not connected by
 suffix-lines or bonds
, as in Fig.\ref{fig4}, we say it is {\it disconnected}. Otherwise
, as in Fig.\ref{fig2}, it is referred to as {\it connected}.

%%%%%%%%%%%%%%%%%%%%%%%%%%%%%%%%%%%%%%%%%%%%%%%%%%%%%%%%%%%%%%%%%%%
%%%%%%%%%%%%%%%%%%%%%%%%%%  Sec. 3       %%%%%%%%%%%%%%%%%%%%%%%%
%%%%%%%%%%%%%%%%%%%%%%%%%%%%%%%%%%%%%%%%%%%%%%%%%%%%%%%%%%%%%%%%%%%
\section{
Classification of $(\pl\pl h)^3$-Invariants
by Bondless Diagrams
        }
\begin{figure}
   \centerline{
$\left(\begin{array}{c}  3 \\ 3  \end{array}\right)_6$:\ 
{\epsfxsize=81pt  \epsfysize=76pt \epsfbox{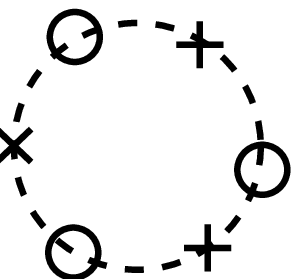}}\q
$\left(\begin{array}{c}  3 \\ 3  \end{array}\right)_4$:\ 
{\epsfxsize=75pt  \epsfysize=87pt \epsfbox{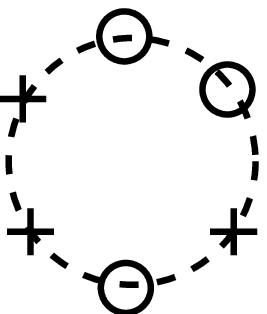}}\q
$\left(\begin{array}{c}  3 \\ 3  \end{array}\right)_2$:\ 
{\epsfxsize=72pt  \epsfysize=84pt \epsfbox{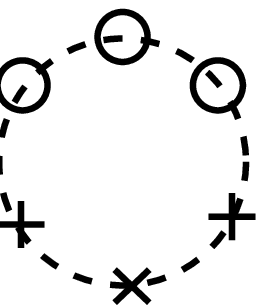}}
              }
\vspace{10pt}
   \centerline{
$\left(\begin{array}{c}  3 \\ 2  \end{array}\right)_4$:\ 
{\epsfxsize=65pt  \epsfysize=65pt \epsfbox{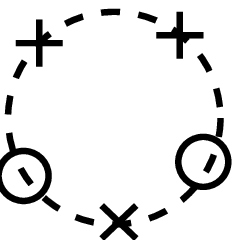}}\q
$\left(\begin{array}{c}  3 \\ 2  \end{array}\right)_2$:\ 
{\epsfxsize=61pt  \epsfysize=65pt \epsfbox{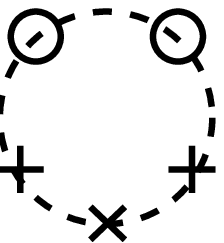}}\q
$\left(\begin{array}{c}  2 \\ 3  \end{array}\right)_4$:\ 
{\epsfxsize=62pt  \epsfysize=68pt \epsfbox{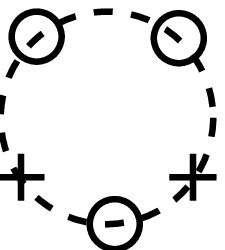}}\q
$\left(\begin{array}{c}  2 \\ 3  \end{array}\right)_2$:\ 
{\epsfxsize=70pt  \epsfysize=68pt \epsfbox{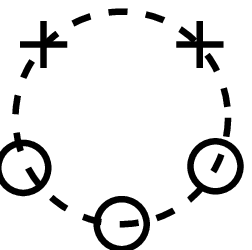}}
              }
\vspace{10pt}
   \centerline{
$\left(\begin{array}{c}  2 \\ 2  \end{array}\right)_4$:\ 
{\epsfxsize=65pt  \epsfysize=61pt \epsfbox{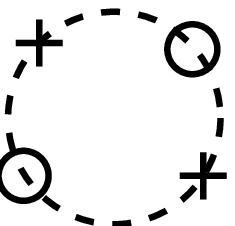}}\q
$\left(\begin{array}{c}  2 \\ 2  \end{array}\right)_2$:\ 
{\epsfxsize=60pt  \epsfysize=60pt \epsfbox{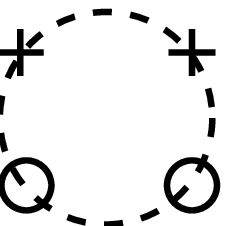}}\q
$\left(\begin{array}{c}  3 \\ 1  \end{array}\right)_2$:\ 
{\epsfxsize=70pt  \epsfysize=73pt \epsfbox{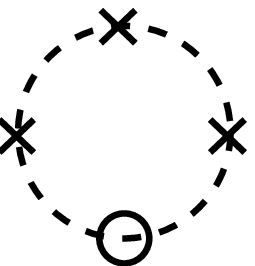}}\q
$\left(\begin{array}{c}  1 \\ 3  \end{array}\right)_2$:\ 
{\epsfxsize=74pt  \epsfysize=70pt \epsfbox{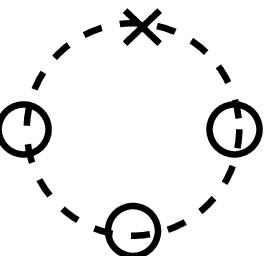}}
              }
\vspace{10pt}
   \centerline{
$\left(\begin{array}{c}  3 \\ 0  \end{array}\right)_0$:\ 
{\epsfxsize=62pt  \epsfysize=66pt \epsfbox{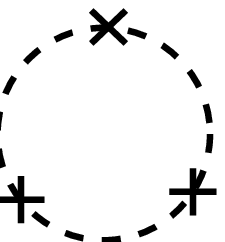}}\q
$\left(\begin{array}{c}  2 \\ 1  \end{array}\right)_2$:\ 
{\epsfxsize=61pt  \epsfysize=68pt \epsfbox{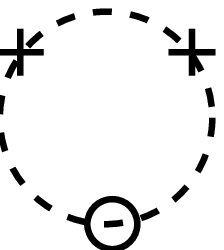}}\q
$\left(\begin{array}{c}  1 \\ 2  \end{array}\right)_2$:\ 
{\epsfxsize=63pt  \epsfysize=66pt \epsfbox{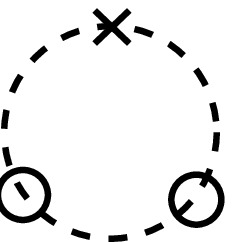}}\q
$\left(\begin{array}{c}  0 \\ 3  \end{array}\right)_0$:\ 
{\epsfxsize=60pt  \epsfysize=66pt \epsfbox{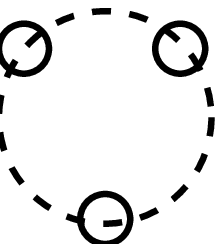}}
              }
\vspace{10pt}
   \centerline{
$\left(\begin{array}{c}  2 \\ 0  \end{array}\right)_0$:\ 
{\epsfxsize=70pt  \epsfysize=60pt \epsfbox{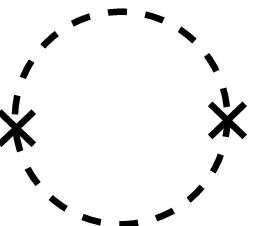}}\q
$\left(\begin{array}{c}  1 \\ 1  \end{array}\right)_2$:\ 
{\epsfxsize=73pt  \epsfysize=61pt \epsfbox{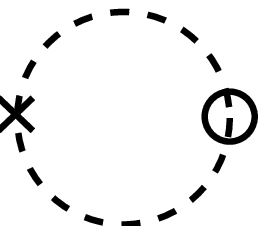}}\q
$\left(\begin{array}{c}  0 \\ 2  \end{array}\right)_0$:\ 
{\epsfxsize=75pt  \epsfysize=60pt \epsfbox{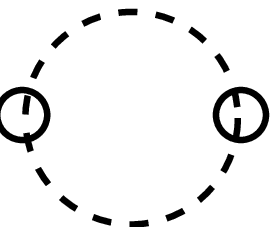}}
              }
\vspace{10pt}
   \centerline{
$\left(\begin{array}{c}  1 \\ 0  \end{array}\right)_0$:\ 
{\epsfxsize=60pt  \epsfysize=66pt \epsfbox{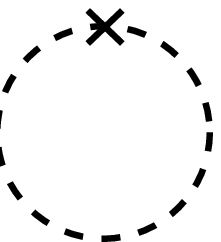}}\q
$\left(\begin{array}{c}  0 \\ 1  \end{array}\right)_0$:\ 
{\epsfxsize=60pt  \epsfysize=67pt \epsfbox{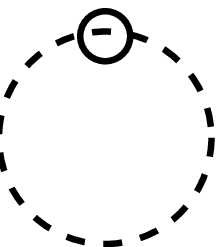}}
              }
%
%\begin{tabular}{p{6cm}p{6cm}p{6cm}}
%$\left(\begin{array}{c}  3 \\ 3  \end{array}\right)_6$:\ 
%{\epsfxsize=4.59cm  \epsfysize=3.67cm \epsfbox{f5n336.ps}}&
%$\left(\begin{array}{c}  3 \\ 3  \end{array}\right)_4$:\ 
%{\epsfxsize=3.77cm  \epsfysize=3.95cm \epsfbox{f5n334.ps}}&
%$\left(\begin{array}{c}  3 \\ 3  \end{array}\right)_2$:\ 
%{\epsfxsize=3.77cm  \epsfysize=3.92cm \epsfbox{f5n332.ps}}
%\\
%$\left(\begin{array}{c}  3 \\ 2  \end{array}\right)_4$:\ 
%{\epsfxsize=2.19cm  \epsfysize=2.29cm \epsfbox{f5n324.ps}}&
%$\left(\begin{array}{c}  3 \\ 2  \end{array}\right)_2$:\ 
%{\epsfxsize=1.98cm  \epsfysize=2.08cm \epsfbox{f5n322.ps}}&
%$\left(\begin{array}{c}  2 \\ 3  \end{array}\right)_4$:\ 
%{\epsfxsize=2.12cm  \epsfysize=2.12cm \epsfbox{f5n234.ps}}
%\end{tabular}
%
\caption{
%*fig5*Fig.5 
Bondless diagrams and values of ($v$, $w$, \ul{vcn}).
}
\label{fig5}
\end{figure}

%%%%%%%%%%%%%%%%%%%  2.1 %%%%%%%%%%%%%%%%%%%%%%%%%%%%%%%%%%%%%%%%
%\subsection{
%Bondless Diagrams and Vertex Changing Number (\ul{vcn})
%            }
Let us first denote a  suffix loop, with $v$  h-vertices, 
$w$ dd-vertices and a vertex changing number \ul{vcn} as
%****(class.1)%%%%%%%%%%%%%%%%%%%%
\begin{eqnarray}
\left(\begin{array}{c}  v \\ w  \end{array}\right)_{\mbox{\ul{vcn}}}
\pr \label{class.1}
\end{eqnarray}
%%%%%%%%%%%%%%%%%%%%%%%%%%%%%
In Fig.\ref{fig5}, all bondless diagrams that appear in 
suffix-loops of
$(\pl\pl h)^3$-invariants, are displayed graphically with 
the above notation.
%%%%%%%%%%%%%%%%%%%%%%%  Fig.5  %%%%%%%%%%%%%%%%%%%%%%%%%%%%%%%%%%%%
%%%%%%%%%%%%%%%%%%%%%%%%%%%%%%%%%%%%%%%%%%%%%%%%%%%%%%%%%%%%%%%%%%%%%
\nl
%%%%%%%%%%%%%%%%%%%%%%%%%%%%%%%%%%%%%%%%%%%%%%%%%%%%%%%%%%%%%%%%%%%%%

In this section,
we classify $(\pl\pl h)^3$-invariants by bondless diagrams.
Taking $s=3$ in (\ref{rev.0}), we list up all cases as follows.
In the following, \ul{vcn} is omitted  when the omission does not
cause ambiguity in specifying a bondless diagram.

%%%%%%%%%%%%%%%%%%%%%%%%%%%%%  l=1 %%%%%%%%%%%%%%%%%%%%%%%%%%%%%%
\flushleft{(i) $\ul{l}=1$}

%****(a.1)%%%%%%%%%%%%%%%%%%%%
\begin{eqnarray}
(1A):
\left(\begin{array}{c}  3 \\ 3  \end{array}\right)_6\com\q
(1B):
\left(\begin{array}{c}  3 \\ 3  \end{array}\right)_4\com\q
(1C):
\left(\begin{array}{c}  3 \\ 3  \end{array}\right)_2\pr
\label{a.1}
\end{eqnarray}
%%%%%%%%%%%%%%%%%%%%%%%%%%%%%

%%%%%%%%%%%%%%%%%%%%%%%%%%%%%  l=2 %%%%%%%%%%%%%%%%%%%%%%%%%%%%%%
\flushleft{(ii) $\ul{l}=2$}

%****(a.2)%%%%%%%%%%%%%%%%%%%%
\begin{eqnarray}
(2A):
\left(\begin{array}{c}  0 \\ 3  \end{array}\right)
\left(\begin{array}{c}  3 \\ 0  \end{array}\right),
(2B):
\left(\begin{array}{c}  0 \\ 2  \end{array}\right)
\left(\begin{array}{c}  3 \\ 1  \end{array}\right),
(2C):
\left(\begin{array}{c}  1 \\ 3  \end{array}\right)
\left(\begin{array}{c}  2 \\ 0  \end{array}\right),
(2D):
\left(\begin{array}{c}  1 \\ 2  \end{array}\right)
\left(\begin{array}{c}  2 \\ 1  \end{array}\right),\nn\\
(2E_a):
\left(\begin{array}{c}  2 \\ 2  \end{array}\right)_2
\left(\begin{array}{c}  1 \\ 1  \end{array}\right),
(2E_b):
\left(\begin{array}{c}  2 \\ 2  \end{array}\right)_4
\left(\begin{array}{c}  1 \\ 1  \end{array}\right), 
(2F_a):
\left(\begin{array}{c}  2 \\ 3  \end{array}\right)_2
\left(\begin{array}{c}  1 \\ 0  \end{array}\right),
(2F_b):
\left(\begin{array}{c}  2 \\ 3  \end{array}\right)_4
\left(\begin{array}{c}  1 \\ 0  \end{array}\right),\nn\\
(2G_a):
\left(\begin{array}{c}  0 \\ 1  \end{array}\right)
\left(\begin{array}{c}  3 \\ 2  \end{array}\right)_2,
(2G_b):
\left(\begin{array}{c}  0 \\ 1  \end{array}\right)
\left(\begin{array}{c}  3 \\ 2  \end{array}\right)_4.\label{a.2}
\end{eqnarray}
%%%%%%%%%%%%%%%%%%%%%%%%%%%%%

%%%%%%%%%%%%%%%%%%%%%%%%%%%%%  l=3 %%%%%%%%%%%%%%%%%%%%%%%%%%%%%%
\flushleft{(iii) $\ul{l}=3$}

%****(a.3)%%%%%%%%%%%%%%%%%%%%
\begin{eqnarray}
(3A):
\left(\begin{array}{c}  3 \\ 1  \end{array}\right)
\left(\begin{array}{c}  0 \\ 1  \end{array}\right)
\left(\begin{array}{c}  0 \\ 1  \end{array}\right),
(3B):
\left(\begin{array}{c}  3 \\ 0  \end{array}\right)
\left(\begin{array}{c}  0 \\ 2  \end{array}\right)
\left(\begin{array}{c}  0 \\ 1  \end{array}\right),
(3C):
\left(\begin{array}{c}  2 \\ 0  \end{array}\right)
\left(\begin{array}{c}  1 \\ 0  \end{array}\right)
\left(\begin{array}{c}  0 \\ 3  \end{array}\right),\nn\\
(3D):
\left(\begin{array}{c}  2 \\ 1  \end{array}\right)
\left(\begin{array}{c}  1 \\ 0  \end{array}\right)
\left(\begin{array}{c}  0 \\ 2  \end{array}\right),
(3E):
\left(\begin{array}{c}  2 \\ 0  \end{array}\right)
\left(\begin{array}{c}  1 \\ 1  \end{array}\right)
\left(\begin{array}{c}  0 \\ 2  \end{array}\right),
(3F_a):
\left(\begin{array}{c}  2 \\ 2  \end{array}\right)_2
\left(\begin{array}{c}  1 \\ 0  \end{array}\right)
\left(\begin{array}{c}  0 \\ 1  \end{array}\right),\nn\\
(3F_b):
\left(\begin{array}{c}  2 \\ 2  \end{array}\right)_4
\left(\begin{array}{c}  1 \\ 0  \end{array}\right)
\left(\begin{array}{c}  0 \\ 1  \end{array}\right),
(3G):
\left(\begin{array}{c}  2 \\ 0  \end{array}\right)
\left(\begin{array}{c}  1 \\ 2  \end{array}\right)
\left(\begin{array}{c}  0 \\ 1  \end{array}\right),
(3H):
\left(\begin{array}{c}  2 \\ 1  \end{array}\right)
\left(\begin{array}{c}  1 \\ 1  \end{array}\right)
\left(\begin{array}{c}  0 \\ 1  \end{array}\right),\label{a.3}\\
(3I):
\left(\begin{array}{c}  1 \\ 3  \end{array}\right)
\left(\begin{array}{c}  1 \\ 0  \end{array}\right)
\left(\begin{array}{c}  1 \\ 0  \end{array}\right),
(3J):
\left(\begin{array}{c}  1 \\ 2  \end{array}\right)
\left(\begin{array}{c}  1 \\ 1  \end{array}\right)
\left(\begin{array}{c}  1 \\ 0  \end{array}\right),
(3K):
\left(\begin{array}{c}  1 \\ 1  \end{array}\right)
\left(\begin{array}{c}  1 \\ 1  \end{array}\right)
\left(\begin{array}{c}  1 \\ 1  \end{array}\right)
.\nn
\end{eqnarray}
%%%%%%%%%%%%%%%%%%%%%%%%%%%%%

%%%%%%%%%%%%%%%%%%%%%%%%%%%%%  l=4 %%%%%%%%%%%%%%%%%%%%%%%%%%%%%%
\flushleft{(iv) $\ul{l}=4$}

%****(a.4)%%%%%%%%%%%%%%%%%%%%
\begin{eqnarray}
(4A):
\left(\begin{array}{c}  3 \\ 0  \end{array}\right)
\left(\begin{array}{c}  0 \\ 1  \end{array}\right)
\left(\begin{array}{c}  0 \\ 1  \end{array}\right)
\left(\begin{array}{c}  0 \\ 1  \end{array}\right),
(4B):
\left(\begin{array}{c}  2 \\ 0  \end{array}\right)
\left(\begin{array}{c}  1 \\ 0  \end{array}\right)
\left(\begin{array}{c}  0 \\ 2  \end{array}\right)
\left(\begin{array}{c}  0 \\ 1  \end{array}\right),\nn\\
(4C):
\left(\begin{array}{c}  2 \\ 1  \end{array}\right)
\left(\begin{array}{c}  1 \\ 0  \end{array}\right)
\left(\begin{array}{c}  0 \\ 1  \end{array}\right)
\left(\begin{array}{c}  0 \\ 1  \end{array}\right),
(4D):
\left(\begin{array}{c}  2 \\ 0  \end{array}\right)
\left(\begin{array}{c}  1 \\ 1  \end{array}\right)
\left(\begin{array}{c}  0 \\ 1  \end{array}\right)
\left(\begin{array}{c}  0 \\ 1  \end{array}\right),\label{a.4}\\
(4E):
\left(\begin{array}{c}  1 \\ 0  \end{array}\right)
\left(\begin{array}{c}  1 \\ 0  \end{array}\right)
\left(\begin{array}{c}  1 \\ 0  \end{array}\right)
\left(\begin{array}{c}  0 \\ 3  \end{array}\right),
(4F):
\left(\begin{array}{c}  1 \\ 1  \end{array}\right)
\left(\begin{array}{c}  1 \\ 0  \end{array}\right)
\left(\begin{array}{c}  1 \\ 0  \end{array}\right)
\left(\begin{array}{c}  0 \\ 2  \end{array}\right),\nn\\
(4G):
\left(\begin{array}{c}  1 \\ 2  \end{array}\right)
\left(\begin{array}{c}  1 \\ 0  \end{array}\right)
\left(\begin{array}{c}  1 \\ 0  \end{array}\right)
\left(\begin{array}{c}  0 \\ 1  \end{array}\right),
(4H):
\left(\begin{array}{c}  1 \\ 1  \end{array}\right)
\left(\begin{array}{c}  1 \\ 1  \end{array}\right)
\left(\begin{array}{c}  1 \\ 0  \end{array}\right)
\left(\begin{array}{c}  0 \\ 1  \end{array}\right).\nn
\end{eqnarray}
%%%%%%%%%%%%%%%%%%%%%%%%%%%%%

%%%%%%%%%%%%%%%%%%%%%%%%%%%%%  l=5 %%%%%%%%%%%%%%%%%%%%%%%%%%%%%%
\flushleft{(iv) $\ul{l}=5$}

%****(a.5)%%%%%%%%%%%%%%%%%%%%
\begin{eqnarray}
(5A):
\left(\begin{array}{c}  2 \\ 0  \end{array}\right)
\left(\begin{array}{c}  1 \\ 0  \end{array}\right)
\left(\begin{array}{c}  0 \\ 1  \end{array}\right)
\left(\begin{array}{c}  0 \\ 1  \end{array}\right)
\left(\begin{array}{c}  0 \\ 1  \end{array}\right),\nn\\
(5B):
\left(\begin{array}{c}  1 \\ 0  \end{array}\right)
\left(\begin{array}{c}  1 \\ 0  \end{array}\right)
\left(\begin{array}{c}  1 \\ 0  \end{array}\right)
\left(\begin{array}{c}  0 \\ 2  \end{array}\right)
\left(\begin{array}{c}  0 \\ 1  \end{array}\right),
(5C):
\left(\begin{array}{c}  1 \\ 1  \end{array}\right)
\left(\begin{array}{c}  1 \\ 0  \end{array}\right)
\left(\begin{array}{c}  1 \\ 0  \end{array}\right)
\left(\begin{array}{c}  0 \\ 1  \end{array}\right)
\left(\begin{array}{c}  0 \\ 1  \end{array}\right).\label{a.5}
\end{eqnarray}
%%%%%%%%%%%%%%%%%%%%%%%%%%%%%

%%%%%%%%%%%%%%%%%%%%%%%%%%%%%  l=6 %%%%%%%%%%%%%%%%%%%%%%%%%%%%%%
\flushleft{(iv) $\ul{l}=6$}

%****(a.6)%%%%%%%%%%%%%%%%%%%%
\begin{eqnarray}
(6A):
\left(\begin{array}{c}1\\0\end{array}\right)
\left(\begin{array}{c}1\\0\end{array}\right)
\left(\begin{array}{c}1\\0\end{array}\right)
\left(\begin{array}{c}0\\1\end{array}\right)
\left(\begin{array}{c}0\\1\end{array}\right)
\left(\begin{array}{c}0\\1\end{array}\right).\label{a.6}
\end{eqnarray}
%%%%%%%%%%%%%%%%%%%%%%%%%%%%%

All these classification names, in addition to another classification
names explained in Sec.IV, will be used  when we label every 
$(\pl\pl h)^3$-graph in App.A.

%%%%%%%%%%%%%%%%%%%%%%%%%%%%%%%%%%%%%%%%%%%%%%%%%%%%%%%%%%%%%%%%%%%
%%%%%%%%%%%%%%%%%%%%%%%%%%  Sec. 4  %%%%%%%%%%%%%%%%%%%%%%%%%%%%%%%
%%%%%%%%%%%%%%%%%%%%%%%%%%%%%%%%%%%%%%%%%%%%%%%%%%%%%%%%%%%%%%%%%%%
\section{
Classification of $(\pl\pl h)^3$-Invariants
by Reduced Graphs
        }
In this section we
classify all $(\pl\pl h)^3$-invariants in a different way from Sec.III.
We introduce two  reduction procedures in graphs, which are used
to classify graphs.

%%%%%%%%%%%%%%%%%%%%%%%%%  (i)  %%%%%%%%%%%%%%%%%%%%%%%%%%%%%%%%%%%
\flushleft{(i)\ First Reduction Procedure}
%\subsection
%{ Classification of $(\pl\pl h)^3$-invariants
%by identifying two vertex-types}

\begin{figure}
   \centerline{
{\epsfxsize=257pt  \epsfysize=63pt \epsfbox{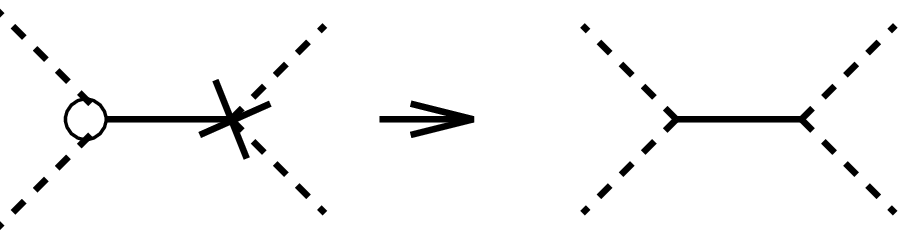}}
               }
\caption{
%*fig6*Fig.6\ 
Reduction procedure of identifying two vertex-types:
dd-vertex and h-vertex.
        }
\label{fig6}
\end{figure}

\begin{figure}
\centerline{
{\epsfxsize=7.48cm  \epsfysize=4.37cm \epsfbox{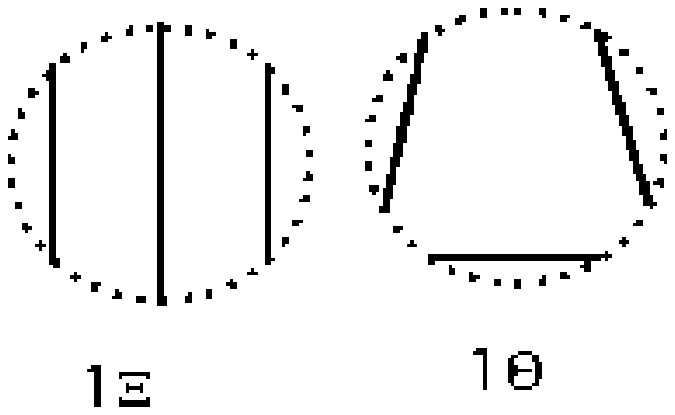}}
           }
\centerline{
{\epsfxsize=7.69cm  \epsfysize=4.23cm \epsfbox{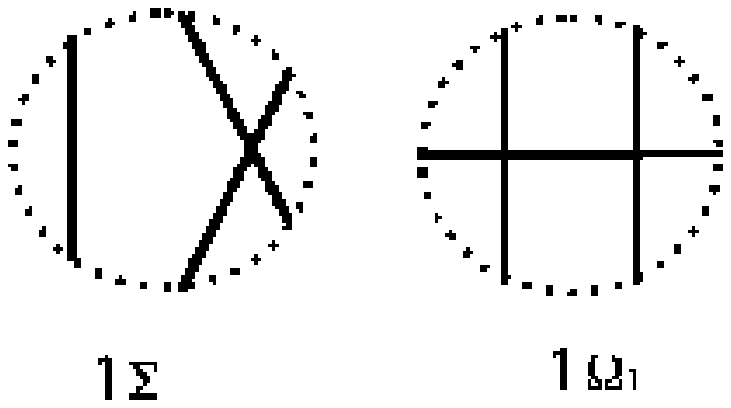}}
{\epsfxsize=3.53cm  \epsfysize=4.13cm \epsfbox{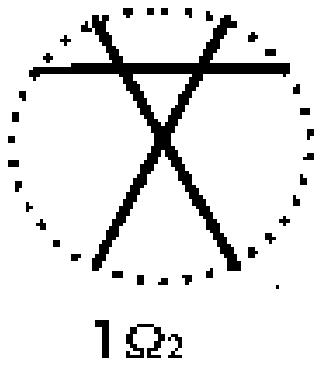}}
           }
\caption{
%*f7p1*Fig.7.1\  
Classification of 
$(\pl\pl h)^3$-graphs by \ul{bcn}[\ ], $\ul{l}=1$.
}
\label{f7p1}
\end{figure}

\begin{figure}
\centerline{
{ \epsfxsize=9.28cm \epsfysize=5.08cm \epsfbox{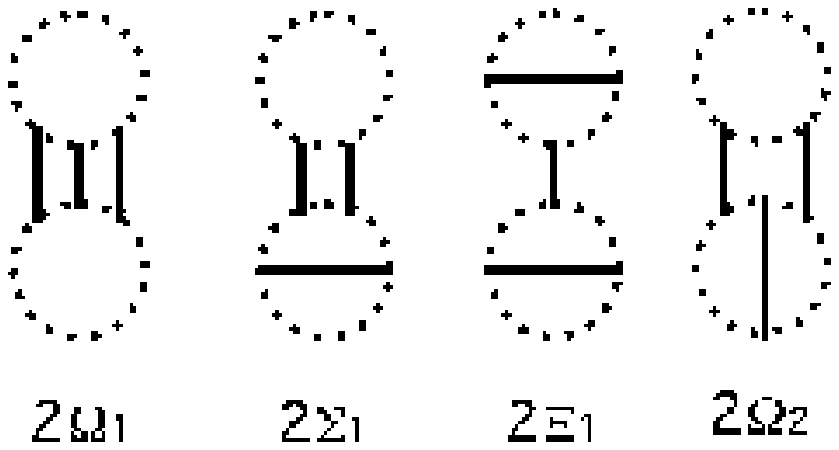}}
           }
\centerline{
{ \epsfxsize=6.56cm \epsfysize=4.97cm \epsfbox{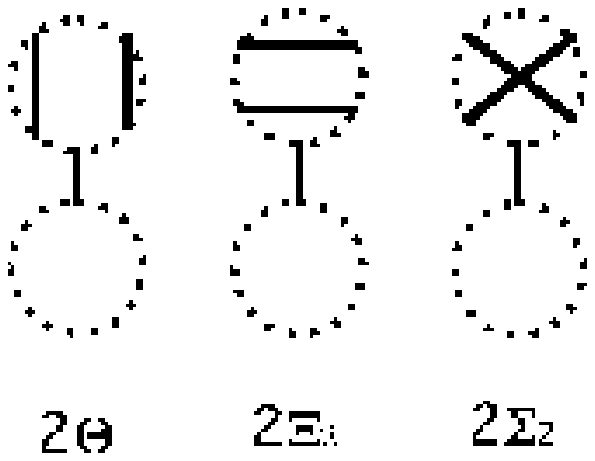}}
           }
\caption{
%*f7p2*Fig.7.2\  
Classification of $(\pl\pl h)^3$-graphs 
by \ul{bcn}[\ ], $\ul{l}=2$.
}
\label{f7p2}
\end{figure}

\begin{figure}
\centerline{
{ \epsfxsize=9.28cm \epsfysize=4.34cm \epsfbox{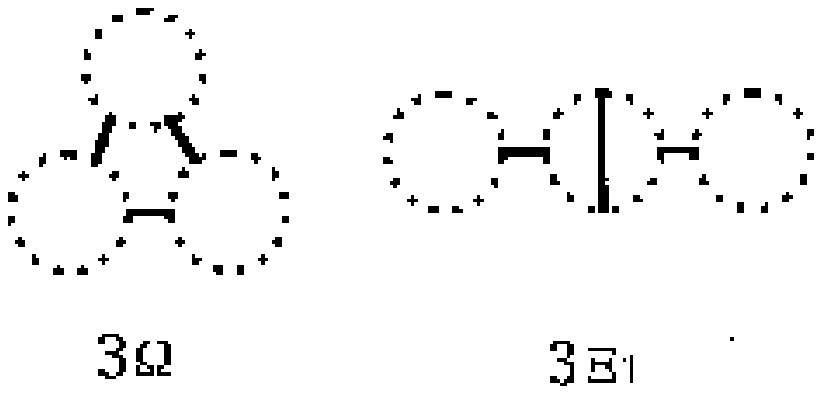}}
           }
\centerline{
{ \epsfxsize=8.93cm \epsfysize=3.99cm \epsfbox{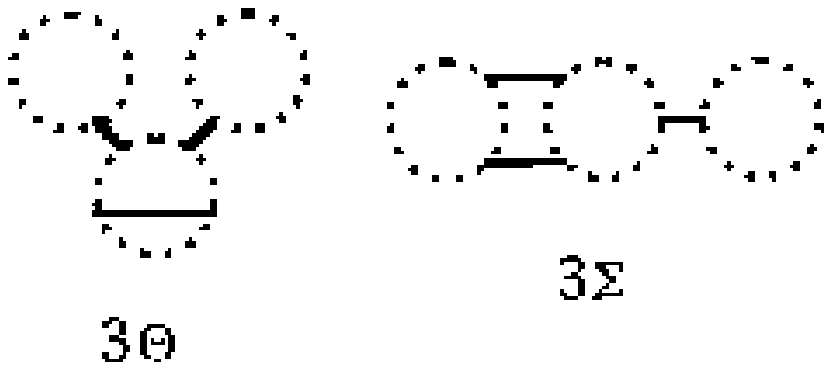}}
           }
\centerline{
{ \epsfxsize=5.04cm \epsfysize=2.65cm \epsfbox{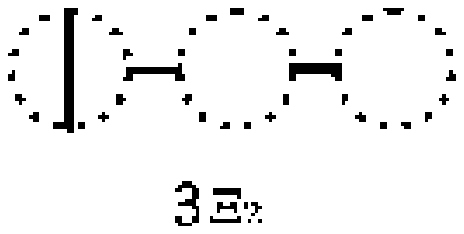}}
           }
\caption{
%*f7p3*Fig.7.3\ 
Classification of $(\pl\pl h)^3$-graphs by \ul{bcn}[\ ], $\ul{l}=3$.
}
\label{f7p3}
\end{figure}

\begin{figure}
\centerline{
{ \epsfxsize=4.30cm \epsfysize=4.52cm \epsfbox{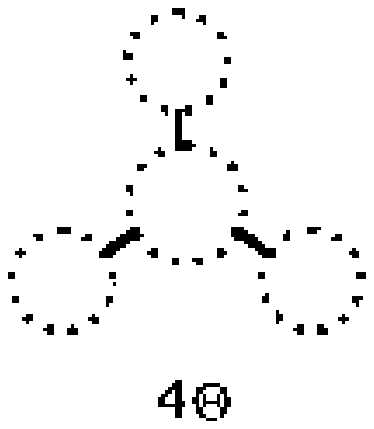}}
{ \epsfxsize=5.68cm \epsfysize=2.68cm \epsfbox{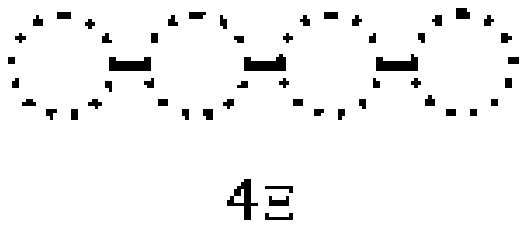}}
           }
\caption{
%*f7p4*Fig.7.4\ 
Classification of $(\pl\pl h)^3$-graphs by \ul{bcn}[\ ], $\ul{l}=4$.
}
\label{f7p4}
\end{figure}

The first reduction procedure is defined by identifying two
vertex-types as shown in Fig.\ref{fig6}.
%%%%%%%%%%%%%%%%%%%%%%%  Fig.6  %%%%%%%%%%%%%%%%%%%%%%%%%%%%%%%%%%%%
%%%%%%%%%%%%%%%%%%%%%%%%%%%%%%%%%%%%%%%%%%%%%%%%%%%%%%%%%%%%%%%%%%%%%
This reduction makes us classify all {\it connected}
$(\pl\pl h)^3$-invariants (totally 19 terms) as follows:\  
(1) $\ul{l}=1$, Fig.\ref{f7p1};\ 
(2) $\ul{l}=2$, Fig.\ref{f7p2};\ 
(3) $\ul{l}=3$, Fig.\ref{f7p3};\ 
(4) $\ul{l}=4$, Fig.\ref{f7p4}. 
%*** Numbers are their bond changing numbers(\ul{bcn})****. 
The classification naming will be explained
in next item (ii).
%

%%%%%%%%%%%%%%%%%%%%%%%%%%%%%  l=1 %%%%%%%%%%%%%%%%%%%%%%%%%%%%%%
%\flushleft{(i) $\ul{l}=1$, Fig.\ref{f7p1}.}
%%%%%%%%%%%%%%%%%%%%%%%%%%%%%%%%%%%%%%%%%%%%%%%%%%%%%%%%%%%%%%%%%%%
%%%%%%%%%%%%%%%%%%%%%%%  Fig.7.1  %%%%%%%%%%%%%%%%%%%%%%%%%%%%%%%%%%%%
%%%%%%%%%%%%%%%%%%%%%%%%%%%%%%%%%%%%%%%%%%%%%%%%%%%%%%%%%%%%%%%%%%%%%
%
%%%%%%%%%%%%%%%%%%%%%%%%%%%%%  l=2 %%%%%%%%%%%%%%%%%%%%%%%%%%%%%%
%\flushleft{(ii) $\ul{l}=2$, Fig.\ref{f7p2}.}
%%%%%%%%%%%%%%%%%%%%%%%%%%%%%%%%%%%%%%%%%%%%%%%%%%%%%%%%%%%%%%%%%%%
%%%%%%%%%%%%%%%%%%%%%%%  Fig.7.2  %%%%%%%%%%%%%%%%%%%%%%%%%%%%%%%%%%%%
%%%%%%%%%%%%%%%%%%%%%%%%%%%%%%%%%%%%%%%%%%%%%%%%%%%%%%%%%%%%%%%%%%%%%
%
%%%%%%%%%%%%%%%%%%%%%%%%%%%%%  l=3 %%%%%%%%%%%%%%%%%%%%%%%%%%%%%%
%\flushleft{(iii) $\ul{l}=3$, Fig.\ref{f7p3}.}
%%%%%%%%%%%%%%%%%%%%%%%%%%%%%%%%%%%%%%%%%%%%%%%%%%%%%%%%%%%%%%%%%%%
%
%%%%%%%%%%%%%%%%%%%%%%%  Fig.7.3  %%%%%%%%%%%%%%%%%%%%%%%%%%%%%%%%%%%%
%%%%%%%%%%%%%%%%%%%%%%%%%%%%%%%%%%%%%%%%%%%%%%%%%%%%%%%%%%%%%%%%%%%%%
%%%%%%%%%%%%%%%%%%%%%%%%%%%%%  l=4 %%%%%%%%%%%%%%%%%%%%%%%%%%%%%%
%\flushleft{(iv) $\ul{l}=4$, Fig.\ref{f7p4}.}
%%%%%%%%%%%%%%%%%%%%%%%%%%%%%%%%%%%%%%%%%%%%%%%%%%%%%%%%%%%%%%%%%%%
%%%%%%%%%%%%%%%%%%%%%%%  Fig.7.4  %%%%%%%%%%%%%%%%%%%%%%%%%%%%%%%%%%%%
%%%%%%%%%%%%%%%%%%%%%%%%%%%%%%%%%%%%%%%%%%%%%%%%%%%%%%%%%%%%%%%%%%%%%
%
%\vs 1
\q For $\ul{l}=5$\ and $6$\ ,\ there is no connected graphs.

%%%%%%%%%%%%%%%%%%%%%%%%%  (ii)  %%%%%%%%%%%%%%%%%%%%%%%%%%%%%%%%%%%
\flushleft{(ii)\ Second Reduction Procedure}

\begin{figure}
   \centerline{
{\epsfxsize=8.57cm  \epsfysize=2.79cm \epsfbox{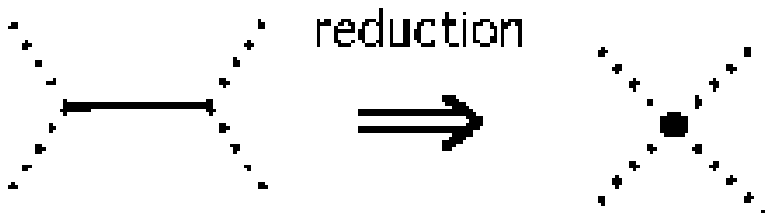}}
               }
\caption{
%*fig8*Fig.8\ 
Reduction of Graphs.
        }
\label{fig8}
\end{figure}

\begin{figure}
\centerline{
{ \epsfxsize=6.21cm \epsfysize=4.13cm \epsfbox{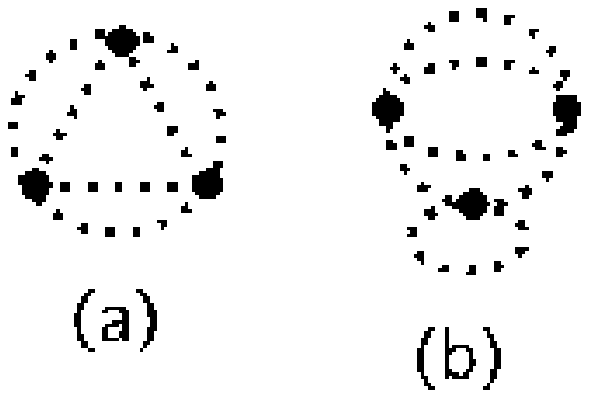}}
{ \epsfxsize=8.22cm \epsfysize=3.63cm \epsfbox{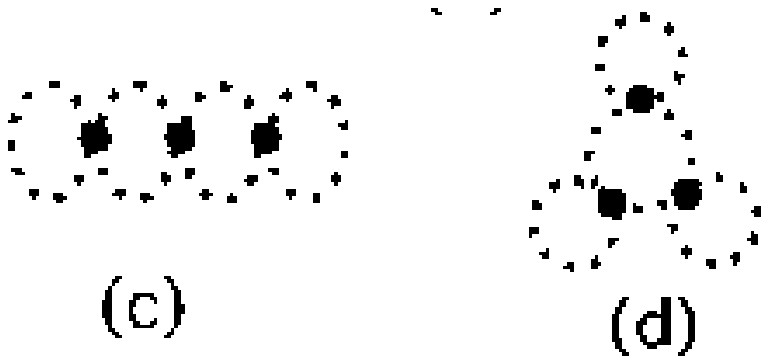}}
           }
\caption{
%*fig9*Fig.9\ 
Reduced Graphs by the procedure Fig.\ref{fig8}.
}
\label{fig9}
\end{figure}

We define the second reduction procedure
by reducing a bond to a vertex, as shown in Fig.\ref{fig8}.
%%%%%%%%%%%%%%%%%%%%%%%  Fig.8  %%%%%%%%%%%%%%%%%%%%%%%%%%%%%%%%%%%%
%%%%%%%%%%%%%%%%%%%%%%%%%%%%%%%%%%%%%%%%%%%%%%%%%%%%%%%%%%%%%%%%%%%%%
%%%%%%%%%%%%%%%%%%%%%%%%%%%%%%%%%%%%%%%%%%%%%%%%%%%%%%%%%%%%%%%%%%%%%
We get the  4 reduced graphs as shown in Fig.\ref{fig9}.
%%%%%%%%%%%%%%%%%%%%%%%  Fig.9  %%%%%%%%%%%%%%%%%%%%%%%%%%%%%%%%%%%%
%%%%%%%%%%%%%%%%%%%%%%%%%%%%%%%%%%%%%%%%%%%%%%%%%%%%%%%%%%%%%%%%%%%%%
%%%%%%%%%%%%%%%%%%%%%%%%%%%%%%%%%%%%%%%%%%%%%%%%%%%%%%%%%%%%%%%%%%%%%
The classification naming of (i) is due to Fig.\ref{fig9}.
We can classify all graphs of (i) as shown in Table \ref{tab1}.

%%%%%%%%%%%%%%%%%%%%%%%%%%%%%%%%%%%%%%%%%%%%%%%%%%%%%%%%%%%%%%%%%%%%%%%%%%
%%%%%%%%%%%%%%%%%%%%%  Table 1   %%%%%%%%%%%%%%%%%%%%%%%%%%%%%%%%%%%%%%%%%
%%%%%%%%%%%%%%%%%%%%%%%%%%%%%%%%%%%%%%%%%%%%%%%%%%%%%%%%%%%%%%%%%%%%%%%%%%
\begin{table}
\begin{tabular}{|c|c|c|c|c|}
\hline
  &  $\Om$          &   $\Si$         &    $\Xi$        &  $\Th$          \\
$\ul{l}$\ / class. &
   Fig.\ref{fig9}(a)& Fig.\ref{fig9}(b) & Fig.\ref{fig9}(c)
                                                   &  Fig.\ref{fig9}(d)      \\
  & 
(No of Tadpole 0) & (No of Tadpole 1) & (No of Tadpole 2) & (No of Tadpole 3)\\
\hline
$\ul{l}=1$ & 
  $1\Om_1\com\q 1\Om_2$ &          $1\Si$   &     $1\Xi$           &  $1\Th$ \\
\hline
$\ul{l}=2$ &
$2\Om_1\com\q 2\Om_2$ & $2\Si_1\com\q 2\Si_2$ 
                                          & $2\Xi_1\com\q 2\Xi_2$ & $2\Th$ \\
\hline
$\ul{l}=3$ &
$3\Om$                & $3\Si$            & $3\Xi_1\com\q 3\Xi_2$ & $3\Th$ \\
\hline
$\ul{l}=4$ &
                      &                   &     $4\Xi$            & $4\Th$ \\
\hline
%\multicolumn{4}{c}{\q}                                                 \\
%\multicolumn{4}{c}{Table 1\ \  Classification of 'vertex-type-less'
%diagrams of Sec.4.1 by }\\
%\multicolumn{4}{c}{ reducing bonds to vertices. }\\
\end{tabular}
\caption{
%*tab1*
Classification of 'vertex-type-less'
diagrams of Fig.\protect\ref{f7p1}-\protect\ref{f7p4} 
by reducing bonds to vertices. }
\label{tab1}
\end{table}
%%%%%%%%%%%%%%%%%%%%%%%%%  END  of  Table 1 %%%%%%%%%%%%%%%%%%%%%%%%%%%%%
%%%%%%%%%%%%%%%%%%%%%%%%%%%%%%%%%%%%%%%%%%%%%%%%%%%%%%%%%%%%%%%%%%%%%%%%%
%
\vs 1
\q The complete list of all $(\pl\pl h)^3$-invariants, totally 90
invariants (66 connected, 24 disconnected), 
are given in App.A,
where graphs are classified in a two-fold way 
using the classification schemes of Sec.III and IV. 
We notice the classification labels in Sec.III refer to 
the distribution of dd- and h-vertices
in suffix-loops, whereas those in Sec.IV refer to the topology of
a graph made of bonds and suffix-loops.
The completeness of the list of App.A 
will be shown in Sec.VI.

%%%%%%%%%%%%%%%%%%%%%%%%%%%%%%%%%%%%%%%%%%%%%%%%%%%%%%%%%%%%%%%%%%%
%%%%%%%%%%%%%%%%%%%%%%%%%%  SEC 5  %%%%%%%%%%%%%%%%%%%%%%%%%%%%%%%
%%%%%%%%%%%%%%%%%%%%%%%%%%%%%%%%%%%%%%%%%%%%%%%%%%%%%%%%%%%%%%%%%%%
\section{
Indices of $(\pl\pl h)^3$-Invariants
        }
Every graph can be characterized by its topological numbers,
such as the number of suffix-loops $\ul{l}$, which we call
{\it indices}. Besides $\ul{l}$, we have already explained
{\it bond changing number} (\ul{bcn}) and {\it vertex changing
number} (\ul{vcn}), which are another good indices. In order
to specify every graph completely, we need to introduce
some other indices.

\q The following points are advantageous when we have such
a set of indices as has one-to-one correspondence with 
a graph (SO(n)-invariant) 
:\ 
1) we can clearly read
the independence of graphs( or SO(n)-invariants)
 because the topologically different
quantities must be distinct;\ 
2) It is indispensable in programming  the  calculation
of quantities expressed by graphs. ( Example: weak-field
expansion calculation of quantum gravity.)

%%%%%%%%%%%%%%%%%%%%%%%%  5.1  %%%%%%%%%%%%%%%%%%%%%%%%%%%%%%%%%%%%
\subsection{Tadpole Number and Type of Tadpole}

\begin{figure}
   \centerline{
\shortstack{
{\epsfxsize=3.42cm  \epsfysize=2.82cm \epsfbox{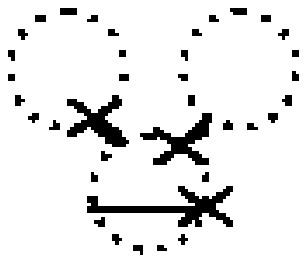}}
\\ G51:\ $3F_a\Th$
            } 
               }
\caption{
%*fig10*Fig.10\ 
G51:\ $3F_a\Th$.
        }
\label{fig10}
\end{figure}

%%%%%%%%%%%%%%%%%%%%%%%%   Def    %%%%%%%%%%%%%%%%%%%%%%%%%%%%%%%%%%%%%%%
\begin{description}
\item[Def]\ 
We call a closed suffix-loop which has only one vertex, a {\it tadpole}.
The number of tadpoles a graph has, is called the {\it tadpole number}
(\ul{tadpoleno})\ of the graph.
When a tadpole has a dd(h)-vertex, 
its {\it tadpole type}, \ul{tadtype}[$t$], is defined
to be 0 (1). \ul{tadtype}[$t$] is assigned for each tadpole 
:$t=1,2,\cdots$,\ul{tadpoleno}.
\end{description}
%%%%%%%%%%%%%%%%%%%%%%%%%%%%%%%%%%%%%%%%%%%%%%%%%%%%%%%%%%%%%%%%%%%%%%%%%%
For example, Fig.\ref{fig10} has \ul{tadpoleno}=2 and \ul{tadtype}[\ ]=0 and 1 for
each tadpole.
%%%%%%%%%%%%%%%%%%%%%%%  Fig.10  %%%%%%%%%%%%%%%%%%%%%%%%%%%%%%%%%%%%

%%%%%%%%%%%%%%%%%%%%%%%%%%%%%%%%%%%%%%%%%%%%%%%%%%%%%%%%%%%%%%%%%%%%%

Generally the indices \ul{tadpoleno} and \ul{tadtype}[\ ] 
are efficient for discriminating
large $\ul{l}$ graphs, whereas \ul{bcn}[\ ] and \ul{vcn}[\ ] 
are efficient for
discriminating small $\ul{l}$ graphs.

%%%%%%%%%%%%%%%%%%%%%%%%  5.2  %%%%%%%%%%%%%%%%%%%%%%%%%%%%%%%%%%%%
\subsection{Indices for Discriminating 'Fine Structure' of
$(\pl\pl h)^3$-Invariants}

%%%%%%%%%%%%%%%%%%%  (i) Vertex-Type Order   %%%%%%%%%%%%%%%%%%%%%%%%
\flushleft{
(i) Vertex-Type Order (\ul{Vorder})
          }

\begin{figure}
     \centerline{
\shortstack{
{\epsfxsize=90pt  \epsfysize=110pt \epsfbox{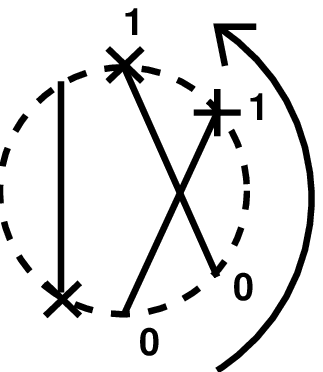}}
\\ G9:\ $1B\Si-c$
            }\q 
\shortstack{
{\epsfxsize=91pt  \epsfysize=105pt \epsfbox{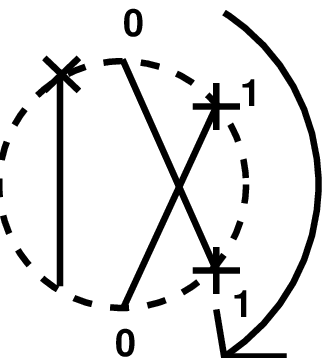}}
\\ G5:\ $1B\Si-a$
            }\q
\shortstack{
{\epsfxsize=90pt  \epsfysize=110pt \epsfbox{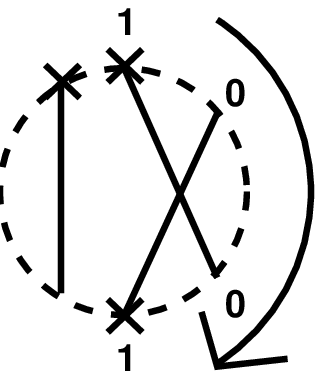}}
\\ G6:\ $1B\Si-b$
            } 
                }
\caption{
%*fig11*Fig.11\ 
Three Graphs with the same $\ul{l}$,\ul{vcn} and \ul{bcn}. \ul{Vorder}
discriminate them.
         }
\label{fig11}
\end{figure}

Let us examine the graphs of Fig.\ref{fig11}.
%%%%%%%%%%%%%%%%%%%%%%%  Fig.11  %%%%%%%%%%%%%%%%%%%%%%%%%%%%%%%%%%%%
%
%%%%%%%%%%%%%%%%%%%%%%%%%%%%%%%%%%%%%%%%%%%%%%%%%%%%%%%%%%%%%%%%%%%%%
We cannot discriminate Graphs G9,G5 and G6 by ($\ul{l}$,\ul{vcn},\ul{bcn}).
It is necessary to introduce a 'relative order' of 4 vertex-types at
the ends of 2 crossed bonds. Here we assign $0$ to a dd-vertex and $1$ to
h-vertex as shown in Fig.\ref{fig11}. Let us define the 
{\it vertex-type order}
(\ul{Vorder}) for each graph
as the sequence of vertex-type numbers in the order,
shown by an arrow in each graph of Fig.\ref{fig11},  
which is uniquely fixed by an 'isolated' bond. 
For example, we have \ul{Vorder}=(0,0,1,1) for Graph G9. Furthermore
we take 1 st ($\Del V$), 2 nd ($\Del \Del V$) and 
3 rd ($\Del\Del\Del V$)
difference of \ul{Vorder}. For this example of G9, we have
$\Del V=(0,1,0)$, $\Del\Del V=(1,-1)$ and $\Del\Del\Del V=-2$.
Instead of the direct use of \ul{Vorder}, $\Del\Del V$ and $\Del\Del\Del V$
are sufficient to discriminate between the three graphs.

\begin{figure}
     \centerline{
\shortstack{
{\epsfxsize=62pt  \epsfysize=123pt \epsfbox{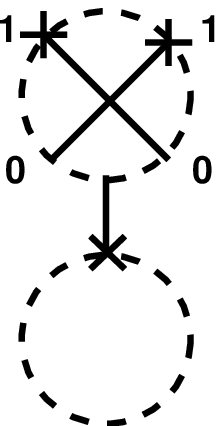}}
\\ G28:\ $2F_a\Si_2-a$
            } 
\q\q
\shortstack{
{\epsfxsize=60pt  \epsfysize=125pt \epsfbox{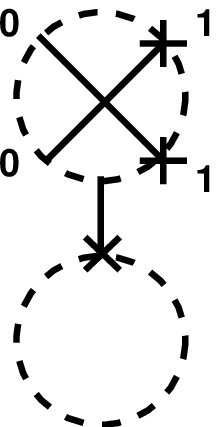}}
\\ G30:\ $2F_a\Si_2-b$
            } 
                }
\caption{
%*fig12*Fig.12\ 
Two Graphs (G28,G30) with the same $\ul{l}$,\ul{vcn}[\ ] 
and \ul{bcn}[\ ]. 
\ul{Vorder} discriminate them.
        }
\label{fig12}
\end{figure}

\q Another case of using \ul{Vorder} is that of Fig.\ref{fig12}. 
%%%%%%%%%%%%%%%%%%%%%%%  Fig.12  %%%%%%%%%%%%%%%%%%%%%%%%%%%%%%%%%%%%
%%%%%%%%%%%%%%%%%%%%%%%%%%%%%%%%%%%%%%%%%%%%%%%%%%%%%%%%%%%%%%%%%%%%%
In this case, we cannot specify the order of vertices because
a 'reference' is a vertex, not a bond. The ambiguity, however, disappears
by taking the value of $|\Del\Del\Del V|$, which is used here for the
discrimination. The same situation occurs for 3 more pairs: G38,G36; G32,G33;
G42,G39.

%%%%%%%%%%%%%%%%%%%  (ii) No of dd-vertices and h-vertices  %%%%%%%%%%%%%%%%
\flushleft{
(ii) Number of dd-vertices (\ul{ddverno}[\ ]) and 
      of h-vertices (\ul{hverno}[\ ])     }

\begin{figure}
     \centerline{
\shortstack{
{\epsfxsize=1.83cm  \epsfysize=3.88cm \epsfbox{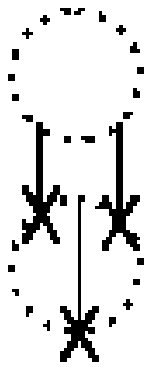}}
\\ G18:\ $2B\Om_2$
            } 
\q\q
\shortstack{
{\epsfxsize=2.01cm  \epsfysize=3.85cm \epsfbox{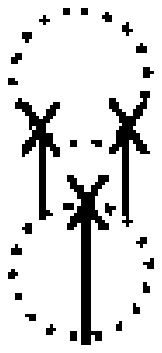}}
\\ G20:\ $2C\Om_2$
            } 
                }
\caption{
%*fig13*Fig.13\ 
Two Graphs with the same $\ul{l}$,\ul{vcn}[\ ] and \ul{bcn}[\ ]. 
\ul{ddverno}[\ ]
and \ul{hverno}[\ ] discriminate them.
        }
\label{fig13}
\end{figure}

In order to discriminate G18 and G20 of Fig.\ref{fig13}, 
we introduce the number
of dd-vertices (\ul{ddverno}[$i$],$i=1,2,\cdots,\ul{l}$) 
and that of h-vertices (\ul{hverno}[$i$],$i=1,2,\cdots,\ul{l}$)
for each loop $i$ as an index\cite{foot4}.
%\footnote{
%The introduced indices here are the same as 
%those explained in Lemma of Sec.2:\ 
%\ul{ddverno}[$i$]=$w_i$,\ul{hverno}[$i$]=$v_i$. 
%}
%%%%%%%%%%%%%%%%%%%%%%%  Fig.13  %%%%%%%%%%%%%%%%%%%%%%%%%%%%%%%%%%%%
%%%%%%%%%%%%%%%%%%%%%%%%%%%%%%%%%%%%%%%%%%%%%%%%%%%%%%%%%%%%%%%%%%%%%
The same situation occurs in G17($2B\Si_1$) and G19($2C\Si_1$).

%%%%%%%%%%%%%%%%%%%  (iii) No of crossing          %%%%%%%%%%%%%%%%
\flushleft{
(iii) Number of crossing (\ul{crossno}[\ ])  }

%%%%%%%%%%%%%%%%%%%%%%%%   Def    %%%%%%%%%%%%%%%%%%%%%%%%%%%%%%%%%%%%%%%
\begin{description}
\item[Def]\ 
When a bond has its both ends ( dd-vertex and h-vertex ) in a
same suffix-loop, we call it {\it loop-bond } of the suffix-loop.
\end{description}
%%%%%%%%%%%%%%%%%%%%%%%%%%%%%%%%%%%%%%%%%%%%%%%%%%%%%%%%%%%%%%%%%%%%%%%%%%
\begin{description}
\item[Def]\ 
We consider $i$-th suffix-loop in a graph of $(\pl\pl h)^s$-invariant
($s\geq 2$). Let the suffix-loop have $r$\ ($0\leq r\leq s$)
loop-bonds.
There are $r(r-1)/2$\ pairs of them. For
each pair, whether they are ``crossed'' or
``not crossed'' is definitely defined by tracing the vertices of
both loop-bonds along the suffix-loop in a fixed direction.
We define, as the total number of the crossed pairs, {\it crossing
number} (\ul{crossno}[$i$];\ $i=1,2,\cdots,\ul{l}$) 
of the $i$-th suffix-loop.
\end{description}
%%%%%%%%%%%%%%%%%%%%%%%%%%%%%%%%%%%%%%%%%%%%%%%%%%%%%%%%%%%%%%%%%%%%%%%%%%
The following are examples.\  
Fig.\ref{fig10}:\ \ul{crossno}[1]=0,\ \ul{crossno}[2]=0,\ 
\ul{crossno}[3]=0;\ 
Fig.\ref{fig11}:\ \ul{crossno}[1]=1 for G9, G5 and G6;\ 
Fig.\ref{fig12}:\ \ul{crossno}[1]=1, \ul{crossno}[2]=0 for G28 and G30. 

\begin{figure}
     \centerline{
\shortstack{
{\epsfxsize=3.81cm  \epsfysize=3.32cm \epsfbox{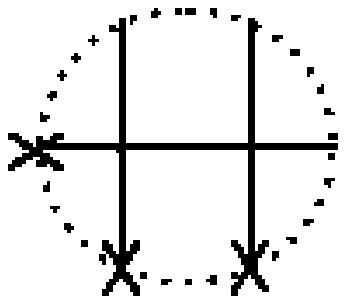}}
\\ G12:\ $1C\Om_1$
            } 
\q\q
\shortstack{
{\epsfxsize=3.77cm  \epsfysize=3.42cm \epsfbox{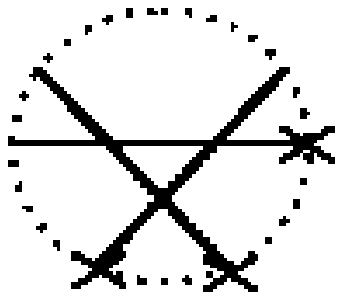}}
\\ G13:\ $1C\Om_2$
            }
                 }
\caption{
%*fig14*Fig.14\ 
Two Graphs with the same $\ul{l}$,\ul{vcn} and \ul{bcn}. \ul{crossno}[\ ]
discriminates them.
        }
\label{fig14}
\end{figure}

G12 and G13 in Fig.\ref{fig14} 
are discriminated by \ul{crossno}[$i$]:\ 
\ul{crossno}[1]=2 for G12, whereas \ul{crossno}[1]=3 for G13.

%%%%%%%%%%%%%%%%%%%%%%%  Fig.14  %%%%%%%%%%%%%%%%%%%%%%%%%%%%%%%%%%%%
%%%%%%%%%%%%%%%%%%%%%%%%%%%%%%%%%%%%%%%%%%%%%%%%%%%%%%%%%%%%%%%%%%%%%

%%%%%%%%%%%%%%%%%%%  (iv) connectivity and disconnectivity  %%%%%%%%%%%%%%%%
\flushleft{
(iv)  \ul{connectivity} and \ul{disconnectivity}  }

%%%%%%%%%%%%%%%%%%%%%%%%   Def    %%%%%%%%%%%%%%%%%%%%%%%%%%%%%%%%%%%%%%%
\begin{description}
\item[Def]\ 
Let us consider a graph of 
$(\pl\pl h)^s$-invariant ($s\geq 2$). There are
$s$\ bonds and $s(s-1)/2$ different pairs of bonds. We define
\ul{connectivity} of the graph as  
the total number of those pairs which are connected by at least
one suffix-line. 
$0\leq \mbox{\ul{connectivity}} \leq s(s-1)/2$. 
\end{description}
%%%%%%%%%%%%%%%%%%%%%%%%%%%%%%%%%%%%%%%%%%%%%%%%%%%%%%%%%%%%%%%%%%%%%%%%%%
As examples, we have the following:\ 
$A1$\ (Fig.\ref{fig2}) for \ul{connectivity}=1;\ 
$PQ$\ (Fig.\ref{fig4}) for \ul{connectivity}=0;\ 
$G51$\ (Fig.\ref{fig10}) for \ul{connectivity}=2;\ 
$G5,G6,G9$\ (Fig.\ref{fig11}) for \ul{connectivity}=3. 

%%%%%%%%%%%%%%%%%%%%%%%%   Def    %%%%%%%%%%%%%%%%%%%%%%%%%%%%%%%%%%%%%%%
\begin{description}
\item[Def]\ 
Let us consider a graph of 
$(\pl\pl h)^s$-invariant ($s\geq 1$). Among
$s$\ bonds, we identify those which are connected
by at least one suffix-line. Let us define, as the
total number of bonds after the identification,
\ul{disconnectivity}+1. 
$0\leq \mbox{\ul{disconnectivity}} \leq s-1$. 
\end{description}
%%%%%%%%%%%%%%%%%%%%%%%%%%%%%%%%%%%%%%%%%%%%%%%%%%%%%%%%%%%%%%%%%%%%%%%%%%
As examples, we have the following:\ 
$A1$\ (Fig.\ref{fig2}) for \ul{disconnectivity}=0;\ 
$PQ$\ (Fig.\ref{fig4}) for \ul{disconnectivity}=1;\ 
$G23=A2Q$\ (App.A) for \ul{disconnectivity}=1;\ 
$G69=QQQ$\ (App.A) for \ul{disconnectivity}=2.

Two graphs $G71$ and $G72$ in App.A are 
examples which are discriminated
by \ul{disconnectivity}.
%%%%%%%%%%%%%%%%%%%  (v) bondno[]  %%%%%%%%%%%%%%%%
%\flushleft{
%(v)  \ul{bondno}[\ ]   }
%need to describe it ??**************
%List of Indices for above 19 terms of Table 1.** Yameru ???

\q The list of indices for all $(\pl\pl h)^3$-invariants is provided
in App.B.

%%%%%%%%%%%%%%%%%%%%%%%%%%%%%%%%%%%%%%%%%%%%%%%%%%%%%%%%%%%%%%%%%%%
%%%%%%%%%%%%%%%%%%%%%%%%%%  SEC 6  %%%%%%%%%%%%%%%%%%%%%%%%%%%%%%%
%%%%%%%%%%%%%%%%%%%%%%%%%%%%%%%%%%%%%%%%%%%%%%%%%%%%%%%%%%%%%%%%%%%
\section{
Identities between Weights
        }

Let us define the {\it weight} of a graph in the present case.
(See paper (I) for a more general case.)
%%%%%%%%%%%%%%%%%%%%%%%%   Def    %%%%%%%%%%%%%%%%%%%%%%%%%%%%%%%%%%%%%%%
\begin{description}
\item[Def]\ 
Let us consider a graph of 
$(\pl\pl h)^s$-invariant ($s\geq 1$). 
There are several ways to obtain the invariant 
from $s$\ different 4-tensors  
($\pl_{\m_1}\pl_{\n_1}h_{\al_1\be_1},
\cdots,\pl_{\m_s}\pl_{\n_s}h_{\al_s\be_s}$) 
by contracting $4s$\ different suffixes.
We define the {\it weight} of the graph
as the number of all possible ways to
obtain the invariant.
\end{description}
%%%%%%%%%%%%%%%%%%%%%%%%%%%%%%%%%%%%%%%%%%%%%%%%%%%%%%%%%%%%%%%%%%%%%%%%%%

In App.B, all independent $(\pl\pl h)^3$-invariants are listed up
with weights\cite{foot5}
%\footnote{
%As a general tendency, we see, from the result of App.B, 
%the weight of a graph decreases as its suffix-loop no (\ul{$l$})
%increases. 
%         }
. 
The total sum of all weights satisfies a
meaningful relation.
%****(weight.1)%%%%%%%%%%%%%%%%%%%%
\begin{eqnarray}
10395\ (=11\times 9\times 7\times 5\times 3\times 1)=\nn\\
 384\times 9\ (G4,G5,G6,G7,G8,G9,G11,G12,G16)\nn\\
+192\times 17\ (G2,G10,G13,G15,G17,G19,G21,G22,G25,\nn\\
G29,G30,G31,G32,G37,G38,G41,G42)\nn\\
+128\times 1\ (G1)
+96\times 26\ (G18,G20,G23,G24,G26,G27,G28,G33,G34,\nn\\
G35,G36,G39,G40,G44,G48,G51,G52,G55,\nn\\
G58,G59,G60,G61,G62,G64,G65,G66)+64\times 2\ (G3,G14)\label{weight.1}\\
+48\times 12\ (G43,G45,G46,G47,G49,G53,G54,G56,G57,G63,G74,G81)\nn\\
+24\times 9\ (G50,G68,G71,G73,G75,G78,G80,G82,G83)+16\times 1\ (G67)\nn\\
+12\times 6\ (G72,G76,G79,G84,G85,G89)+8\times 3\ (G69,G70,G77)\nn\\
+6\times 3\ (G86,G87,G88)
+1\times 1\ (G90)\pr\nn
\end{eqnarray}
%%%%%%%%%%%%%%%%%%%%%%%%%%%%%
This relation shows the completeness of the listing of App.A.

\q Furthermore we can see the structure of classification in 
relations between weights. In Sec.IV we have used two reduction
procedures,Fig.\ref{fig6} and Fig.\ref{fig8}. 
The procedure of Fig.\ref{fig6} reduce 66 connected graphs
(see App.A) to
19 ones cited in Fig.\ref{f7p1}-\ref{f7p4}. 
That of Fig.\ref{fig8} reduces the 19 graphs
to 4 ones cited in Fig.\ref{fig9}.
The following examples of relations between
weights clearly show the classification structure.
%*******Weight Structure 66 -$>$ 19 -$>$ 4   ******
%****(weight.2a,2b,2c,2d)%%%%%%%%%%%%%%%%%%%%
\begin{eqnarray}
\mbox{(i) 66 $\ra$ 19} & \nn\\
 &
4^3\times 4\times 2\ (1\Th)\ =512 
                     =128\ (1A\Th)\ +384\ (1B\Th)\ ,\label{weight.2a}\\
 &
2^2\times 3\times 2\times 2^2\ (4\Xi)\ =96\nn\\ 
              &       =24\ (4B\Xi)\ +24\ (4D\Xi)+24\ (4F\Xi)
                               +24\ (4H\Xi)\ .\label{weight.2b}\\
&\nn \\
\mbox{(ii) 19 $\ra$ 4}& \nn\\
 &
(\ _4C_2)^3\times 2^3\ (\Th)\ =1728 
          =512\ (1\Th)\ +768\ (2\Th)+384\ (3\Th)
                              +64\ (4\Th)\ ,\label{weight.2c}\\
 &
3^3\times (4\times 2)\times 2^3\ (\Om)\ =1728\nn\\ 
              &     =768\ (1\Om_1)\ +256\ (1\Om_2)
            +256\ (2\Om_1)+384\ (2\Om_2)\ +64\ (3\Om)\ .
                                           \label{weight.2d}\\
&\nn\\
\mbox{(iii) 66 $\ra$ 4} & \nn\\
 &
10395-891(\mbox{discon}) =9504
               =1728(\Om)+3456(\Si)+2592(\Xi)+1728(\Th)\ ,
                                               \label{weight.2e}
\end{eqnarray}
%%%%%%%%%%%%%%%%%%%%%%%%%%%%%
where the total weight for the disconnected part (891) will be
explained in Sec.VII.

\begin{figure}
   \centerline{
{\epsfxsize=88pt  \epsfysize=85pt \epsfbox{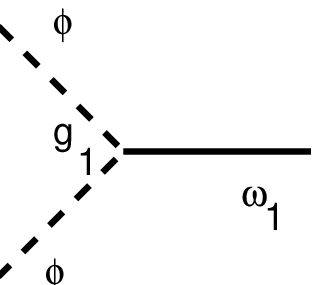}}\qqq
{\epsfxsize=87pt  \epsfysize=87pt \epsfbox{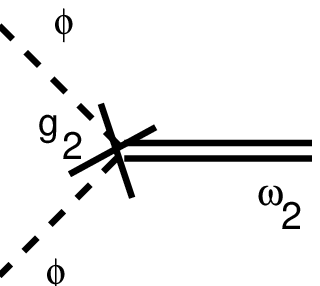}}\qqq
{\epsfxsize=75pt  \epsfysize=91pt \epsfbox{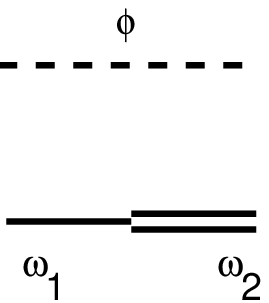}}
               }
\caption{
%*fig15p1*Fig.15.1\ 
Vertices and propagators of (\protect\ref{weight.3}).
        }
\label{fig15p1}
\end{figure}

\begin{figure}
   \centerline{
{\epsfxsize=90pt  \epsfysize=85pt \epsfbox{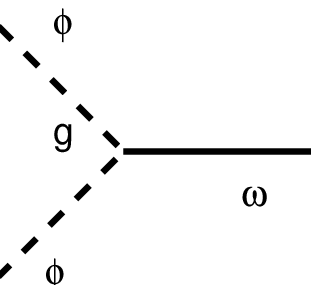}}\qqq
{\epsfxsize=53pt  \epsfysize=77pt \epsfbox{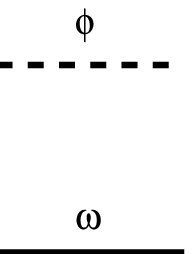}}
               }
\caption{
%*fig15p2*Fig.15.2\ 
Vertices and propagators of (\protect\ref{weight.4}).
        }
\label{fig15p2}
\end{figure}

\begin{figure}
   \centerline{
{\epsfxsize=53pt  \epsfysize=43pt \epsfbox{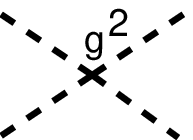}}
               }
\caption{
%*fig15p3*Fig.15.3\ 
The vertex  of (\protect\ref{weight.5}).
        }
\label{fig15p3}
\end{figure}

\q We can simply understand the above relations in the field theory language.
The initial 66 connected diagrams are produced by connected Feynman diagrams
of the following lagrangian. (See a general field theory text book.)
%****(weight.3)%%%%%%%%%%%%%%%%%%%%
\begin{eqnarray}
\Lcal=\Lcal_0+\Lcal_I\com\nn\\
\Lcal_0=\half\p^2+\om_1\om_2\com\nn\\
\Lcal_I=g_1\p^2\om_1+g_2\p^2\om_2\pr
     \label{weight.3}
\end{eqnarray}
%%%%%%%%%%%%%%%%%%%%%%%%%%%%%
(Of course the 24 disconnected ones (see App.A) are produced as
the disconnected Feynman diagrams.)
The vertices and propagators are shown in Fig.\ref{fig15p1}.
%%%%%%%%%%%%%%%%%%%%%%%  Fig.15.1  %%%%%%%%%%%%%%%%%%%%%%%%%%%%%%%%%%%%
%%%%%%%%%%%%%%%%%%%%%%%%%%%%%%%%%%%%%%%%%%%%%%%%%%%%%%%%%%%%%%%%%%%%%
The first reduction procedure (Fig.\ref{fig6}) corresponds to
taking $g_1=g_2=g/\sqrt{2},\ \om_1=\om_2\equiv\om/\sqrt{2}$ in
(\ref{weight.3}).
%****(weight.4)%%%%%%%%%%%%%%%%%%%%
\begin{eqnarray}
\Lcal'={\Lcal_0}'+{\Lcal_I}'\com\nn\\
{\Lcal_0}'=\half\p^2+\half\om^2\com\nn\\
{\Lcal_I}'=g\p^2\om\pr
     \label{weight.4}
\end{eqnarray}
%%%%%%%%%%%%%%%%%%%%%%%%%%%%%
The vertices and propagators are shown in Fig.\ref{fig15p2}.
%%%%%%%%%%%%%%%%%%%%%%%  Fig.15.2  %%%%%%%%%%%%%%%%%%%%%%%%%%%%%%%%%%%%
%%%%%%%%%%%%%%%%%%%%%%%%%%%%%%%%%%%%%%%%%%%%%%%%%%%%%%%%%%%%%%%%%%%%%
The 19 diagrams of Fig.\ref{f7p1}-\ref{f7p4}
are produced from the above lagrangian.
Integrating out the $\om$-integral, we obtain an effective action
$\Lcal^{\mbox{eff}}$.
%****(weight.5)%%%%%%%%%%%%%%%%%%%%
\begin{eqnarray}
\int\Dcal\om \exp\{\int d^n x(
\half\p^2+\half\om^2+g\p^2\om) \}\sim
\exp\{\int d^n x(\half\p^2-\half g^2\p^4)\}\nn\\
\equiv\exp\{\int d^n x \Lcal^{\mbox{eff}}\}
\pr     \label{weight.5}
\end{eqnarray}
%%%%%%%%%%%%%%%%%%%%%%%%%%%%%
Fig.\ref{fig15p3} shows the vertex graphically.
%%%%%%%%%%%%%%%%%%%%%%%  Fig.15.3  %%%%%%%%%%%%%%%%%%%%%%%%%%%%%%%%%%%%
%%%%%%%%%%%%%%%%%%%%%%%%%%%%%%%%%%%%%%%%%%%%%%%%%%%%%%%%%%%%%%%%%%%%%
This corresponds to the second reduction procedure of Fig.\ref{fig8}. 
In fact
the $g^6$-order connected Feynman diagrams of (\ref{weight.5})
produces the diagrams of Fig.\ref{fig9}.

\q This field theory approach is important 
when we treat general invariants with
higher mass-dimensional ( $M^8$,$M^{10}$, ...) cases.
We comment on further generalization in Sec.X. 

%%%%%%%%%%%%%%%%%%%%%%%%%%%%%%%%%%%%%%%%%%%%%%%%%%%%%%%%%%%%%%%%%%%
%%%%%%%%%%%%%%%%%%%%%%%%%%  SEC 7  %%%%%%%%%%%%%%%%%%%%%%%%%%%%%%%
%%%%%%%%%%%%%%%%%%%%%%%%%%%%%%%%%%%%%%%%%%%%%%%%%%%%%%%%%%%%%%%%%%%
\section{
Disconnected part
         }

Among the 90 invariants listed in App.A, there are 24 disconnected graphs.
They are composed of lower dimensional invariants ($\pl\pl h$,
$(\pl\pl h)^2$) treated in paper (I). The disconnected invariants
are classified by \ul{disconnectivity}.
\flushleft{i) \ul{disconnectivity}=2,\q 4 terms}

We have the 4 terms as listed in Table \ref{tab2}.
%%%%%%%%%%%%%%%%%%%%%%%%%%%%%%%%%%%%%%%%%%%%%%%%%%%%%%%%%%%%%%%%%%%%%%%%%%
%%%%%%%%%%%%%%%%%%%%%  Table 2   %%%%%%%%%%%%%%%%%%%%%%%%%%%%%%%%%%%%%%%%%
%%%%%%%%%%%%%%%%%%%%%%%%%%%%%%%%%%%%%%%%%%%%%%%%%%%%%%%%%%%%%%%%%%%%%%%%%%
\begin{table}
\begin{tabular}{|c|c|c|c|c|}
\hline
        & G69: QQQ    & G85: PQQ               & G88: PPQ    & G90: PPP \\
\hline
weight  & $2^3=8$    & $1\times 2^2\times 3=12$ & $1^2\times 2\times 3=6$ 
                                                             & $1^3=1$      \\
\hline
%\multicolumn{5}{c}{\q}                                                 \\
%\multicolumn{5}{c}{Table 2\ \  $(\pl\pl h)^3$-terms of disconnectivity=2
%and their weights}\\
%\multicolumn{5}{c}{}\\
\end{tabular}
\caption{
%*tab2*
$(\pl\pl h)^3$-terms of disconnectivity=2
and their weights}
\label{tab2}
\end{table}
%%%%%%%%%%%%%%%%%%%%%%%%%  END  of  Table 2 %%%%%%%%%%%%%%%%%%%%%%%%%%%%%
%%%%%%%%%%%%%%%%%%%%%%%%%%%%%%%%%%%%%%%%%%%%%%%%%%%%%%%%%%%%%%%%%%%%%%%%%
The total weight for \ul{disconnectivity}=2 is 27.

\flushleft{ii) \ul{disconnectivity}=1,\q 20 terms}

We have the 20 terms as listed in Table \ref{tab3}.
%%%%%%%%%%%%%%%%%%%%%%%%%%%%%%%%%%%%%%%%%%%%%%%%%%%%%%%%%%%%%%%%%%%%%%%%%%
%%%%%%%%%%%%%%%%%%%%%  Table 3   %%%%%%%%%%%%%%%%%%%%%%%%%%%%%%%%%%%%%%%%%
%%%%%%%%%%%%%%%%%%%%%%%%%%%%%%%%%%%%%%%%%%%%%%%%%%%%%%%%%%%%%%%%%%%%%%%%%%
\begin{table}
\begin{tabular}{|c|c|c|}
\hline
           &  Q\com\q 2              &  P\com\q 1 \\
\hline
A1\com 16  & G24:\ A1Q\com\q$16\times 2\times 3=96$
                                     & G54:\ A1P\com\q$16\times 3=48$      \\
\hline
A2\com 16  & G23:\ A2Q\com\q$16\times 2\times 3=96$
                                     & G53:\ A2P\com\q$16\times 3=48$      \\
\hline
A3\com 16  & G26:\ A3Q\com\q$16\times 2\times 3=96$
                                     & G56:\ A3P\com\q$16\times 3=48$      \\
\hline
\hline
B1\com 16  & G66:\ B1Q\com\q$16\times 2\times 3=96$
                                     & G81:\ B1P\com\q$16\times 3=48$      \\
\hline
B2\com 16  & G61:\ B2Q\com\q$16\times 2\times 3=96$
                                     & G74:\ B2P\com\q$16\times 3=48$      \\
\hline
B3\com 4   & G50:\ B3Q\com\q$4\times 2\times 3=24$
                                     & G72:\ B3P\com\q$4\times 3=12$      \\
\hline
B4\com 4   & G68:\ B4Q\com\q$4\times 2\times 3=24$
                                     & G84:\ B4P\com\q$4\times 3=12$      \\
\hline
\hline
C1\com 2   & G79:\ C1Q\com\q$2\times 2\times 3=12$
                                     & G87:\ C1P\com\q$2\times 3=6$      \\
\hline
C2\com 2   & G76:\ C2Q\com\q$2\times 2\times 3=12$
                                     & G86:\ C2P\com\q$2\times 3=6$      \\
\hline
C3\com 4   & G83:\ C3Q\com\q$4\times 2\times 3=24$
                                     & G89:\ C3P\com\q$4\times 3=12$      \\
\hline
%\multicolumn{3}{c}{\q}                                                 \\
%\multicolumn{3}{c}{Table 3\ \  $(\pl\pl h)^3$-terms of disconnectivity=1.
%Numbers are weights.
%}\\
%\multicolumn{3}{c}{(A1$\sim$C3) are connected $(\pl\pl h)^2$-invariants.
%Q and P are $\pl\pl h$-invariants.}\\
\end{tabular}
\caption{
%*tab3*
$(\pl\pl h)^3$-terms of disconnectivity=1.
Numbers are weights.
(A1$\sim$C3) are connected $(\pl\pl h)^2$-invariants.
Q and P are $\pl\pl h$-invariants.}
\label{tab3}
\end{table}
%%%%%%%%%%%%%%%%%%%%%%%%%  END  of  Table 3 %%%%%%%%%%%%%%%%%%%%%%%%%%%%%
%%%%%%%%%%%%%%%%%%%%%%%%%%%%%%%%%%%%%%%%%%%%%%%%%%%%%%%%%%%%%%%%%%%%%%%%%
In Table \ref{tab3}, A1-C3 are $(\pl\pl h)^2$-invariants and Q and P are
$\pl\pl h$-invariants ( Sec.II) .
The total weight for graphs with \ul{disconnectivity}=1 is 864.

\vs {0.5}
Summing (i) and (ii), we see the total weight for the disconnected
graphs is 891.

\vs 1
\q In sections from II to VII, we have explained the classification
of $(\pl\pl h)^3$-invariants only. 
Other types of SO(n)-invariants
are classified in App.C (for $\pl^4h\cdot \pl^2h$-invariants)
and in App.D (for $\pl^3h\cdot\pl^3h$-invariants).
%%%%%%%%%%%%%%%%%%%%%%%%%%%%%%%%%%%%%%%%%%%%%%%%%%%%%%%%%%%%%%%%%%%
%%%%%%%%%%%%%%%%%%%%%%%%%%  SEC 8  %%%%%%%%%%%%%%%%%%%%%%%%%%%%%%%%%%%%
%%%%%%%%%%%%%%%%%%%%%%%%%%%%%%%%%%%%%%%%%%%%%%%%%%%%%%%%%%%%%%%%%%%
\section{Independence of General Invariants}

So far we have discussed the global SO(n)-invariants which appear in
the weak-field perturbation of gravity. In this section we discuss
properties of general invariants themselves. 
We consider the general space-dimension. Therefore ``independence''
in this section means that in the general space-dimension.
%%%%%%%%%%%%%%%%%%%%%%%%%  8.1   %%%%%%%%%%%%%%%%%%%%%%%%%%%%%%
\subsection{Graphical Representation of General Tensors
and Invariants}
\begin{figure}
     \centerline{
{\epsfxsize=105pt  \epsfysize=96pt \epsfbox{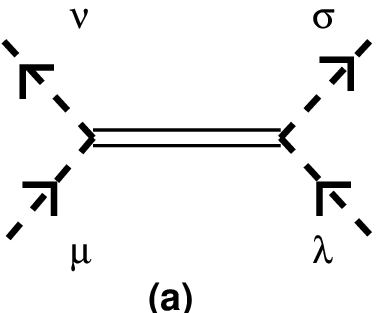}}
\q
{\epsfxsize=107pt  \epsfysize=105pt \epsfbox{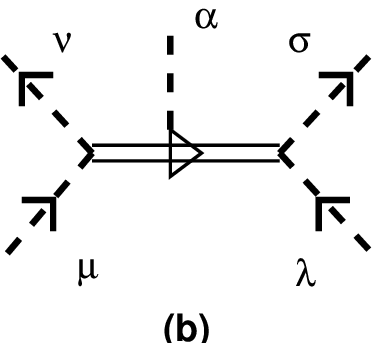}}
\q
{\epsfxsize=107pt  \epsfysize=107pt \epsfbox{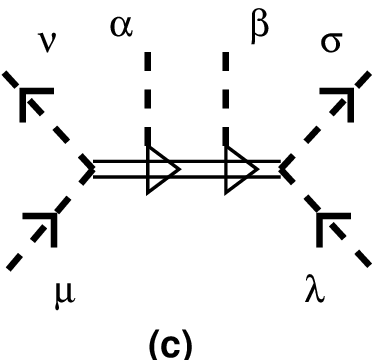}}
                 }
\caption{
%*fig8p1*Fig.8.1\
Graphical representations for
\ (a)\ $R_{\mn\ls}$,\ (b)\ $\na_\al R_{\mn\ls}$\  
and\ (c)\ $\na_\al\na_\be R_{\mn\ls}$.
        }
\label{fig8p1}
\end{figure}

\begin{figure}
     \centerline{
{\epsfxsize=71pt  \epsfysize=71pt \epsfbox{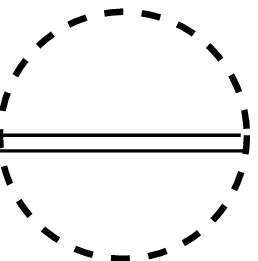}}
                 }
\caption{
%*fig8p2*Fig.8.2\ 
Graphical representation for the Riemann scalar $R$.
        }
\label{fig8p2}
\end{figure}

\begin{figure}
     \centerline{
{\epsfxsize=71pt  \epsfysize=96pt \epsfbox{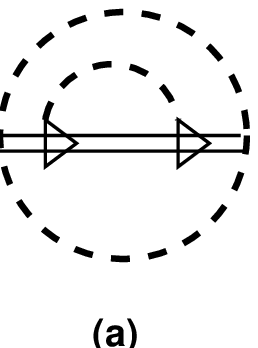}}
{\epsfxsize=115pt  \epsfysize=68pt \epsfbox{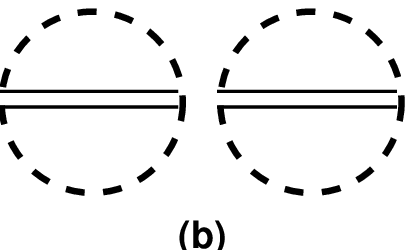}}
{\epsfxsize=71pt  \epsfysize=95pt \epsfbox{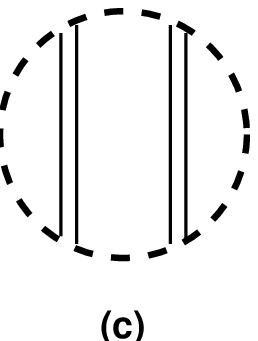}}
{\epsfxsize=131pt  \epsfysize=68pt \epsfbox{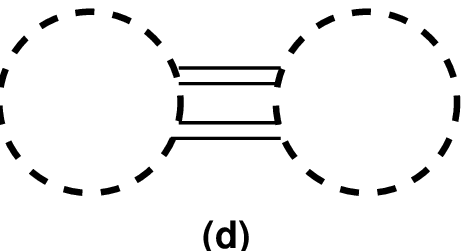}}
                 }
\caption{
%*fig8p3*Fig.8.3\ 
Graphical representation for
\ (a)\ $\na^2R$, (b)\ $R^2$, (c)\ $R_\mn R^\mn$,  
and (d)\ $R_{\mn\ls}R^{\mn\ls}$.
        }
\label{fig8p3}
\end{figure}

In this case also,
a graphical representation is very useful\cite{SI}.
We briefly explain the representation necessary for the present
explanation. 
We can express
\ $R_{\mn\ls},\ \na_\al R_{\mn\ls} \mbox{and}\ \na_\al\na_\be
R_{\mn\ls}$\ 
as in Fig.\ref{fig8p1}\cite{foot6}.
%\footnote{
%In contrast with \cite{SI},
%we take here double solid lines to express the 'bond' of $R_{\mn\ls}$
%in order to avoid the confusion with $\pl_\m\pl_\n h_\ab$ of 
%Fig.\ref{fig1}.
%         }
%%%%%%%%%%%%%%%%%%%%%%%  Fig.8.1  %%%%%%%%%%%%%%%%%%%%%%%%%%%%%%%%%%%%
%%%%%%%%%%%%%%%%%%%%%%%%%%%%%%%%%%%%%%%%%%%%%%%%%%%%%%%%%%%%%%%%%%%%%
%
They represent their own suffix-permutation symmetries.
The arrows are introduced there, 
and we have a simple rule: when we change the direction of arrows,
we change the sign of the overall factor. 
This expresses the
(anti-)symmetric properties:
$R_{\mn\ls}=-R_{\n\m\ls}=-R_{\mn\si\la}=+R_{\n\m\si\la}$.
%(see \cite{SI} for detail).
Relations
between general invariants, like the Bianchi identity, are
introduced as graphical rules. 
%They are so useful to treat general invariants. 
We now examine {\it local} general invariants which are made of
$\na_\m, R_{\mn\ls}$ and $g_\mn$. As for those with lower mass dimensions,
the independent ones are well known due to much experience in the past
literature. For $M^2$-dimension, we have
%****(ind.1)%%%%%%%%%%%%%%%%%%%%
\begin{eqnarray}
R\com                                \label{ind.1}
\end{eqnarray}
%%%%%%%%%%%%%%%%%%%%%%%%%%%%%
as a unique general invariant ( except a cosmological constant). 
It is graphically
represented as in Fig.\ref{fig8p2}. Generally suffix-lines(dotted lines)
are closed for general invariants.
%%%%%%%%%%%%%%%%%%%%%%%  Fig.8.2  %%%%%%%%%%%%%%%%%%%%%%%%%%%%%%%%%%%%
%%%%%%%%%%%%%%%%%%%%%%%%%%%%%%%%%%%%%%%%%%%%%%%%%%%%%%%%%%%%%%%%%%%%%
When a closed suffix-loop has an even number of vertices, the graph is
invariant under the change of the direction of  arrows. 
In this case
we may drop the arrow in the graph ( see Fig.\ref{fig8p2}) .
For $M^4$-dimension, we have 4 independent
ones.
%****(ind.2)%%%%%%%%%%%%%%%%%%%%
\begin{eqnarray}
\na^2R\com\q R^2\com\q R_\mn R^\mn\com\q R_{\mn\ls}R^{\mn\ls}\pr \label{ind.2}
\end{eqnarray}
%%%%%%%%%%%%%%%%%%%%%%%%%%%%%
They are graphically represented as in Fig.\ref{fig8p3}.
%%%%%%%%%%%%%%%%%%%%%%%  Fig.8.3  %%%%%%%%%%%%%%%%%%%%%%%%%%%%%%%%%%%%
%%%%%%%%%%%%%%%%%%%%%%%%%%%%%%%%%%%%%%%%%%%%%%%%%%%%%%%%%%%%%%%%%%%%%
The independence of 4 terms of (\ref{ind.2}) is well known.
They are those terms which appear in the Weyl anomaly in
4 dim gravity-matter theories.
A proper proof of their independence, 
in terms of weak-expansion and its graphical representation, 
is given in paper (I).
%%%%%%%%%%%%%%%%%%%%%%%%%  8.2  %%%%%%%%%%%%%%%%%%%%%%%%%%%%%%%%%%%
\subsection{Independent General Invariants as Local Functions}

As for higher mass-dimensional cases,
listing up all independent invariants is not easy. 
We must take into account all relations such as the Bianchi identity
and the cyclic relation.
The graphical
representation helps greatly\cite{SI}. 
Using this method,
we can easily list up the following 17 invariants 
(which were obtained, by the ordinary method, in
\cite{G, BPB})
as the finally reduced $M^6$-invariants\cite{foot7,foot8}.
%\footnote{
%Note that in \cite{SI}, total derivative terms are neglected, whereas
%they are not neglected in the present case.
%          }
%\footnote{
%\{$P_1-P_6$\},$A_1$,$B_1$ and \{$O_1-O_2$\} are the same notation 
%as that in \cite{SI}.
%$O_3$ and $O_4$ in (\ref{ind.3}) correspond to $Q_2$ and $Q_{10}$ in \cite{SI}
%respectively.(There is a typographical mistake in the figure caption
%of Fig.15 of \cite{SI} where a minus sign is missing in the literal
%(not graphical) definition of $P_5$. The one given in (\ref{ind.3})
%of the present text is the right one.)
%          }
%****(ind.3)%%%%%%%%%%%%%%%%%%%%
\begin{eqnarray}
P_1=RRR\com\ \ P_2=RR_\mn R^\mn\com\ \ P_3=RR_{\mn\ls}R^{\mn\ls}\com \nn\\
P_4=R_\mn R^{\n\la}R_\la^{~\mu}\com\ \ P_5=-R_{\mn\ls}R^{\mu\la}R^{\nu\si}
\com\ \ P_6=R_{\mn\ls}R_\tau^{~\nu\ls}R^{\mu\tau}\com\nn\\
A_1=R_{\mn\ls}R^{\si\la}_{~~~\tau\om}R^{\om\tau\n\m}\com\ \ 
B_1=R_{\mn\tau\si}R^{\n~~~\tau}_{~\la\om}R^{\la\mu\si\om}\com\nn\\
O_1=\na^\mu R\cdot \na_\mu R\com\ \ 
O_2=\na^\mu R_\ls\cdot \na_\mu R^\ls\com\nn\\ 
O_3=\na^\mu R^{\la\rho\si\tau}\cdot \na_\mu R_{\la\rho\si\tau}\com\ \ 
O_4=\na^\mu R_{\la\n}\cdot \na^\n R^\la_{~\m}\com\nn\\ 
T_1=\na^2R\cdot R\com\ \ T_2=\na^2R_\ls\cdot R^\ls\com\ \ 
T_3=\na^2R_{\la\rho\si\tau}\cdot R^{\la\rho\si\tau}\com\ \ \nn\\
T_4=\na^\m\na^\n R\cdot R_\mn\com\ \nn\\ 
%T_5=\na^\m\na^\n R^{\la\rho}\cdot R_{\mu\la\n\rho}\com\nn\\
S=\na^2\na^2R\pr
                               \label{ind.3}
\end{eqnarray}
%%%%%%%%%%%%%%%%%%%%%%%%%%%%%
The above 17 terms are graphically given in App.F.
The above listing, however,  
%guarantees only that
%the all invariants are given by a sum of
%some terms among (\ref{ind.3}).
does not guarantee that  all terms of (\ref{ind.3}) are independent each
other. 
We do not have  a proper basis in the 'full
metric' treatment, which makes it difficult to show the independence. 
%This is the point which has not been examined so much.
%The independence check has been relied only on 
%results of specific problems or specific geometries.
%It is not good when we make general analysis on the gravitational system.
As an application of the results 
about the classification of SO(n)-invariants (Sec.II-VII),
we can prove the independence of the above 17 terms of (\ref{ind.3})
for a general geometry in a general space-dimension. 
In order to show the independence as a {\it local} function,
we can safely use the weak field expansion:\ 
$g_\mn=\del_\mn+h_\mn\ ,\ |h_\mn|\ll 1$.

%%%%%%%%%%%%%%%%%%%%%%%%%%%% (i)  S  %%%%%%%%%%%%%%%%%%%%%%%%%%%%%%%%%%%%%%%%
\flushleft{(i)\ $S=\na^2\na^2 R$}
\q The leading order is given by $\pl^6 h\sim O(h)$.
%****(ind.4)%%%%%%%%%%%%%%%%%%%%
\begin{eqnarray}
S=\pl^2\pl^2(\pl^2h-\pl_\m\pl_\n h_\mn)+O(h^2)\pr   \label{ind.4}
\end{eqnarray}
%%%%%%%%%%%%%%%%%%%%%%%%%%%%%
Other terms do not have $O(h)$ contribution, therefore $S$ is independent
from others.
%%%%%%%%%%%%%%%%%%%%%%%%%%%% (ii)  T1-T4  %%%%%%%%%%%%%%%%%%%%%%%%%%%%%%%%%%%%%
\flushleft{(ii)\ $T_1\sim T_4$}
\q
The leading order of every term is $\pl^4h\times\pl^2h\sim O(h^2)$.
The classification of $\pl^4 h\cdot\pl^2 h$-invariants
are given in App.C.
The expansions of $T_1\sim T_4$, in terms of 
$\pl^4 h\cdot\pl^2 h$-invariants 
are also given there. The explicit forms of their expansions show 
that the 4 terms are independent. 
Because other terms, except $S$, do not
contribute to terms of this type,
we see they can be taken as independent terms.
%%%%%%%%%%%%%%%%%%%%%%%%%%%% (iii)  O1-O4  %%%%%%%%%%%%%%%%%%%%%%%%%%%%%%%%%%%
\flushleft{(iii)\ $O_1\sim O_4$}
\q
The leading order of every term is given by $\pl^3h\times\pl^3h\sim O(h^2)$.
In App.D, the classification of $\pl^3 h\cdot \pl^3 h$-invariants 
are given. The expansions of $O_1\sim O_4$, in terms of
them, are also obtained explicitly. Their explicit forms show the 4
terms are independent. 
In the similar way to (ii), we see
they can be taken as independent terms.
%%%%%%%%%%%%%%%%%%%%%%%%%%%% (iv)  P1-P6, A1,B1   %%%%%%%%%%%%%%%%%%%%%%%%%%%%
\flushleft{(iv)\ $P_1\sim P_6,A_1,B_1$}
\q
The leading order of every term is given by $(\pl\pl h)^3\sim O(h^3)$.
$(\pl\pl h)^3$-invariants 
are completely classified in the text, and the results 
(especially the set of indices) allow us to easily
calculate (by computer) the weak-field expansion.
This shows the power of the present classification.
The result is given in App.E, which shows the 8 terms $\{ P_1\sim P_6,
A_1,B_1\}$ are  independent each other.
Furthermore they are  ``orthogonal'' in the
space of 90 terms except the G3 and G13 ``directions''.($A_1$ and $B_1$
have common components to G3 and G13 ``directions''. 
If some inner product can be defined in this ``vector'' space, 
orthogonal ones could be chosen by taking some linear combinations
of $A_1$ and $B_1$. )

\vspace{1cm}

From (i)$\sim$(iv), we may say the 17 terms of (\ref{ind.3}) are
independent each other as local functions, so far as symmetries valid
for general space dimension are concerned.

%%%%%%%%%%%%%%%%%%%%%%%%%%%%%%%%%%%%%%%%%%%%%%%%%%%%%%%%%%%%%%%%%%%
%%%%%%%%%%%%%%%%%%%%%%%%%%  SEC 9  %%%%%%%%%%%%%%%%%%%%%%%%%%%%%%%%%%%%
%%%%%%%%%%%%%%%%%%%%%%%%%%%%%%%%%%%%%%%%%%%%%%%%%%%%%%%%%%%%%%%%%%%
\section{Relations Valid for Only Each Dimension}

It is known that,
for each fixed space-dimension, there generally appear additional relations
among general invariants and topological quantities
(say, \cite{GS,FKWC,DS93,SI}). 
This kind of relations have been noticed rather fragmentally
in specific situations so far. Here
we explicitly derive them in a systematic way. We still keep a
general space-dimension $n$ for a while.

\begin{figure}
     \centerline{
{\epsfxsize=210pt  \epsfysize=160pt \epsfbox{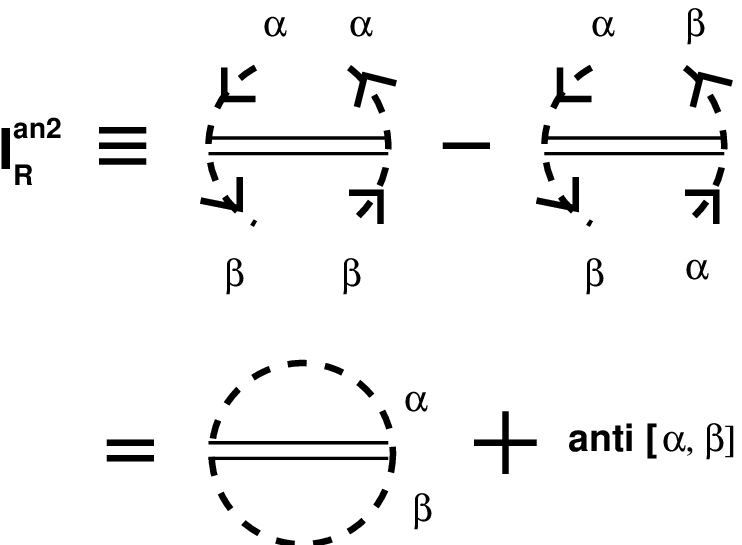}}
                 }
\caption{
%*fig9p1*Fig.9.1\ 
Graphical representation for $I^{an2}_R\equiv R^{\ab}_{~\be\al}
-R^\ab_{~\ab}$. 
In the figure, anti[$\al,\be$] means anti-symmetrization
w.r.t. $\al$ and $\be$.
The second-line figure demonstrates the present
notation used in the following.
        }
\label{fig9p1}
\end{figure}

\begin{figure}
     \centerline{
{\epsfxsize=286pt  \epsfysize=57pt \epsfbox{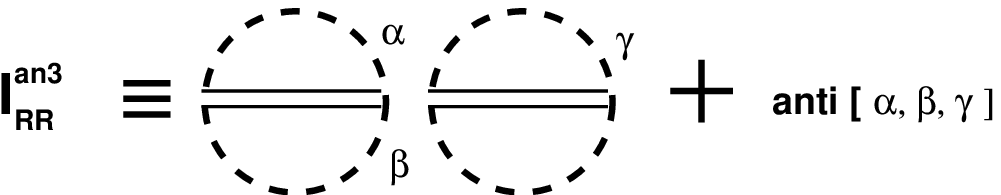}}
                 }
\vspace{10pt}
     \centerline{
{\epsfxsize=293pt  \epsfysize=57pt \epsfbox{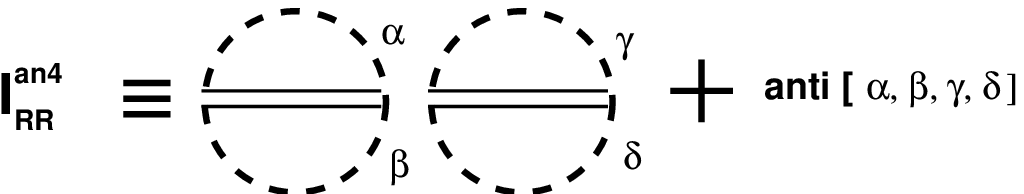}}
                 }
\caption{
%*fig9p2*Fig.9.2\ 
Graphical definition for $I^{an3}_{RR}$\ and $I^{an4}_{RR}$.
In the figure, anti[$\al,\be,\ga$]
and anti[$\al,\be,\ga,\del$] mean totally anti-symmetrization
w.r.t. ($\al,\be,\ga$) and ($\al,\be,\ga,\del$) respectively.
        }
\label{fig9p2}
\end{figure}

\q Let us introduce the quantity, $I^{an2}_R$, 
graphically defined in Fig.\ref{fig9p1}, where a convenient
notation is introduced and is used in the following. 
In the figure, anti[$\al,\be$] means anti-symmetrization
w.r.t. $\al$ and $\be$.
%%%%%%%%%%%%%%%%%%%%%%%% Fig.9.1 %%%%%%%%%%%%%%%%%%%%%%%%%%%%%%%%%
%%%%%%%%%%%%%%%%%%%%%%%%%%%%%%%%%%%%%%%%%%%%%%%%%%%%%%%%%%%%%%%%%%%%%
We define similar quantities in Fig.\ref{fig9p2} where anti[$\al,\be,\ga$]
and anti[$\al,\be,\ga,\del$] mean total anti-symmetrization
w.r.t. ($\al,\be,\ga$) and ($\al,\be,\ga,\del$) respectively.
We can easily compute them by the use of
algebraic calculation, and we obtain
as follows.
%%%%%%%%%%%%%%%%%%%%%%%% Fig.9.2 %%%%%%%%%%%%%%%%%%%%%%%%%%%%%%%%%
%%%%%%%%%%%%%%%%%%%%%%%%%%%%%%%%%%%%%%%%%%%%%%%%%%%%%%%%%%%%%%%%%%%%%

%****(anti.1)%%%%%%%%%%%%%%%%%%%%
\begin{eqnarray}
I^{an2}_R=2R\com\nn\\
I^{an3}_{R^2}=2(-2R_\mn R^\mn +R^2)\com\nn\\
I^{an4}_{R^2}
=4(R^2-4R_\mn R^\mn+R_{\mn\ab}R^{\mn\ab})
\com   \label{anti.1}
\end{eqnarray}
%%%%%%%%%%%%%%%%%%%%%%%%%%%%%
When we anti-symmetrize some suffixes in a given graph (general
invariant) $G$, we notice the following general facts.
\begin{enumerate}
\item
We generally obtain different anti-symmetrized quantities by taking
different choices of the starting graph $G$ and of the number of
anti-symmetrized suffixes ($\equiv N_{an}$).
\item
$N_{an}$ is maximized at the number of internal dotted lines
($\equiv N_I$):$N_{an}\leq N_I$. (Ex. In Fig.\ref{fig9p2}, 
$N_{an}=3,N_I=4$ for $I^{an3}_{R^2}$; 
$N_{an}=4,N_I=4$ for $I^{an4}_{R^2}$.)
When $N_{an}<N_I$, there generally
appear some choices of suffixes
to be anti-symmetrized.
\item
There are two cases when we calculate anti-symmetrized
quantities.
    \begin{enumerate}
    \item
It identically vanishes. In this case we do not have relations
among invariants.
    \item
It gives a sum of some invariants listed in Sec.VIII.
    \end{enumerate}
\end{enumerate}

For $M^2$- and $M^4$- invariants, all possible non-zero
anti-symmetrized quantities are given by (\ref{anti.1}).

\q We can do the same thing for $M^6$-invariants. Anti-symmetrized
quantities are defined in App.G.
%Fig.9.3 ($R\times R\times R$),
%Fig.9.4 ($\na\na R\times R$) and Fig.9.5($\na R\times \na R$).
They are all linearly-independent non-zero ones 
( except a relation in (\ref{anti.3}), which is written for 
an interest ) and are 
computed as follows. 
From Fig.\ref{fig9p3a} of App.G, we have 
%****(anti.2)%%%%%%%%%%%%%%%%%%%%
\begin{eqnarray}
I^{an3}_{P_1}=P_1-3P_2+2P_4\com\nn\\
I^{an3}_{A_1}=P_4-3P_6+\frac{5}{4} A_1-B_1\com\nn\\
I^{an3}_{P_3}=-P_2+P_3+2P_5-2P_6\com\nn\\
I^{an3}_{B_1(a)}=4B_1\com\nn\\
I^{an3}_{B_1(b)}=-\half A_1+2B_1+P_6\com\nn\\
I^{an3}_{P_5}=P_5-\half P_6\com
                                       \label{anti.2}
\end{eqnarray}
%%%%%%%%%%%%%%%%%%%%%%%%%%%%%
The expression of $I^{an3}_{B_1(a)}$ implies $B_1$ is
``self-dual''. 
From the Fig.\ref{fig9p3b} of App.G, we have
%****(anti.2b)%%%%%%%%%%%%%%%%%%%%
\begin{eqnarray}
I^{an4}_{P_1}=2P_1-10P_2+8P_4+4P_5\com\nn\\
I^{an4}_{A_1}=4P_4+4P_5-10P_6+3A_1-4B_1\com\nn\\
I^{an4}_{P_3}=-2P_2+2P_3+8P_5-8P_6+A_1-4B_1\com\nn\\
I^{an4}_{P_6}=-2P_5+2P_6-\half A_1+2B_1\com\nn\\
                                       \label{anti.2b}
\end{eqnarray}
%%%%%%%%%%%%%%%%%%%%%%%%%%%%%
From the Fig.\ref{fig9p3c} of App.G, we have 
%****(anti.3)%%%%%%%%%%%%%%%%%%%%
\begin{eqnarray}
I^{an5}_{P_1}=4(P_1-8P_2+P_3+8P_4+8P_5-4P_6)
=2(I^{an6}_{P_1}-4I^{an5}_{A_1})
\com\nn\\
I^{an5}_{A_1}=4(-2P_2+P_3+4P_4+8P_5-10P_6+2A_1-4B_1)\com\nn\\
I^{an6}_{P_1}=4(2P_1-24P_2+6P_3+32P_4+48P_5-48P_6+8A_1-16B_1)\com
                                       \label{anti.3}
\end{eqnarray}
%%%%%%%%%%%%%%%%%%%%%%%%%%%%%
Eqs.(\ref{anti.2})-(\ref{anti.3}) show $RRR$-type invariants
are closed within themselves for the anti-symmetrization.
From the Fig.\ref{fig9p4} of App.G, we have 
%****(anti.4)%%%%%%%%%%%%%%%%%%%%
\begin{eqnarray}
I^{an3}_{T_1(a)}=2(T_1-2T_2)\com\nn\\
I^{an3}_{T_1(b)}=-P_4+P_5+\half T_1-T_2\com\nn\\
I^{an3}_{T_1(c)}=2(T_1-2T_4)\com\nn\\
I^{an3}_{T_3}=-P_6+\half A_1+2B_1\com\nn\\
I^{an4}_{T_1(a)}=4(T_1-4T_2+T_3)\com\nn\\
I^{an4}_{T_1(b)}=-4P_4+4P_5+2P_6-A_1-4B_1+T_1-4T_2+T_3\com
                                       \label{anti.4}
\end{eqnarray}
%%%%%%%%%%%%%%%%%%%%%%%%%%%%%
These relations show $\na\na R\times R$ type invariants are not
closed within themselves. In particular,$I^{an3}_{T_3}$ does not
have $T_3$. 
From the Fig.\ref{fig9p5} of App.G, we have 
%****(anti.5)%%%%%%%%%%%%%%%%%%%%
\begin{eqnarray}
I^{an3}_{O_1(a)}=2(O_1-2O_2)\com\nn\\
I^{an3}_{O_1(b)}=\fourth O_1-O_2+O_4\com\nn\\
I^{an4}_{O_1}=4(O_1-4O_2+O_3)\com
                                       \label{anti.5}
\end{eqnarray}
%%%%%%%%%%%%%%%%%%%%%%%%%%%%%
$\na R\times \na R$ type invariants are closed within themselves.
\vs{1}

\q Let us examine relations for each space-dimension.

%%%%%%%%%%%%%%%%%%%%%%%%%%%% (i)  n=2   %%%%%%%%%%%%%%%%%%%%%%%%%%%%
\flushleft{(i)\ n=2}

For the $n=2$ space dimension, $I^{an2}_R$ gives 
the Gauss-Bonnet relation,
%****(ind.5)%%%%%%%%%%%%%%%%%%%%
\begin{eqnarray}
\int d^2x \sqg I^{an2}_R=\mbox{topological invariant}
\pr   \label{ind.5}
\end{eqnarray}
%%%%%%%%%%%%%%%%%%%%%%%%%%%%%
The invariant $R$, (\ref{ind.1}), remains as the unique
general $M^2$-invariant although $\sqg R$ is a surface term
(total derivative).
$I^{an~i}_*=0\ (i\geq 3)$ give us relations. From 
$I^{an3}_{R^2}=0$ and $I^{an4}_{R^2}=0$, we have the following
ones between $M^4$-invariants.
%****(ind.6)%%%%%%%%%%%%%%%%%%%%
\begin{eqnarray}
R_{\mn\ls}R^{\mn^\ls}=2R_\mn R^\mn=R^2
\pr   \label{ind.6}
\end{eqnarray}
%%%%%%%%%%%%%%%%%%%%%%%%%%%%%
Therefore we can take, as all independent $M^4$-invariants
in 2 space dimension,  
%****(ind.6b)%%%%%%%%%%%%%%%%%%%%
\begin{eqnarray}
\na^2R\com\q R^2
\pr   \label{ind.6b}
\end{eqnarray}
%%%%%%%%%%%%%%%%%%%%%%%%%%%%%
Relations (\ref{ind.6}) deduce the following ones between
$M^6$-invariants. 
%****(ind.6c)%%%%%%%%%%%%%%%%%%%%
\begin{eqnarray}
P_1=2P_2=P_3
\pr   \label{ind.6c}
\end{eqnarray}
%%%%%%%%%%%%%%%%%%%%%%%%%%%%%
From the vanishing of all quantities of (\ref{anti.2})-(\ref{anti.3}),
we have
%****(ind.6d)%%%%%%%%%%%%%%%%%%%%
\begin{eqnarray}
P_1=4P_4=4P_5=2P_6=A_1\com\q B_1=0
\pr   \label{ind.6d}
\end{eqnarray}
%%%%%%%%%%%%%%%%%%%%%%%%%%%%%
From the vanishing of all quantities of (\ref{anti.4}),
we have
%****(ind.6e)%%%%%%%%%%%%%%%%%%%%
\begin{eqnarray}
T_1=2T_2=2T_4=T_3
\pr   \label{ind.6e}
\end{eqnarray}
%%%%%%%%%%%%%%%%%%%%%%%%%%%%%
From the vanishing of all quantities of (\ref{anti.5}),
we have
%****(ind.7)%%%%%%%%%%%%%%%%%%%%
\begin{eqnarray}
O_1=2O_2=O_3=4O_4
\pr   \label{ind.7}
\end{eqnarray}
%%%%%%%%%%%%%%%%%%%%%%%%%%%%%
Therefore
we have the following 4 terms as independent $M^6$-invariants.
%****(ind.8)%%%%%%%%%%%%%%%%%%%%
\begin{eqnarray}
P_1=RRR\com\ \ O_1=\na^\mu R\cdot \na_\mu R\com\ \ 
T_1=\na^2R\cdot R\com\ \ S=\na^2\na^2R\pr
                               \label{ind.8}
\end{eqnarray}
%%%%%%%%%%%%%%%%%%%%%%%%%%%%%
\q We should note here that the above relations are derived
without the use of the well-known relation between
Riemann tensors which is valid only in 2 space-dimension.
%****(ind.8b)%%%%%%%%%%%%%%%%%%%%
\begin{eqnarray}
R_{\mn\ls}=\half (g_{\m\si}g_{\n\la}-g_{\m\la}g_{\n\si})R
\pr
                               \label{ind.8b}
\end{eqnarray}
%%%%%%%%%%%%%%%%%%%%%%%%%%%%%
(Of course, the obtained relations are consistent with the above
relation. This is a strong check of the present approach.)
Because the degree of local freedom of the Riemann tensor
in n-dim space is $f(n)=n^2(n^2-1)/12, n\geq 4$, 
we do not have such simple
relations as (\ref{ind.8b}) in higher space dimension.
Hence the present approach is indispensable to obtain
all relations in higher space-dimension.

%%%%%%%%%%%%%%%%%%%%%%%%%%%% (ii)  n=4   %%%%%%%%%%%%%%%%%%%%%%%%%%%%
\flushleft{(ii)\ n=4}

In the $n=4$ space dimension, $I^{an4}_{RR}$ gives 
the Gauss-Bonnet relation,
%****(ind.9)%%%%%%%%%%%%%%%%%%%%
\begin{eqnarray}
\int d^4x \sqg I^{an4}_{RR}=\mbox{topological invariant}
\pr   \label{ind.9}
\end{eqnarray}
%%%%%%%%%%%%%%%%%%%%%%%%%%%%%
The four invariants (\ref{ind.2}) remain as independent 
general $M^4$-invariants.
From the vanishing of all quantities of (\ref{anti.3}), we have
two independent relations between $M^6$-invariants.
%****(ind.10)%%%%%%%%%%%%%%%%%%%%
\begin{eqnarray}
I^{an5}_{P_1}=0\com\q I^{an5}_{A_1}=0
\pr   \label{ind.10}
\end{eqnarray}
%%%%%%%%%%%%%%%%%%%%%%%%%%%%%
There exist no relations between $T_i$'s and $O_i$'s.
Therefore we have 17-2=15 terms
 as independent $M^6$-invariants, say,
%****(ind.11)%%%%%%%%%%%%%%%%%%%%
\begin{eqnarray}
P_1\com\q P_2\com\q P_3\com\q P_4\com\q P_5\com\q A_1\com\nn\\
T_1\com\q T_2\com\q T_3\com\q T_4\com\q
O_1\com\q O_2\com\q O_3\com\q O_4\com\q S
\pr   \label{ind.11}
\end{eqnarray}
%%%%%%%%%%%%%%%%%%%%%%%%%%%%%
They are considered to appear in 
the higher-order of the Weyl anomaly due to the graviton-loop effect
if they can be properly defined. 
(In the usual(1-loop) Weyl anomaly, 4 terms of (\ref{ind.2}) appear).

%%%%%%%%%%%%%%%%%%%%%%%%%%%% (iii)  n=6   %%%%%%%%%%%%%%%%%%%%%%%%%%%%
\flushleft{(iii)\ n=6}

In the $n=6$ space-dimension, $I^{an6}_{P_1}$ gives 
the Gauss-Bonnet relation,
%****(ind.12)%%%%%%%%%%%%%%%%%%%%
\begin{eqnarray}
\int d^6x \sqg I^{an6}_{P_1}=\mbox{topological invariant}
\pr   \label{ind.12}
\end{eqnarray}
%%%%%%%%%%%%%%%%%%%%%%%%%%%%%
The 17 invariants (\ref{ind.3}) remain as independent general
invariants.

\vs 1

\q Therefore we have confirmed that, 
in n space-dimension, 
all independent $M^n$-invariants ( the (1-loop) Weyl anomaly
is given by them ) are given in Sec.VIII :\ 
(\ref{ind.1}) for n=2, (\ref{ind.2}) for n=4 and (\ref{ind.3}) for n=6.
Only for general invariants with higher mass-dimension $M^m, m>n$,
the number of independent ones reduces from those given in Sec.VIII
due to relations valid only for each dimension.

\q As a comparison, it is interesting to examine the situation in
independent general invariants as counterterms.
Generally the counterterms $\Del\Lcal$ are defined in a space
integral in such a way that the action
%****(ind.13)%%%%%%%%%%%%%%%%%%%%
\begin{eqnarray}
\int\sqg~\Del\Lcal~d^nx
\com   \label{ind.13}
\end{eqnarray}
%%%%%%%%%%%%%%%%%%%%%%%%%%%%%
cancels (ultra-violet) divergences due to the quantum fluctuation.
$\Del\Lcal$ is a sum of general invariants 
with 'divergent'-constant coefficients. 
Here we have interest
in what terms could appear 
as independent ones. We {\it may neglect total derivative
terms} because the fields are usually assumed to damp sufficiently
rapidly at the boundary. As a choice, we give a complete list
of independent counterterms in the following.

%%%%%%%%%%%%%%%%%%%%%%%%%%%% (i)  n=2   %%%%%%%%%%%%%%%%%%%%%%%%%%%%
\flushleft{(i)\ n=2}

$M^2$-invariants(1-loop):\ no terms\nl
$M^4$-invariants(2-loop):\ $R^2$\nl
$M^6$-invariants(3-loop):\ $P_1=RRR,O_1=\na^\mu R\cdot \na_\mu R$

%%%%%%%%%%%%%%%%%%%%%%%%%%%% (iii)  n=4   %%%%%%%%%%%%%%%%%%%%%%%%%%%%
\flushleft{(ii)\ n=4}

$M^4$-invariants(1-loop): $R^2, R_\mn R^\mn$\nl
$M^6$-invariants(2-loop): $P_1,P_2,P_3,P_4,P_5,A_1,O_1,O_2,O_3,O_4$

%%%%%%%%%%%%%%%%%%%%%%%%%%%% (iii)  n=6   %%%%%%%%%%%%%%%%%%%%%%%%%%%%
\flushleft{(iii)\ n=6}

$M^6$-invariants(1-loop): $P_1,P_2,P_3,P_4,P_5,P_6,A_1,O_1,O_2,O_3,O_4$

\vs 1
If we consider pure gravity and impose the S-matrix condition 
(on-shell condition, Ricci flat condition) 
$R_\mn=0$ on the above results, we see 2 dim
pure gravity is finite, 
4 dim case is not finite at higher-loops from 2-loop, 
6 dim case is not finite at higher-loops from 1-loop. 
In the latter two cases, non-finite term appears as $A_1$ term.
This is well known from the 
divergence problem in the S-matrix in
perturbative quantum gravity\cite{K,VW}.

%%%%%%%%%%%%%%%%%%%%%%%%%%%%%%%%%%%%%%%%%%%%%%%%%%%%%%%%%%%%%%%%%%%
%%%%%%%%%%%%%%%%%%%%%%%%%%  SEC 10  %%%%%%%%%%%%%%%%%%%%%%%%%%%%%%%%%%%%
%%%%%%%%%%%%%%%%%%%%%%%%%%%%%%%%%%%%%%%%%%%%%%%%%%%%%%%%%%%%%%%%%%%
\section{Discussions and Conclusions}

We have presented a way to classify SO(n)-invariants which generally
appear in weak-field perturbations of (quantum) gravity. Taking
the explicit example of $(\pl\pl h)^3$-invariants, we have presented 
the general way of classification. 
The following important items have been explained
:\ 1) the graphical
representation of global SO(n)-tensors and invariants,
2) the weight of a graph, 
3) indices characterizing a graph, 
4) reduction procedures of graphs, and  
5) bondless diagrams. 
In the higher dimensional cases, such as $(\pl\pl h)^4$ and $(\pl\pl h)^5$
( which appear, for example,  
in the Weyl anomaly in 8 dim and 10 dim gravity, respectively) 
the same procedure can be applied except some additional indices
might be required. 

\q We have mainly discussed $(\pl\pl h)^3$-invariants in the text,
$\pl^4 h\cdot \pl^2 h$-invariants in App.C and
$(\pl^3 h)^3$-invariants in App.D. 
$\pl\pl h$- and $(\pl\pl h)^2$-invariants have been treated
in paper (I).
Clearly it must be generalized to treat all SO(n)-invariants
which appear in the weak expansion of all general invariants.
For such direction, we comment on the generalization of
the field theory approach proposed in Sec.VI. Let us consider
the following Lagrangian in $2$ space-dimension\cite{foot9}. 
%\footnote{
%The space-dimension is taken to be 2 in order to
%obtain the mass-dimensions of the couplings, (\ref{conc.3}), 
%without
%introducing any additional mass parameters. Note that
%the present purpose is the graph classification.
%There the important thing is the topological structure
%of graphs. 
%The space-dimension of the field theory is irrelevant. 
%}
%****(conc.1)%%%%%%%%%%%%%%%%%%%%
\begin{eqnarray}
\Lcal[\p,\om_1,\om_2]=\Lcal_0+\Lcal_I\com\nn\\
\Lcal_0=\half\p^2+\om_1\om_2\com\nn\\
\Lcal_I[\p,\om_1,\om_2]
=\sum_{i=1}^{\infty}g_i\p^i\om_1+\la\p^2\om_2\pr
     \label{conc.1}
\end{eqnarray}
%%%%%%%%%%%%%%%%%%%%%%%%%%%%%
We assign mass-dimension  as follows.
%****(conc.2)%%%%%%%%%%%%%%%%%%%%
\begin{eqnarray}
[\Lcal]=M^2\com\q
[\p]=M\com\q
[\om_1]=M^2\com\q
[\om_2]=M^0\pr
     \label{conc.2}
\end{eqnarray}
%%%%%%%%%%%%%%%%%%%%%%%%%%%%%
Then we obtain
%****(conc.3)%%%%%%%%%%%%%%%%%%%%
\begin{eqnarray}
[g_i]=M^{-i}\com\q
[\la]=M^0\pr
     \label{conc.3}
\end{eqnarray}
%%%%%%%%%%%%%%%%%%%%%%%%%%%%%
This result turns out to give the mass-dimension
of each expanded term in the following.
The generating functional of all graphs ( SO(n)-invariants,
SO(n)-tensors ) is given by
%****(conc.4)%%%%%%%%%%%%%%%%%%%%
\begin{eqnarray}
W[J,K_1,K_2]=\exp^{\Ga[J,K_1,K_2]}\nn\\
=\int\Dcal\p\Dcal\om_1\Dcal\om_2\exp \left[
\int d^2x (\Lcal[\p,\om_1,\om_2]+J\p+K_1\om_1+K_2\om_2)
                                    \right]             \nn\\
=\sum_{r=0}^{\infty}\frac{1}{r!}\left[
\int d^2x \Lcal_I(\frac{\del}{\del J(x)},\frac{\del}{\del K_1(x)},
\frac{\del}{\del K_2(x)})       \right]^r
\exp \int d^2x (-\half J(x)J(x)-K_1(x)K_2(x) )\pr
                                     \label{conc.4}
\end{eqnarray}
%%%%%%%%%%%%%%%%%%%%%%%%%%%%%
All graphs of connected n-tensors appear in the n-point
Green function.
%****(conc.5)%%%%%%%%%%%%%%%%%%%%
\begin{eqnarray}
\frac{1}{n!}\left.
\frac{\del}{\del J(x_1)}\frac{\del}{\del J(x_2)}\cdots
\frac{\del}{\del J(x_n)}\Ga[J,K_1,K_2]\right|_{J=0,K_1=0,K_2=0}\pr
                                     \label{conc.5}
\end{eqnarray}
%%%%%%%%%%%%%%%%%%%%%%%%%%%%%
In particular all SO(n)-invariants appear in
%****(conc.6)%%%%%%%%%%%%%%%%%%%%
\begin{eqnarray}
\left.
\Ga[J,K_1,K_2]\right|_{J=0,K_1=0,K_2=0}\pr
                                     \label{conc.6}
\end{eqnarray}
%%%%%%%%%%%%%%%%%%%%%%%%%%%%%
These quantities are given by perturbation with respect to
the couplings ($g_1,g_2,\cdots; \la$) in $\Lcal_I$.
For example, $(\pl\pl h)^s$-invariants ($s=1,2,\cdots$) are
given by $(g_2\la)^s$-terms ($r=2s$) in (\ref{conc.6}). 
$\pl^4 h\cdot\pl^2 h$-invariants (App.C) and
$\pl^3 h\cdot\pl^3 h$-invariants (App.D) are given
by $g_4\cdot g_2\cdot \la^2$-terms ($r=4$) and
$g_3\cdot g_3\cdot \la^2$-terms ($r=4$) respectively.
From the coupling-dependence, the mass-dimension of each graph
is given. For example  
$[(g_2\la)^s]=M^{-2s}$, 
$[g_4\cdot g_2\cdot \la^2]=M^{-6}$ and 
$[g_3\cdot g_3\cdot \la^2]=M^{-6}$.
They are the inverse of their mass-dimensions. 
The coefficient in front of each expanded term are related
with the weight of the corresponding graph.
The generalization using this field theory approach
is useful for classification of general SO(n)-invariants.

\q The result is not only interesting as the
mathematical (graphical) structure by itself, but also provides
a very efficient computer-algorithm for the tensor calculation.
As an example of a computer calculation, 
we have presented some results of weak-perturbation
of general $M^6$-invariants in App.E.
They are used to prove the independence of general $M^6$-invariants
in Sec.VIII. Further
important applications of the present result are the anomaly
and the (1-loop) counterterm calculation in 6 dim quantum gravity. 
Generally
in n-dim gravity, the Weyl anomaly is given by some combination of 
general invariants with dimension $M^{n}$, and  
%For example in 6-dim gravity, it is given
%by some combination of $\na\na\na\na R$-, $\na\na RR$- and $RRR$-type
%invariants(see \cite{II1}). 
$L$-loop counterterms 
are given by some combination of invariants with dimension
$M^{n+2L-2}$. In both cases, all coefficients can be fixed by
the weak-field perturbation\cite{II1}.

\q So far, we have been annoyed by the complicated tensor calculation
in the analysis of (quantum) gravity.
It is serious especially in a higher-dimensional case
or in a higher-order case. 
This is because we have not known an efficient way to manipulate
tensors.
It is not an exaggeration to say
that the complication has been a hinderence to understanding the
theory of gravity. 
We believe the present approach provides a new possibility in
analysing (quantum) gravity in such cases. 

\vs 1
\q The results of Sec.IX, App.B, App.E and
some others are obtained by computer calculation
(FORM, C-language program).

\vs 2
\begin{flushleft}
{\bf Acknowledgement}
\end{flushleft}
The authors thank Prof. K.Murota (RIMS,Kyoto Univ.,Japan)
for an important suggestion in Sec.IV. 
They express
gratitude to Prof. H.Osborn for reading the manuscript
and some comments. The check of the manuscript by
Dr. H.Shanahan and the kind help in the computer 
manipulation by Dr. C.Houghton are greatly appreciated.
A part of
this work has been done when one of the authors (S.I.)
are staying at DAMTP, Univ. of Cambridge. He thanks
all members of DAMTP for hospitality. 
One of the authors (S.I.) thanks Japanese ministry of
education for the partial financial support (Researcher
No: 40193445, Institution No: 23803).
\vs 2

\newpage
%%%%%%%%%%%%%%%%%%%%%%%%%%  App. A  %%%%%%%%%%%%%%%%%%%%%%%%%%%%%%%
%%%%%%%%%%%%%%%%%%%%%%%%%%%%%%%%%%%%%%%%%%%%%%%%%%%%%%%%%%%%%%%%%%%
\begin{flushleft}
{\Large\bf Appendix A.\ Full List of $(\pl\pl h)^3$-Invariants}
\end{flushleft}

In this appendix we list up graphs of all independent 
$(\pl\pl h)^3$-invariants. Every graph is named according to
the classification scheme explained in Sec.III and IV. They
are grouped with respect to the number of suffix-loops $\ul{l}$ as
follows. (``con'' means ``connected graphs'' and ``discon'' means
``disconnected graphs''.)

\begin{figure}
\begin{tabular}{p{5cm}p{5cm}p{5cm}}

\shortstack{
{\epsfxsize=3.81cm  \epsfysize=3.49cm \epsfbox{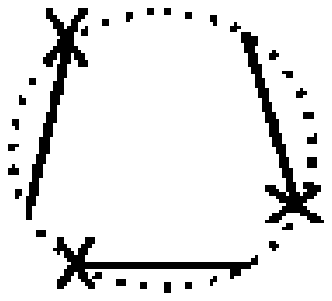}}
\\ G1:\ $1A\Th$
            } &
\shortstack{
{\epsfxsize=3.53cm  \epsfysize=3.56cm \epsfbox{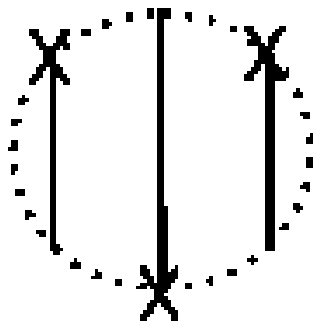}}
\\ G2:\ $1A\Xi$
            } &
\shortstack{
{\epsfxsize=3.95cm  \epsfysize=3.39cm \epsfbox{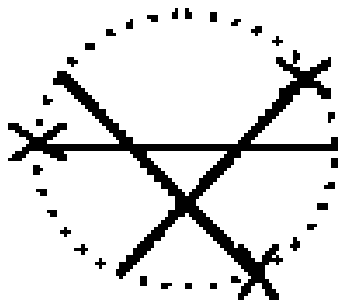}}
\\ G3:\ $1A\Om_2$
            } 
\\
\shortstack{
{\epsfxsize=3.77cm  \epsfysize=3.32cm \epsfbox{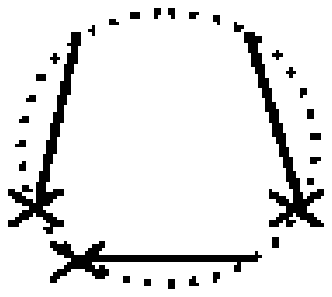}}
\\ G4:\ $1B\Th$
            } &
\shortstack{
{\epsfxsize=3.85cm  \epsfysize=3.35cm \epsfbox{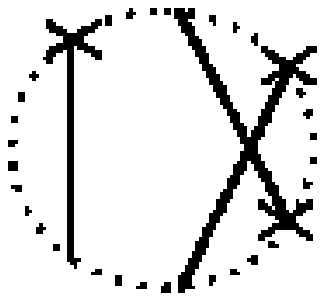}}
\\ G5:\ $1B\Si-a$
            } &
\shortstack{
{\epsfxsize=3.63cm  \epsfysize=3.70cm \epsfbox{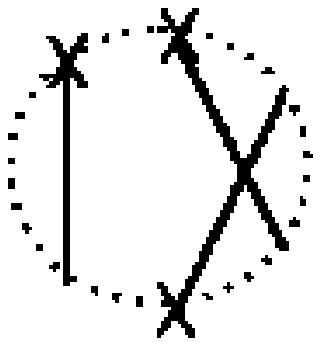}}
\\ G6:\ $1B\Si-b$
            } 
\\
\shortstack{
{\epsfxsize=3.81cm  \epsfysize=3.60cm \epsfbox{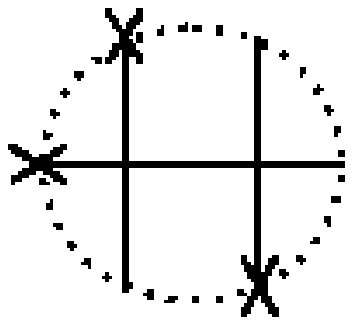}}
\\ G7:\ $1B\Om_1$
            } &
\shortstack{
{\epsfxsize=3.56cm  \epsfysize=3.53cm \epsfbox{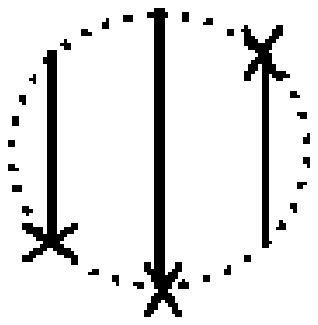}}
\\ G8:\ $1B\Xi$
            } &
\shortstack{
{\epsfxsize=3.70cm  \epsfysize=3.53cm \epsfbox{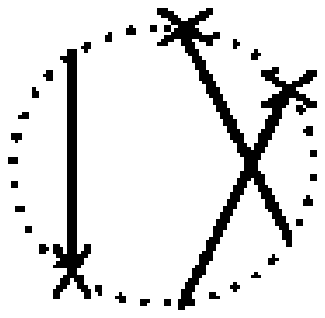}}
\\ G9:\ $1B\Si-c$
            } 
\\
\shortstack{
{\epsfxsize=3.67cm  \epsfysize=3.63cm \epsfbox{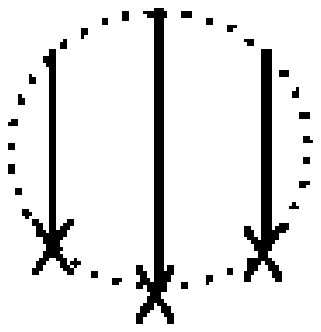}}
\\ G10:\ $1C\Xi$
            } &
\shortstack{
{\epsfxsize=3.60cm  \epsfysize=3.42cm \epsfbox{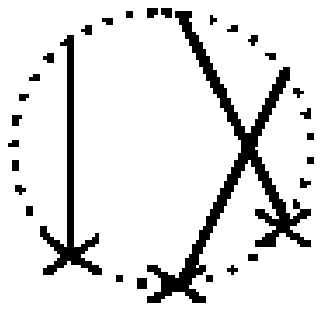}}
\\ G11:\ $1C\Si$
            } &
\shortstack{
{\epsfxsize=3.81cm  \epsfysize=3.32cm \epsfbox{G12.ps}}
\\ G12:\ $1C\Om_1$
            } 
\\
\shortstack{
{\epsfxsize=3.77cm  \epsfysize=3.42cm \epsfbox{G13.ps}}
\\ G13:\ $1C\Om_2$
            } & &

\end{tabular}
\caption{
%*G1-13*
(i) $\ul{l}=1$\ (\ 13(con)+0(discon)=13 terms
G1-13}
\label{G1-13}
\end{figure}

\begin{figure}
\begin{tabular}{p{5cm}p{5cm}p{5cm}}

\shortstack{
{\epsfxsize=2.05cm  \epsfysize=3.81cm \epsfbox{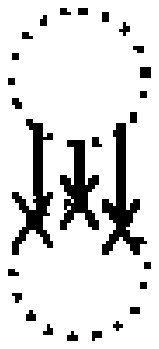}}
\\ G14:\ $2A\Om_1$
            } &
\shortstack{
{\epsfxsize=1.87cm  \epsfysize=3.77cm \epsfbox{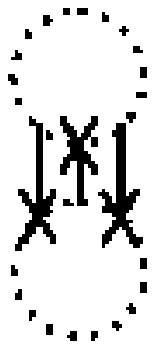}}
\\ G15:\ $2D\Om_1$
            } &
\shortstack{
{\epsfxsize=2.15cm  \epsfysize=3.81cm \epsfbox{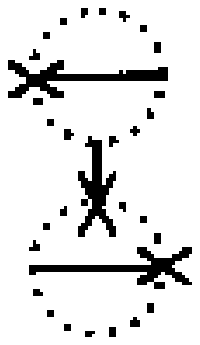}}
\\ G16:\ $2D\Xi_1$
            } 
\\
%\end{tabular}
%\end{figure}
%\newpage
%\begin{figure}
%\begin{tabular}{p{5cm}p{5cm}p{5cm}}
\shortstack{
{\epsfxsize=2.08cm  \epsfysize=3.74cm \epsfbox{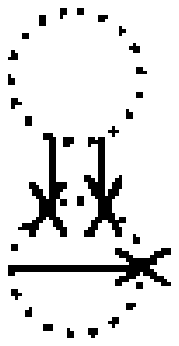}}
\\ G17:\ $2B\Si_1$
            } &
\shortstack{
{\epsfxsize=1.83cm  \epsfysize=3.88cm \epsfbox{G18.ps}}
\\ G18:\ $2B\Om_2$
            } &
\shortstack{
{\epsfxsize=2.08cm  \epsfysize=3.74cm \epsfbox{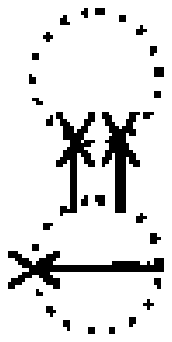}}
\\ G19:\ $2C\Si_1$
            } 
\\
\shortstack{
{\epsfxsize=2.01cm  \epsfysize=3.85cm \epsfbox{G20.ps}}
\\ G20:\ $2C\Om_2$
            } &
\shortstack{
{\epsfxsize=2.15cm  \epsfysize=3.70cm \epsfbox{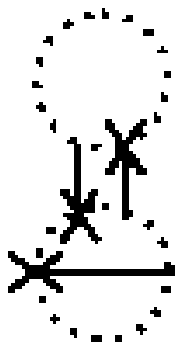}}
\\ G21:\ $2E_a\Si_1$
            } &
\shortstack{
{\epsfxsize=1.94cm  \epsfysize=4.02cm \epsfbox{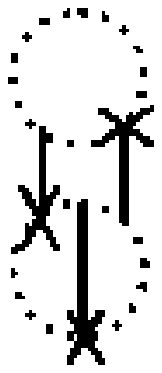}}
\\ G22:\ $2E_a\Om_2$
            } 
\\
\shortstack{
{\epsfxsize=2.15cm  \epsfysize=3.81cm \epsfbox{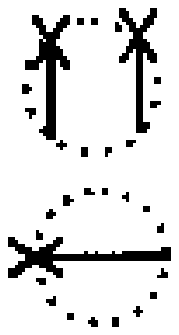}}
\\ G23:\ $A2Q$
            } &
\shortstack{
{\epsfxsize=2.08cm  \epsfysize=3.67cm \epsfbox{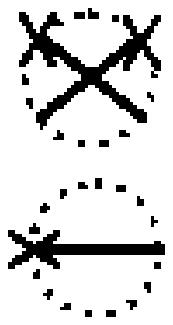}}
\\ G24:\ $A1Q$
            } &
\shortstack{
{\epsfxsize=2.01cm  \epsfysize=3.70cm \epsfbox{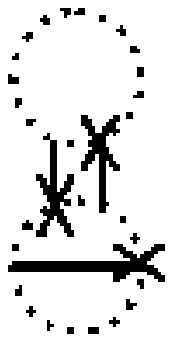}}
\\ G25:\ $2E_b\Si_1$
            } 
\\
\shortstack{
{\epsfxsize=2.26cm  \epsfysize=3.63cm \epsfbox{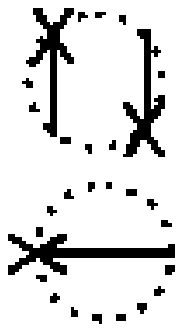}}
\\ G26:\ $A3Q$
            } &
\shortstack{
{\epsfxsize=1.94cm  \epsfysize=3.85cm \epsfbox{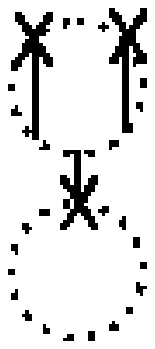}}
\\ G27:\ $2F_a\Th$
            } &
\shortstack{
{\epsfxsize=2.08cm  \epsfysize=3.85cm \epsfbox{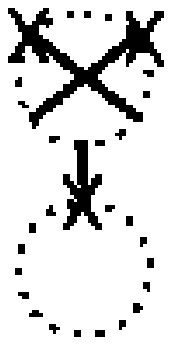}}
\\ G28:\ $2F_a\Si_2-a$
            } 
\end{tabular}
\caption{
%*G14-42a*
(ii) $\ul{l}=2$\ (\ 26(con)+3(discon)=29\ terms \ G14-42 \ No.1}
\label{G14-42a}
\end{figure}

\begin{figure}
\begin{tabular}{p{5cm}p{5cm}p{5cm}}
\shortstack{
{\epsfxsize=2.01cm  \epsfysize=3.67cm \epsfbox{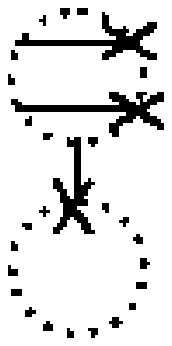}}
\\ G29:\ $2F_a\Xi_2$
            } &
\shortstack{
{\epsfxsize=1.87cm  \epsfysize=3.81cm \epsfbox{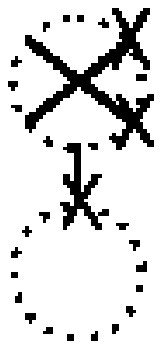}}
\\ G30:\ $2F_a\Si_2-b$
            } &
\shortstack{
{\epsfxsize=2.12cm  \epsfysize=3.60cm \epsfbox{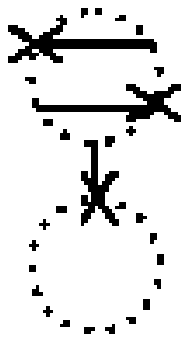}}
\\ G31:\ $2F_b\Xi_2$
            } 
\\
\shortstack{
{\epsfxsize=2.05cm  \epsfysize=3.74cm \epsfbox{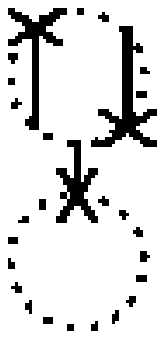}}
\\ G32:\ $2F_b\Th-a$
            } &
\shortstack{
{\epsfxsize=1.83cm  \epsfysize=3.67cm \epsfbox{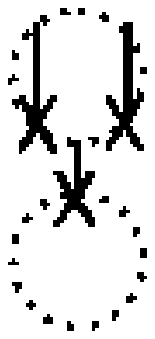}}
\\ G33:\ $2F_b\Th-b$
            } &
\shortstack{
{\epsfxsize=1.91cm  \epsfysize=3.67cm \epsfbox{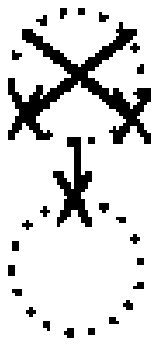}}
\\ G34:\ $2F_b\Si_2$
            } 
\\
\shortstack{
{\epsfxsize=1.87cm  \epsfysize=3.60cm \epsfbox{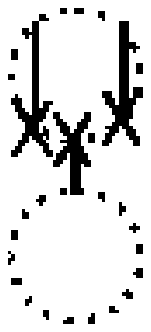}}
\\ G35:\ $2G_a\Th$
            } &
\shortstack{
{\epsfxsize=1.83cm  \epsfysize=3.56cm \epsfbox{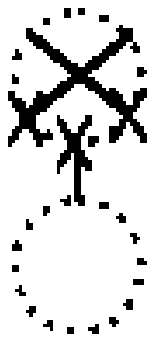}}
\\ G36:\ $2G_a\Si_2-a$
            } &
\shortstack{
{\epsfxsize=1.94cm  \epsfysize=3.70cm \epsfbox{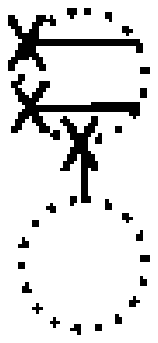}}
\\ G37:\ $2G_a\Xi_2$
            } 
\\
\shortstack{
{\epsfxsize=1.91cm  \epsfysize=3.85cm \epsfbox{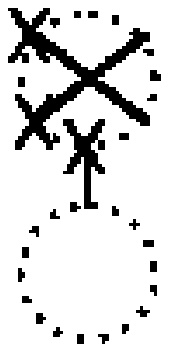}}
\\ G38:\ $2G_a\Si_2-b$
            } &
\shortstack{
{\epsfxsize=1.94cm  \epsfysize=3.70cm \epsfbox{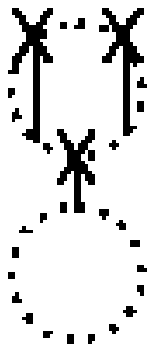}}
\\ G39:\ $2G_b\Th-a$
            } &
\shortstack{
{\epsfxsize=1.94cm  \epsfysize=3.70cm \epsfbox{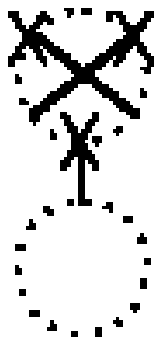}}
\\ G40:\ $2G_b\Si_2$
            } 
\\
\shortstack{
{\epsfxsize=2.12cm  \epsfysize=3.81cm \epsfbox{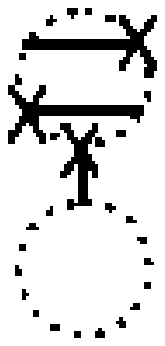}}
\\ G41:\ $2G_b\Xi_2$
            } &
\shortstack{
{\epsfxsize=2.01cm  \epsfysize=3.63cm \epsfbox{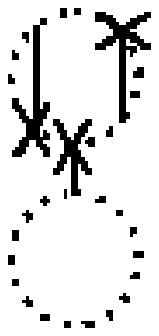}}
\\ G42:\ $2G_b\Th-b$
            } &

\end{tabular}
\caption{
%*G14-42b*
(ii) $\ul{l}=2$\ (\ 26(con)+3(discon)=29\ terms \ G14-42 \ No.2}
\label{G14-42b}
\end{figure}

\begin{figure}
\begin{tabular}{p{5cm}p{5cm}p{5cm}}

\shortstack{
{\epsfxsize=4.66cm  \epsfysize=1.87cm \epsfbox{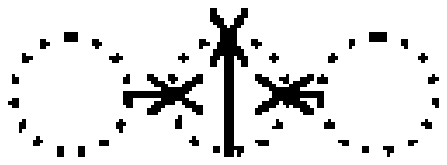}} 
\\ G43:\ $3A\Xi_1$
            }&
\shortstack{
{\epsfxsize=3.21cm  \epsfysize=2.75cm \epsfbox{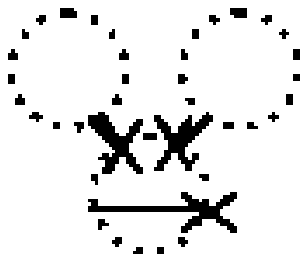}} 
\\ G44:\ $3A\Th$
            }&
\shortstack{
{\epsfxsize=4.80cm  \epsfysize=1.80cm \epsfbox{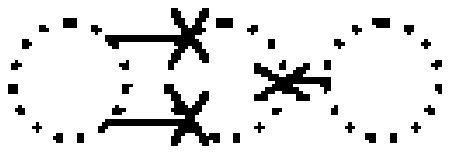}}
\\ G45:\ $3B\Si$
            } 
\\
\shortstack{
{\epsfxsize=4.76cm  \epsfysize=1.76cm \epsfbox{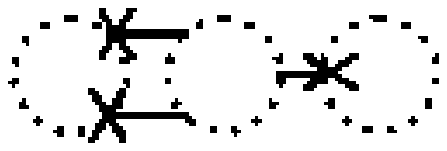}}
\\ G46:\ $3C\Si$
            } &
\shortstack{
{\epsfxsize=4.80cm  \epsfysize=1.69cm \epsfbox{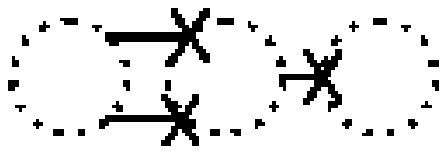}}
\\ G47:\ $3D\Si$
            } &
\shortstack{
{\epsfxsize=4.83cm  \epsfysize=1.83cm \epsfbox{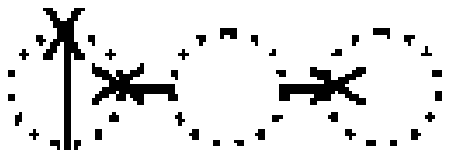}}
\\ G48:\ $3D\Xi_2$
            } 
\\
\shortstack{
{\epsfxsize=3.42cm  \epsfysize=3.17cm \epsfbox{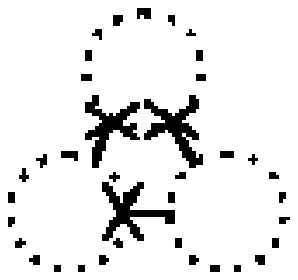}}
\\ G49:\ $3E\Om$
            } &
\shortstack{
{\epsfxsize=4.76cm  \epsfysize=1.83cm \epsfbox{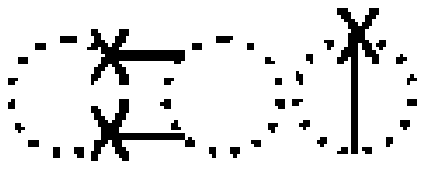}}
\\ G50:\ $B3Q$
            } &
\shortstack{
{\epsfxsize=3.42cm  \epsfysize=2.82cm \epsfbox{G51.ps}}
\\ G51:\ $3F_a\Th$
            } 
\\
\shortstack{
{\epsfxsize=4.73cm  \epsfysize=1.80cm \epsfbox{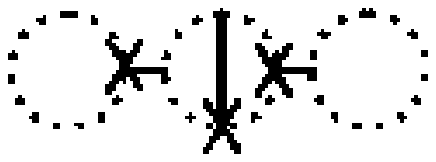}}
\\ G52:\ $3F_a\Xi_1$
            } &
\shortstack{
{\epsfxsize=5.08cm  \epsfysize=1.87cm \epsfbox{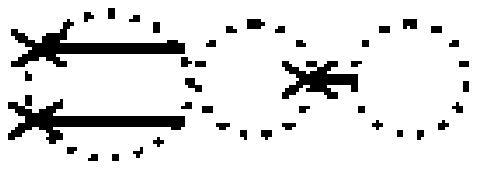}}
\\ G53:\ $A2P$
            } &
\shortstack{
{\epsfxsize=4.62cm  \epsfysize=1.69cm \epsfbox{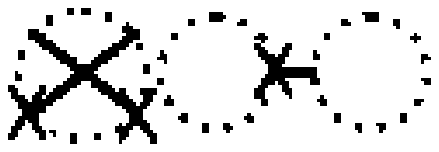}}
\\ G54:\ $A1P$
            } 
\\
\shortstack{
{\epsfxsize=3.77cm  \epsfysize=3.14cm \epsfbox{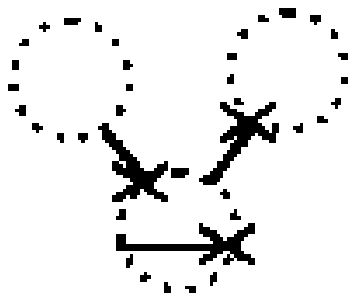}}
\\ G55:\ $3F_b\Th$
            } &
\shortstack{
{\epsfxsize=5.29cm  \epsfysize=1.83cm \epsfbox{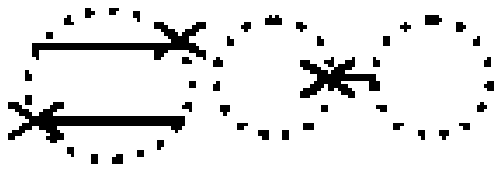}}
\\ G56:\ $A3P$
            } &
\shortstack{
{\epsfxsize=4.80cm  \epsfysize=1.66cm \epsfbox{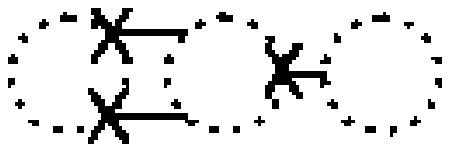}}
\\ G57:\ $3G\Si$
            } 
\\
\shortstack{
{\epsfxsize=4.76cm  \epsfysize=1.76cm \epsfbox{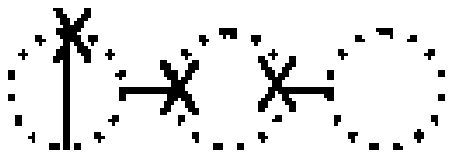}}
\\ G58:\ $3G\Xi_2$
            } &
\shortstack{
{\epsfxsize=4.66cm  \epsfysize=1.62cm \epsfbox{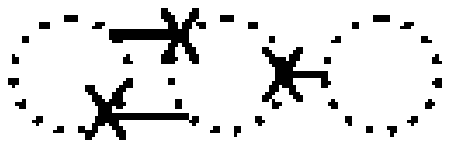}}
\\ G59:\ $3H\Si$
            } &
\shortstack{
{\epsfxsize=4.69cm  \epsfysize=1.76cm \epsfbox{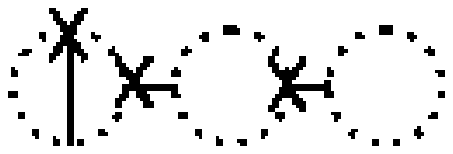}}
\\ G60:\ $3H\Xi_2$
            } 
\\
\shortstack{
{\epsfxsize=2.93cm  \epsfysize=3.21cm \epsfbox{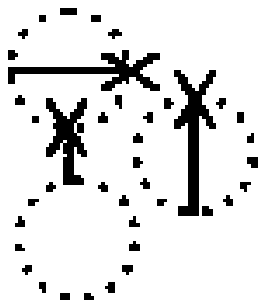}}
\\ G61:\ $B2Q$
            } &
\shortstack{
{\epsfxsize=3.77cm  \epsfysize=2.96cm \epsfbox{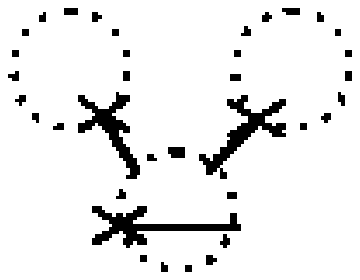}}
\\ G62:\ $3I\Th$
            } &
\shortstack{
{\epsfxsize=4.62cm  \epsfysize=1.73cm \epsfbox{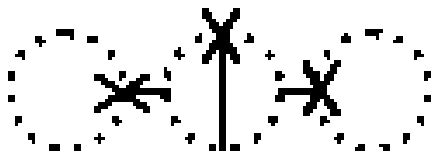}}
\\ G63:\ $3I\Xi_1$
            } 
\end{tabular}
\caption{
%*G43-69a*
(iii) $\ul{l}=3$\ (\ 19(con)+8(discon)=27 terms\ G43-69\ No.1}
\label{G43-69a}
\end{figure}

\begin{figure}
\begin{tabular}{p{5cm}p{5cm}p{5cm}}

\shortstack{
{\epsfxsize=4.62cm  \epsfysize=1.66cm \epsfbox{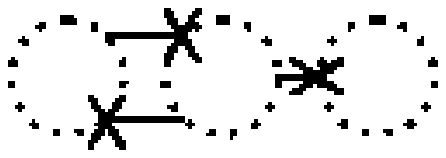}}
\\ G64:\ $3J\Si$
            } &
\shortstack{
{\epsfxsize=4.87cm  \epsfysize=1.76cm \epsfbox{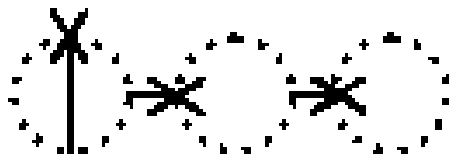}}
\\ G65:\ $3J\Xi_2$
            } &
\shortstack{
{\epsfxsize=2.96cm  \epsfysize=3.14cm \epsfbox{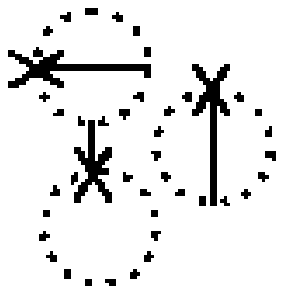}}
\\ G66:\ $B1Q$
            } 
\\
\shortstack{
{\epsfxsize=3.32cm  \epsfysize=3.03cm \epsfbox{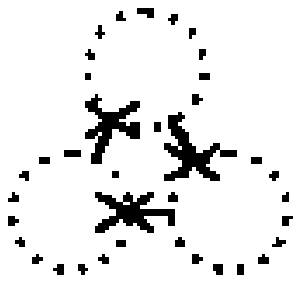}}
\\ G67:\ $3K\Om$
            } &
\shortstack{
{\epsfxsize=4.83cm  \epsfysize=1.83cm \epsfbox{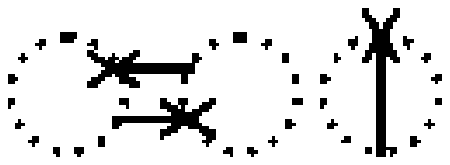}}
\\ G68:\ $B4Q$
            } &
\shortstack{
{\epsfxsize=4.41cm  \epsfysize=1.83cm \epsfbox{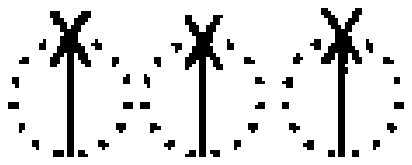}}
\\ G69:\ $QQQ$
            } 
\end{tabular}
\caption{
%*G43-69b*
(iii) $\ul{l}=3$\ (\ 19(con)+8(discon)=27 terms\ G43-69\ No.2}
\label{G43-69b}
\end{figure}

\begin{figure}
\begin{tabular}{p{5cm}p{5cm}p{5cm}}

\shortstack{
{\epsfxsize=3.95cm  \epsfysize=3.63cm \epsfbox{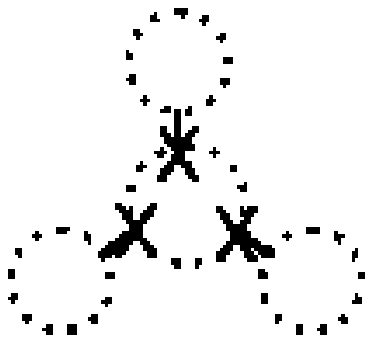}}
\\ G70:\ $4A\Th$
            } &
\shortstack{
{\epsfxsize=5.36cm  \epsfysize=1.48cm \epsfbox{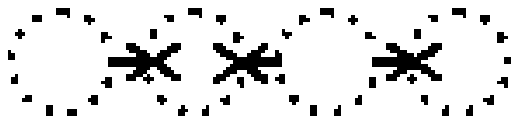}}
\\ G71:\ $4B\Xi$
            } &
\shortstack{
{\epsfxsize=3.39cm  \epsfysize=2.86cm \epsfbox{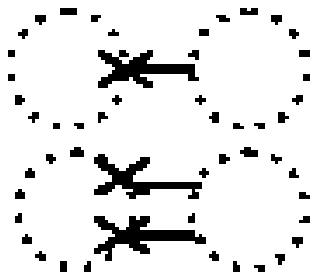}}
\\ G72:\ $B3P$
            } 
\\
\shortstack{
{\epsfxsize=4.30cm  \epsfysize=3.92cm \epsfbox{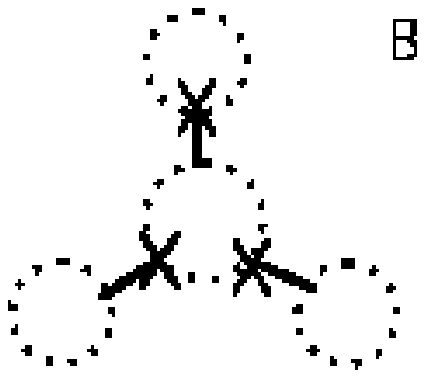}}
\\ G73:\ $4C\Th$
            } &
\shortstack{
{\epsfxsize=3.63cm  \epsfysize=3.17cm \epsfbox{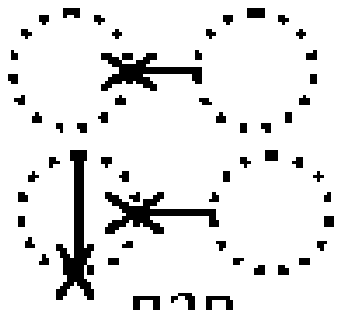}}
\\ G74:\ $B2P$
            } &
\shortstack{
{\epsfxsize=5.47cm  \epsfysize=1.34cm \epsfbox{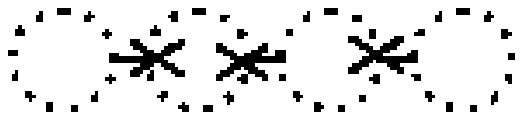}}
\\ G75:\ $4D\Xi$
            } 
\\
\shortstack{
{\epsfxsize=5.75cm  \epsfysize=1.38cm \epsfbox{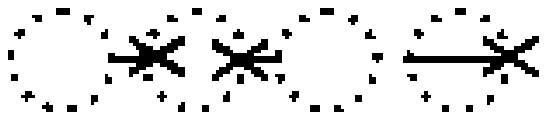}}
\\ G76:\ $C2Q$
            } &
\shortstack{
{\epsfxsize=4.23cm  \epsfysize=3.60cm \epsfbox{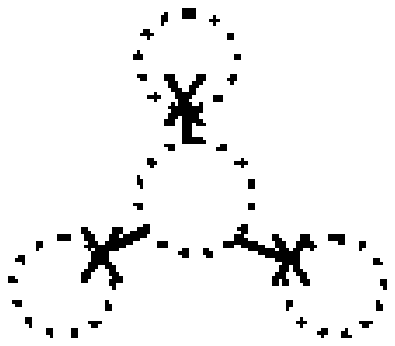}}
\\ G77:\ $4E\Th$
            } &
\shortstack{
{\epsfxsize=5.43cm  \epsfysize=1.41cm \epsfbox{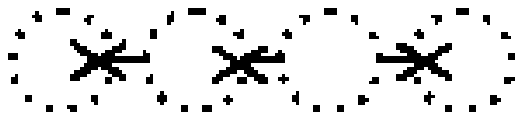}}
\\ G78:\ $4F\Xi$
            } 
\\
\shortstack{
{\epsfxsize=5.54cm  \epsfysize=1.38cm \epsfbox{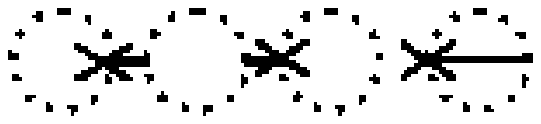}}
\\ G79:\ $C1Q$
            } &
\shortstack{
{\epsfxsize=3.92cm  \epsfysize=3.49cm \epsfbox{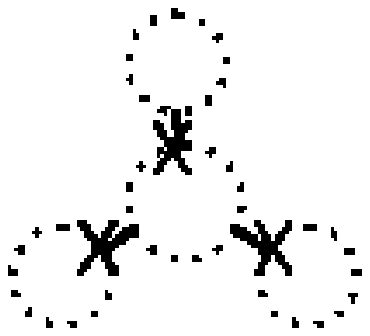}}
\\ G80:\ $4G\Th$
            } &
\shortstack{
{\epsfxsize=5.43cm  \epsfysize=1.69cm \epsfbox{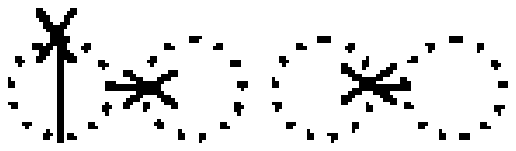}}
\\ G81:\ $B1P$
            }
\end{tabular}
\caption{
%*G70-85a*
(iv) $\ul{l}=4$\ (\ 8(con)+8(discon)=16 terms\ G70-85, No.1}
\label{G70-85a}
\end{figure}

\begin{figure}
\begin{tabular}{p{5cm}p{5cm}p{5cm}}

\shortstack{
{\epsfxsize=5.50cm  \epsfysize=1.41cm \epsfbox{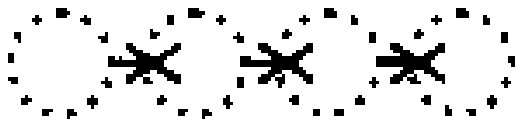}}
\\ G82:\ $4H\Xi$
            } &
\shortstack{
{\epsfxsize=5.57cm  \epsfysize=1.41cm \epsfbox{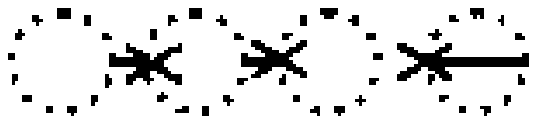}}
\\ G83:\ $C3Q$
            } &
\shortstack{
{\epsfxsize=3.39cm  \epsfysize=3.00cm \epsfbox{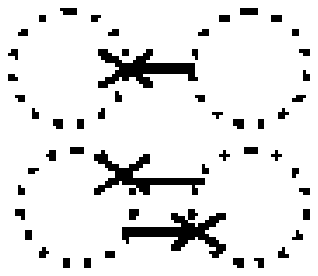}}
\\ G84:\ $B4P$
            } 
\\
\shortstack{
{\epsfxsize=3.35cm  \epsfysize=3.00cm \epsfbox{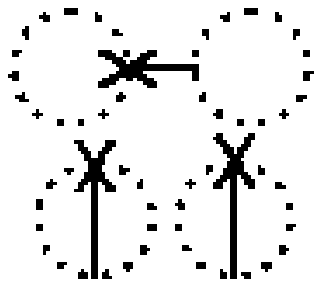}}
\\ G85:\ $PQQ$
            } & &

\end{tabular}
\caption{
%*G70-85b*
(iv) $\ul{l}=4$\ (\ 8(con)+8(discon)=16 terms\ G70-85, No.2}
\label{G70-85b}
\end{figure}

\begin{figure}
\begin{tabular}{p{5cm}p{5cm}p{5cm}}

\shortstack{
{\epsfxsize=4.16cm  \epsfysize=2.68cm \epsfbox{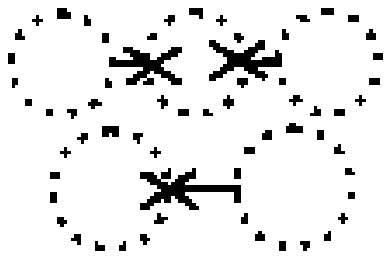}}
\\ G86:\ $C2P$
            } &
\shortstack{
{\epsfxsize=4.02cm  \epsfysize=2.68cm \epsfbox{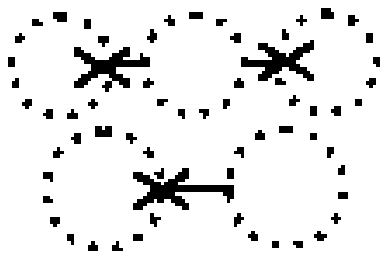}}
\\ G87:\ $C1P$
            } &
\shortstack{
{\epsfxsize=5.01cm  \epsfysize=2.82cm \epsfbox{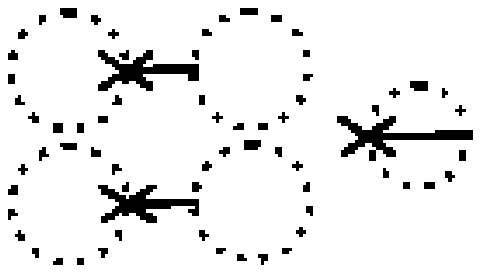}}
\\ G88:\ $PPQ$
            } 
\\
\shortstack{
{\epsfxsize=4.06cm  \epsfysize=2.86cm \epsfbox{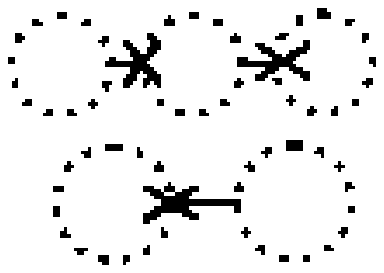}}
\\ G89:\ $C3P$
            } & &

\end{tabular}
\caption{
%*G86-89*
(v) $\ul{l}=5$\ (\ 0(con)+4(discon)=4 terms,\ G86-89}
\label{G86-89}
\end{figure}

\begin{figure}
\begin{tabular}{p{5cm}p{5cm}p{5cm}}

\shortstack{
{\epsfxsize=3.56cm  \epsfysize=4.27cm \epsfbox{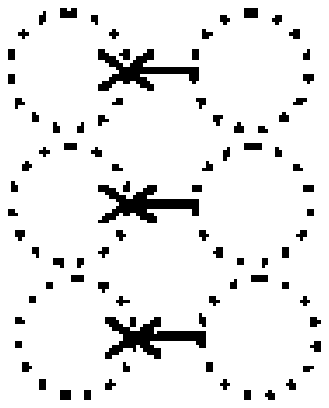}}
\\ G90:\ $PPP$
            } & &

\end{tabular}
\caption{
%*G90*
(vi) $\ul{l}=6$\ (\ 0(con)+1(discon)=1 term,\ G90}
\label{G90}

\end{figure}

%%%%%%%%%%%%%%%%%%%%%%%%%%%%%  l=1 %%%%%%%%%%%%%%%%%%%%%%%%%%%%%%
\flushleft{(i) $\ul{l}=1$\ (\ 13(con)+0(discon)=13 terms\ ),
Fig.\ref{G1-13}.}
%%%%%%%%%%%%%%%%%%%%%%%%%%%%%%%%%%%%%%%%%%%%%%%%%%%%%%%%%%%%%%%%%%%
%%%%%%%%%%%%%%%%%%%%%%%%%%%%%  l=2 %%%%%%%%%%%%%%%%%%%%%%%%%%%%%%
\flushleft{(ii) $\ul{l}=2$\ (\ 26(con)+3(discon)=29\ terms\ ),
Fig.\ref{G14-42a} and Fig.\ref{G14-42b}.}
%%%%%%%%%%%%%%%%%%%%%%%%%%%%%%%%%%%%%%%%%%%%%%%%%%%%%%%%%%%%%%%%%%%
%%%%%%%%%%%%%%%%%%%%%%%%%%%%%  l=3 %%%%%%%%%%%%%%%%%%%%%%%%%%%%%%
\flushleft{(iii) $\ul{l}=3$\ (\ 19(con)+8(discon)=27 terms\ ),
Fig.\ref{G43-69a} and Fig.\ref{G43-69b}.}
%%%%%%%%%%%%%%%%%%%%%%%%%%%%%%%%%%%%%%%%%%%%%%%%%%%%%%%%%%%%%%%%%%%
%%%%%%%%%%%%%%%%%%%%%%%%%%%%%  l=4 %%%%%%%%%%%%%%%%%%%%%%%%%%%%%%
\flushleft{(iv) $\ul{l}=4$\ (\ 8(con)+8(discon)=16 terms\ ),
Fig.\ref{G70-85a} and Fig.\ref{G70-85b}.}
%%%%%%%%%%%%%%%%%%%%%%%%%%%%%%%%%%%%%%%%%%%%%%%%%%%%%%%%%%%%%%%%%%%
%%%%%%%%%%%%%%%%%%%%%%%%%%%%%  l=5 %%%%%%%%%%%%%%%%%%%%%%%%%%%%%%
\flushleft{(v) $\ul{l}=5$\ (\ 0(con)+4(discon)=4 terms\ ),
Fig.\ref{G86-89}.}
%%%%%%%%%%%%%%%%%%%%%%%%%%%%%%%%%%%%%%%%%%%%%%%%%%%%%%%%%%%%%%%%%%%
%%%%%%%%%%%%%%%%%%%%%%%%%%%%%  l=6 %%%%%%%%%%%%%%%%%%%%%%%%%%%%%%
\flushleft{(vi) $\ul{l}=6$\ (\ 0(con)+1(discon)=1 term\ ),
Fig.\ref{G90}.}
%%%%%%%%%%%%%%%%%%%%%%%%%%%%%%%%%%%%%%%%%%%%%%%%%%%%%%%%%%%%%%%%%%%
\vs 1

%%%%%%%%%%%%%%%%%%%%%%%%%%%%%%%%%%%%%%%%%%%%%%%%%%%%%%%%%%%%%%%%%%%
%%%%%%%%%%%%%%%%%%%%%%%  Fig.G1-13  %%%%%%%%%%%%%%%%%%%%%%%%%%%%%%%%%%%%
%%%%%%%%%%%%%%%%%%%%%%%%%%%%%%%%%%%%%%%%%%%%%%%%%%%%%%%%%%%%%%%%%%%%%

%%%%%%%%%%%%%%%%%%%%%%%%%%%%%  l=2  No.1  %%%%%%%%%%%%%%%%%%%%%%%%%%%%%%
%\flushleft{(ii) $\ul{l}=2$\ (\ 26(con)+3(discon)=29\ terms\ ),Fig.A.2.}
%%%%%%%%%%%%%%%%%%%%%%%%%%%%%%%%%%%%%%%%%%%%%%%%%%%%%%%%%%%%%%%%%%%
%%%%%%%%%%%%%%%%%%%%%%%  Fig.G14-42  %%%%%%%%%%%%%%%%%%%%%%%%%%%%%%%%%%%%
%\newpage

%%%%%%%%%%%%%%%%%%%%%%%%%  l=2, No.2  %%%%%%%%%%%%%%%%%%%%
%%%%%%%%%%%%%%%%%%%%%%%%%%%%%%%%%%%%%%%%%%%%%%%%%%%%%%%%%%%%%%%%%%%%%
%\newpage

%%%%%%%%%%%%%%%%%%%%%%%%%%%%%  l=3  No.1 %%%%%%%%%%%%%%%%%%%%%%%%%%%%%%
%\flushleft{(iii) $l=3$\ (\ 19(con)+8(discon)=27 terms\ ),Fig.A.3.}
%%%%%%%%%%%%%%%%%%%%%%%%%%%%%%%%%%%%%%%%%%%%%%%%%%%%%%%%%%%%%%%%%%

%%%%%%%%%%%%%%%%%%%%%%%  Fig.G43-69  %%%%%%%%%%%%%%%%%%%%%%%%%%%%%%%%%%%%

%%%%%%%%%%%%%%%%%%%%%%%%%%  l=3  No.2  %%%%%%%%%%%%%%%%%
%\newpage
%%%%%%%%%%%%%%%%%%%%%%%%%%%%%%%%%%%%%%%%%%%%%%%%%%%%%%%%%%%%%%%%%%%%%
\vs{12}

%%%%%%%%%%%%%%%%%%%%%%%%%%%%%  l=4  No 1  %%%%%%%%%%%%%%%%%%%%%%%%%%%%%%
%\flushleft{(iv) $l=4$\ (\ 8(con)+8(discon)=16 terms\ ),Fig.A.4.}
%%%%%%%%%%%%%%%%%%%%%%%%%%%%%%%%%%%%%%%%%%%%%%%%%%%%%%%%%%%%%%%%%%%

%%%%%%%%%%%%%%%%%%%%%%%  Fig.G70-85  %%%%%%%%%%%%%%%%%%%%%%%%%%%%%%%%%%%%

%%%%%%%%%%%%%%%%%%%%%%% l=4  No2  %%%%%%%%%%%%%%%%%%%%%%%%%%%%%%%%%%%%
\newpage
%%%%%%%%%%%%%%%%%%%%%%%%%%%%%%%%%%%%%%%%%%%%%%%%%%%%%%%%%%%%%%%%%%%%%

%%%%%%%%%%%%%%%%%%%%%%%%%%%%%  l=5 %%%%%%%%%%%%%%%%%%%%%%%%%%%%%%
%\flushleft{(v) $l=5$\ (\ 0(con)+4(discon)=4 terms\ ),Fig.A.5.}
%%%%%%%%%%%%%%%%%%%%%%%%%%%%%%%%%%%%%%%%%%%%%%%%%%%%%%%%%%%%%%%%%%%

%%%%%%%%%%%%%%%%%%%%%%%  Fig.G86-89  %%%%%%%%%%%%%%%%%%%%%%%%%%%%%%%%%%%%
%%%%%%%%%%%%%%%%%%%%%%%%%%%%%%%%%%%%%%%%%%%%%%%%%%%%%%%%%%%%%%%%%%%%%

%%%%%%%%%%%%%%%%%%%%%%%%%%%%%  l=6 %%%%%%%%%%%%%%%%%%%%%%%%%%%%%%
%\flushleft{(vi) $l=6$\ (\ 0(con)+1(discon)=1 term\ ),Fig.A.6.}
%%%%%%%%%%%%%%%%%%%%%%%%%%%%%%%%%%%%%%%%%%%%%%%%%%%%%%%%%%%%%%%%%%%

%%%%%%%%%%%%%%%%%%%%%%%  Fig.G90  %%%%%%%%%%%%%%%%%%%%%%%%%%%%%%%%%%%%
%%%%%%%%%%%%%%%%%%%%%%%%%%%%%%%%%%%%%%%%%%%%%%%%%%%%%%%%%%%%%%%%%%%%%

\newpage
%%%%%%%%%%%%%%%%%%%%%%%%%%  App. B  %%%%%%%%%%%%%%%%%%%%%%%%%%%%%%%
%%%%%%%%%%%%%%%%%%%%%%%%%%%%%%%%%%%%%%%%%%%%%%%%%%%%%%%%%%%%%%%%%%%
\begin{flushleft}
{\Large\bf Appendix B.\ List of Indices of All $(\pl\pl h)^3$-Invariants}
\end{flushleft}

Every graph can be specified completely by a set of indices
which expresses its topology( see Sec.II and V).
Lists of indices ($\ul{l}$,\ul{tadpoleno},\ul{tadtype},\ul{bcn},\ul{vcn})
and weight
are given in Table \ref{tab4} for G1-G13, 
in Table \ref{tab5} for G14-G42, in Table \ref{tab6} for G43-G69
and in Table \ref{tab7} for G70-90. This result is coded into the
program, which enables the present computer calculation.
In the tables, there is a column of 'fine splitting'.
These boxes show how to discriminate topologically quite similar
graphs which are those graphs with the same \#-number. It is explained
in Subsec.V.ii of the text.

\vs 2
%%%%%%%%%%%%%%%%%%%%%%%%%%%%%%%%%%%%%%%%%%%%%%%%%%%%%%%%%%%%%%%%%%%%%%%%%%
%%%%%%%%%%%%%%%%%%%%%  Table G1-G13 %%%%%%%%%%%%%%%%%%%%%%%%%%%%%
%%%%%%%%%%%%%%%%%%%%%%%%%%%%%%%%%%%%%%%%%%%%%%%%%%%%%%%%%%%%%%%%%%%%%%%%%%
\begin{table}
\begin{tabular}{|l||c|c|c|c|l|}
\hline
Graph No:Graph  
      & $\ul{l}$ & \ul{tadpo} & \ul{tadtype} & (\ul{bcn},\ul{vcn})& `fine \\
Name,weight  &   & \ul{leno}  &              &                 & splitting'\\
\hline
$G1:1A\Th,\ 128$  
 &  $1$  & $0$   & n   & $(3,6)$   & \\
\hline
$G2:1A\Xi,\ 192$  
 &  $1$  & $0$   & n   & $(4,6)$   & \\
\hline
$G3:1A\Om_2,\ 64$  \nl
 &  $1$  & $0$   & n   & $(6,6)$   & \\
\hline
$G4:1B\Th,\ 384$
 &  $1$  & $0$   & n   & $(3,4)$   & \\
\hline
$G5:1B\Si-a,\ 384$ 
 &  $1$  & $0$   & n   & $(5,4)$   &\#1\ \ul{Vorder} \\
\hline
$G6:1B\Si-b,\ 384$
 &  $1$  & $0$   & n   & $(5,4)$   &\#1\ \ul{Vorder} \\
\hline
$G7:1B\Om_1,\ 384$
 &  $1$  & $0$   & n   & $(6,4)$   & \\
\hline
$G8:1B\Xi,\ 384$ 
 &  $1$  & $0$   & n   & $(4,4)$   & \\
\hline
$G9:1B\Si-c,\ 384$  \nl
 &  $1$  & $0$   & n   & $(5,4)$   &\#1\ \ul{Vorder} \\
\hline
$G10:1C\Xi,\ 192$
 &  $1$  & $0$   & n   & $(4,2)$  & \\
\hline
$G11:1C\Si,\ 384$  
 &  $1$  & $0$   & n   & $(5,2)$  & \\
\hline
$G12:1C\Om_1,\ 384$
 &  $1$  & $0$   & n   & $(6,2)$   &\#2\ \ul{crossno} \\
\hline
$G13:1C\Om_2,\ 192$
 &  $1$  & $0$   & n   & $(6,2)$   &\#2\ \ul{crossno} \\
\hline
%\multicolumn{6}{c}{\q}                                                 \\
%\multicolumn{6}{c}{Table B.1\ \ Index list of SO(n)-Invariants 
%$(\pl\pl h)^3$. }\\
%\multicolumn{6}{c}{G1-G13(\ul{l}=1)    }\\
\end{tabular}
\caption{ 
%*tab4*
Index list of SO(n)-Invariants 
$(\pl\pl h)^3$. G1-G13($\ul{l}$=1)    }
\label{tab4}
\end{table}
%%%%%%%%%%%%%%%%%%%%%%%%%  END  of  Table G1-G13 %%%%%%%%%%%%%%%%%%%%%%%%%%%%%
%%%%%%%%%%%%%%%%%%%%%%%%%%%%%%%%%%%%%%%%%%%%%%%%%%%%%%%%%%%%%%%%%%%%%%%%%

\newpage
%%%%%%%%%%%%%%%%%%%%%%%%%%%%%%%%%%%%%%%%%%%%%%%%%%%%%%%%%%%%%%%%%%%%%%%%%%
%%%%%%%%%%%%%%%%%%%%%  Table G14-G42 %%%%%%%%%%%%%%%%%%%%%%%%%%%%%
%%%%%%%%%%%%%%%%%%%%%%%%%%%%%%%%%%%%%%%%%%%%%%%%%%%%%%%%%%%%%%%%%%%%%%%%%%
\begin{table}
\begin{tabular}{|l||c|c|c|c|l|}
\hline
Graph No:Graph  
      & $\ul{l}$ & \ul{tadpo} & \ul{tadtype} & (\ul{bcn},\ul{vcn})& `fine \\
Name,weight  &  & \ul{leno}  &              &                  & splitting'\\
\hline
$G14:2A\Om_1,\ 64$
 &  $2$  & $0$   & n   & $(3,0),(3,0)$   & \\
\hline
$G15:2D\Om_1,\ 192$  
 &  $2$  & $0$   & n   & $(3,2),(3,2)$   & \\
\hline
$G16:2D\Xi_1,\ 384$
 &  $2$  & $0$   & n   & $(2,2),(2,2)$   & \\
\hline
$G17:2B\Si_1,\ 192$
 &  $2$  & $0$   & n   & $(2,0),(3,2)$   &\#3\ \ul{dd(h)verno} \\
\hline
$G18:2B\Om_2,\ 96$  
 &  $2$  & $0$   & n   & $(2,0),(4,2)$   &\#4\ \ul{dd(h)verno} \\
\hline
$G19:2C\Si_1,\ 192$
 &  $2$  & $0$   & n   & $(2,0),(3,2)$   &\#3\ \ul{dd(h)verno} \\
\hline
$G20:2C\Om_2,\ 96$
 &  $2$  & $0$   & n   & $(2,0),(4,2)$   &\#4\ \ul{dd(h)verno} \\
\hline
$G21:2E_a\Si_1,\ 192$  
 &  $2$  & $0$   & n   & $(2,2),(3,2)$   & \\
\hline
$G22:2E_a\Om_2,\ 192$
 &  $2$  & $0$   & n   & $(2,2),(4,2)$   & \\
\hline
$G23:A2Q,\ 96$
 &  $2$  & $0$   & n   & $(0,2),(2,2)$   & \\
\hline
$G24:A1Q,\ 96$
 &  $2$  & $0$   & n   & $(0,2),(4,2)$   & \\
\hline
$G25:2E_b\Si_1,\ 192$
 &  $2$  & $0$   & n   & $(2,2),(3,4)$   & \\
\hline
$G26:A3Q,\ 96$
 &  $2$  & $0$   & n   & $(0,2),(2,4)$   & \\
\hline
$G27:2F_a\Th,\ 96$  
 &  $2$  & $1$   & $1$   & $(3,2)$   & \\
\hline
$G28:2F_a\Si_2-a,\ 96$
 &  $2$  & $1$   & $1$   & $(5,2)$   &\#5\ \ul{Vorder} \\
\hline
$G29:2F_a\Xi_2,\ 192$
 &  $2$  & $1$   & $1$   & $(4,2)$   & \\
\hline
$G30:2F_a\Si_2-b,\ 192$  
 &  $2$  & $1$   & $1$   & $(5,2)$   &\#5\ \ul{Vorder} \\
\hline
$G31:2F_b\Xi_2,\ 192$
 &  $2$  & $1$   & $1$   & $(4,4)$   & \\
\hline
$G32:2F_b\Th-a,\ 192$
 &  $2$  & $1$   & $1$   & $(3,4)$   &\#6\ \ul{Vorder} \\
\hline
$G33:2F_b\Th-b,\ 96$  
 &  $2$  & $1$   & $1$   & $(3,4)$   &\#6\ \ul{Vorder} \\
\hline
$G34:2F_b\Si_2,\ 96$
 &  $2$  & $1$   & $1$   & $(5,4)$   & \\
\hline
$G35:2G_a\Th,\ 96$
 &  $2$  & $1$   & $0$   & $(3,2)$   & \\
\hline
$G36:2G_a\Si_2-a,\ 96$  
 &  $2$  & $1$   & $0$   & $(5,2)$   &\#7\ \ul{Vorder} \\
\hline
$G37:2G_a\Xi_2,\ 192$
 &  $2$  & $1$   & $0$   & $(4,2)$   & \\
\hline
$G38:2G_a\Si_2-b,\ 192$
 &  $2$  & $1$   & $0$   & $(5,2)$   &\#7\ \ul{Vorder} \\
\hline
$G39:2G_b\Th-a,\ 96$  
 &  $2$  & $1$   & $0$   & $(3,4)$   &\#8\ \ul{Vorder} \\
\hline
$G40:2G_b\Si_2,\ 96$
 &  $2$  & $1$   & $0$   & $(5,4)$   & \\
\hline
$G41:2G_b\Xi_2,\ 192$
 &  $2$  & $1$   & $0$   & $(4,4)$   & \\
\hline
$G42:2G_b\Th-b,\ 192$  
 &  $2$  & $1$   & $0$   & $(3,4)$   &\#8\ \ul{Vorder} \\
\hline
%\multicolumn{6}{c}{\q}                                                 \\
%\multicolumn{6}{c}{Table B.2\ \ Index list of SO(n)-Invariants 
%$(\pl\pl h)^3$. }\\
%\multicolumn{6}{c}{G14-G42(\ul{l}=2)    }\\
\end{tabular}
\caption{
%*tab5*
Index list of SO(n)-Invariants 
$(\pl\pl h)^3$. G14-G42($\ul{l}$=2)    }
\label{tab5}
\end{table}
%%%%%%%%%%%%%%%%%%%%%%%%%  END  of  Table G14-G42 %%%%%%%%%%%%%%%%%%%%%%%%%%%%%
%%%%%%%%%%%%%%%%%%%%%%%%%%%%%%%%%%%%%%%%%%%%%%%%%%%%%%%%%%%%%%%%%%%%%%%%%

\newpage
%%%%%%%%%%%%%%%%%%%%%%%%%%%%%%%%%%%%%%%%%%%%%%%%%%%%%%%%%%%%%%%%%%%%%%%%%%
%%%%%%%%%%%%%%%%%%%%%  Table G43-G69 %%%%%%%%%%%%%%%%%%%%%%%%%%%%%
%%%%%%%%%%%%%%%%%%%%%%%%%%%%%%%%%%%%%%%%%%%%%%%%%%%%%%%%%%%%%%%%%%%%%%%%%%
\begin{table}
\begin{tabular}{|l||c|c|c|c|l|}
\hline
Graph No:Graph  
      & $\ul{l}$ & \ul{tadpo} & \ul{tadtype} & (\ul{bcn},\ul{vcn})& `fine \\
Name,weight  &   & \ul{leno}  &              &                 & splitting'\\
\hline
$G43:3A\Xi_1,\ 48$
 &  $3$  & $2$   & $0,0$   & $(4,2)$   & \\
\hline
$G44:3A\Th,\ 96$  
 &  $3$  & $2$   & $0,0$   & $(3,2)$   & \\
\hline
$G45:3B\Si,\ 48$
 &  $3$  & $1$   & $0$   & $(2,0),(3,0)$   & \\
\hline
$G46:3C\Si,\ 48$
 &  $3$  & $1$   & $1$   & $(2,0),(3,0)$   & \\
\hline
$G47:3D\Si,\ 48$  
 &  $3$  & $1$   & $1$   & $(2,0),(3,2)$   & \\
\hline
$G48:3D\Xi_2,\ 96$
 &  $3$  & $1$   & $1$   & $(2,0),(2,2)$   & \\
\hline
$G49:3E\Om,\ 48$
 &  $3$  & $0$   &  n  & $(2,0),(2,0),(2,2)$   & \\
\hline
$G50:B3Q,\ 24$  
 &  $3$  & $0$   & n   & $(0,2),(2,0),(2,0)$   & \\
\hline
$G51:3F_a\Th,\ 96$
 &  $3$  & $2$   & $1,0$   & $(3,2)$   & \\
\hline
$G52:3F_a\Xi_1,\ 96$
 &  $3$  & $2$   & $1,0$   & $(4,2)$   & \\
\hline
$G53:A2P,\ 48$  
 &  $3$  & $2$   & $0,1$   & $(2,2)$   & \\
\hline
$G54:A1P,\ 48$
 &  $3$  & $2$   & $0,1$   & $(4,2)$   & \\
\hline
$G55:3F_b\Th,\ 96$
 &  $3$  & $2$   & $0,1$   & $(3,4)$   & \\
\hline
$G56:A3P,\ 48$  
 &  $3$  & $2$   & $0,1$   & $(2,4)$   & \\
\hline
$G57:3G\Si,\ 48$
 &  $3$  & $1$   & $0$   & $(2,0),(3,2)$   & \\
\hline
$G58:3G\Xi_2,\ 96$
 &  $3$  & $1$   & $0$   & $(2,0),(2,2)$   & \\
\hline
$G59:3H\Si,\ 96$  
 &  $3$  & $1$   & $0$   & $(2,2),(3,2)$   & \\
\hline
$G60:3H\Xi_2,\ 96$
 &  $3$  & $1$   & $0$   & $(2,2),(2,2)$   & \\
\hline
$G61:B2Q,\ 96$
 &  $3$  & $1$   & $0$   & $(0,2),(2,2)$   & \\
\hline
$G62:3I\Th,\ 96$  
 &  $3$  & $2$   & $1,1$   & $(3,2)$   & \\
\hline
$G63:3I\Xi_1,\ 48$
 &  $3$  & $2$   & $1,1$   & $(4,2)$   & \\
\hline
$G64:3J\Si,\ 96$
 &  $3$  & $1$   & $1$   & $(2,2),(3,2)$   & \\
\hline
$G65:3J\Xi_2,\ 96$  
 &  $3$  & $1$   & $1$   & $(2,2),(2,2)$   & \\
\hline
$G66:B1Q,\ 96$
 &  $3$  & $1$   & $1$   & $(0,2),(2,2)$   & \\
\hline
$G67:3K\Om,\ 16$
 &  $3$  & $0$   & n   & $(2,2),(2,2),(2,2)$   & \\
\hline
$G68:B4Q,\ 24$  
 &  $3$  & $0$   & n   & $(0,2),(2,2),(2,2)$   & \\
\hline
$G69:QQQ,\ 8$
 &  $3$  & $0$   & n   & $(0,2),(0,2),(0,2)$   & \\
\hline
%\multicolumn{6}{c}{\q}                                                 \\
%\multicolumn{6}{c}{Table B.3\ \ Index list of SO(n)-Invariants 
%$(\pl\pl h)^3$. }\\
%\multicolumn{6}{c}{G43-G69(\ul{l}=3)    }\\
\end{tabular}
\caption{
%*tab6*
Index list of SO(n)-Invariants 
$(\pl\pl h)^3$. G43-G69($\ul{l}$=3)    }
\label{tab6}
\end{table}
%%%%%%%%%%%%%%%%%%%%%%%%%  END  of  Table G43-G69 %%%%%%%%%%%%%%%%%%%%%%%%%%%%%
%%%%%%%%%%%%%%%%%%%%%%%%%%%%%%%%%%%%%%%%%%%%%%%%%%%%%%%%%%%%%%%%%%%%%%%%%

\newpage
%%%%%%%%%%%%%%%%%%%%%%%%%%%%%%%%%%%%%%%%%%%%%%%%%%%%%%%%%%%%%%%%%%%%%%%%%%
%%%%%%%%%%%%%%%%%%%%%  Table G70-G85,G86-G89,G90 %%%%%%%%%%%%%%%%%%%%%%%%%%%%%
%%%%%%%%%%%%%%%%%%%%%%%%%%%%%%%%%%%%%%%%%%%%%%%%%%%%%%%%%%%%%%%%%%%%%%%%%%
\begin{table}
\begin{tabular}{|l||c|c|c|c|l|}
\hline
Graph No:Graph  
      & $\ul{l}$ & \ul{tadpo} & \ul{tadtype} & (\ul{bcn},\ul{vcn})& `fine \\
Name,weight  &   & \ul{leno}  &              &                  & splitting'\\
\hline
$G70:4A\Th,\ 8$
 &  $4$  & $3$   & $0,0,0$    & $(3,0)$   & \\
\hline
$G71:4B\Xi,\ 24$  
 &  $4$  & $2$   & $0,1$   & $(2,0),(2,0)$   &\#9\ \ul{disconnect} \\
\hline
$G72:B3P,\ 12$
 &  $4$  & $2$   & $0,1$   & $(2,0),(2,0)$   &\#9\ \ul{disconnect} \\
\hline
$G73:4C\Th,\ 24$
 &  $4$  & $3$   & $0,0,1$   & $(3,2)$   & \\
\hline
$G74:B2P,\ 48$  
 &  $4$  & $3$   & $0,0,1$   & $(2,2)$   & \\
\hline
$G75:4D\Xi,\ 24$
 &  $4$  & $2$   & $0,0$   & $(2,0),(2,2)$   & \\
\hline
$G76:C2Q,\ 12$
 &  $4$  & $2$   & $0,0$   & $(0,2),(2,0)$   & \\
\hline
$G77:4E\Th,\ 8$  
 &  $4$  & $3$   & $1,1,1$   & $(3,0)$   & \\
\hline
$G78:4F\Xi,\ 24$
 &  $4$  & $2$   & $1,1$   & $(2,0),(2,2)$   & \\
\hline
$G79:C1Q,\ 12$
 &  $4$  & $2$   & $1,1$   & $(0,2),(2,0)$   & \\
\hline
$G80:4G\Th,\ 24$  
 &  $4$  & $3$   & $0,1,1$   & $(3,2)$   & \\
\hline
$G81:B1P,\ 48$
 &  $4$  & $3$   & $0,1,1$   & $(2,2)$   & \\
\hline
$G82:4H\Xi,\ 24$
 &  $4$  & $2$   & $0,1$   & $(2,2),(2,2)$   &\#10\ \ul{disconnect} \\
\hline
$G83:C3Q,\ 24$  
 &  $4$  & $2$   & $0,1$   & $(0,2),(2,2)$   & \\
\hline
$G84:B4P,\ 12$
 &  $4$  & $2$   & $0,1$   & $(2,2),(2,2)$   &\#10\ \ul{disconnect} \\
\hline
$G85:PQQ,\ 12$
 &  $4$  & $2$   & $0,1$   & $(0,2),(0,2)$   & \\
\hline
\hline
$G86:C2P,\ 6$
 &  $5$  & $4$   & $0,0,0,1$   & $(2,0)$   & \\
\hline
$G87:C1P,\ 6$  
 &  $5$  & $4$   & $0,1,1,1$   & $(2,0)$   & \\
\hline
$G88:PPQ,\ 6$
 &  $5$  & $4$   & $0,0,1,1$   & $(0,2)$   & \\
\hline
$G89:C3P,\ 12$
 &  $5$  & $4$   & $0,0,1,1$   & $(2,2)$   & \\
\hline
\hline
$G90:PPP,\ 1$
 &  $6$  & $6$   & $0,0,0,1,1,1$   & n   & \\
\hline
%\multicolumn{6}{c}{\q}                                                 \\
%\multicolumn{6}{c}{Table B.4\ \ Index list of SO(n)-Invariants 
%$(\pl\pl h)^3$. }\\
%\multicolumn{6}{c}{G70-G85(\ul{l}=4),G86-G89(\ul{l}=5),G90(\ul{l}=6)    }\\
\end{tabular}
\caption{
%*tab7*
Index list of SO(n)-Invariants 
$(\pl\pl h)^3$. G70-G85($\ul{l}$=4),G86-G89($\ul{l}$=5),G90($\ul{l}$=6) }
\label{tab7}
\end{table}
%%%%%%%%%%%%%%%%%%%%%%%%%  END  of  Table G70-G85,G86-G89,G90  %%%%%%%%
%%%%%%%%%%%%%%%%%%%%%%%%%%%%%%%%%%%%%%%%%%%%%%%%%%%%%%%%%%%%%%%%%%%%%%%%%

\newpage
%%%%%%%%%%%%%%%%%%%%%%%%%%  App. C  %%%%%%%%%%%%%%%%%%%%%%%%%%%%%%%
%%%%%%%%%%%%%%%%%%%%%%%%%%%%%%%%%%%%%%%%%%%%%%%%%%%%%%%%%%%%%%%%%%%
\begin{flushleft}
{\Large\bf Appendix C.\ Classification of 
$\pl^4 h\cdot\pl^2 h$-Invariants
and Weak-Expansion of $\na\na R\times R$-terms}
\end{flushleft}

\begin{figure}
   \centerline{
{\epsfxsize=105pt  \epsfysize=100pt \epsfbox{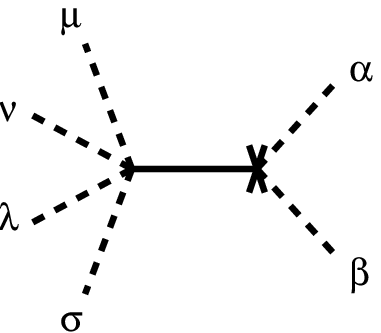}}
               }
\caption{
%*figCp1*Fig.C.1\ 
Graphs of 6-tensor $\pl_\m\pl_\n\pl_\la\pl_\si h_\ab$
        }
\label{figCp1}
\end{figure}

\begin{figure}
   \centerline{
{\epsfxsize=215pt  \epsfysize=100pt \epsfbox{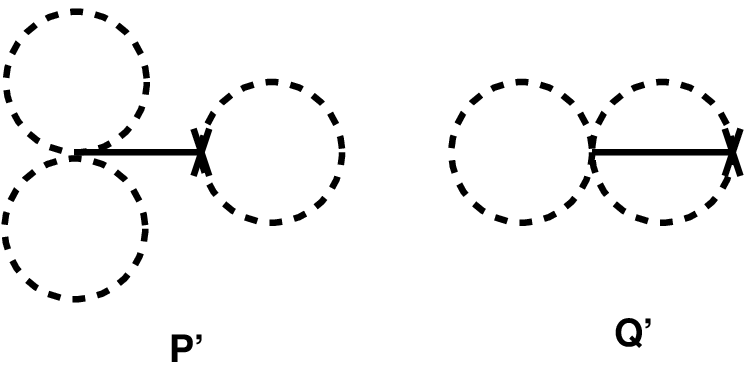}}
               }
\caption{
%*figCp2*Fig.C.2\ 
Graphs for
$P'\equiv \pl^2\pl^2 h_{\m\m}$ and $Q'\equiv \pl^2\pl_\m\pl_\n h_\mn$.
        }
\label{figCp2}
\end{figure}

\begin{figure}
\centerline{
{\epsfxsize=135pt  \epsfysize=55pt \epsfbox{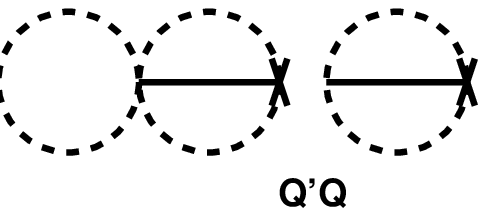}}
           }
\vspace{10pt}
\centerline{
{\epsfxsize=195pt  \epsfysize=60pt \epsfbox{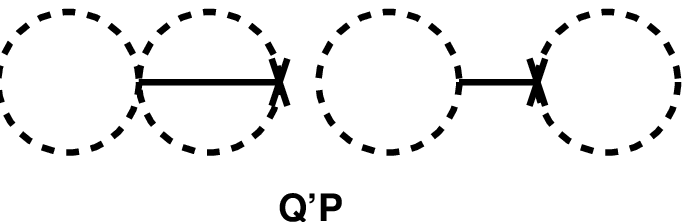}}\qq
{\epsfxsize=150pt  \epsfysize=90pt \epsfbox{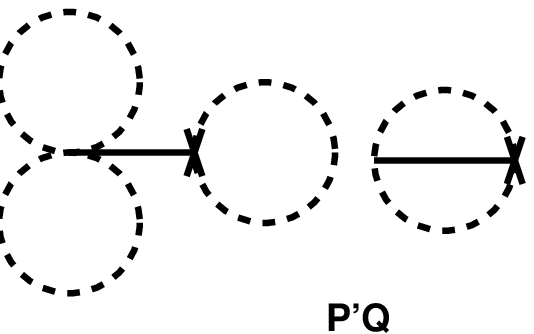}}
           }
\vspace{10pt}
\centerline{
{\epsfxsize=210pt  \epsfysize=95pt \epsfbox{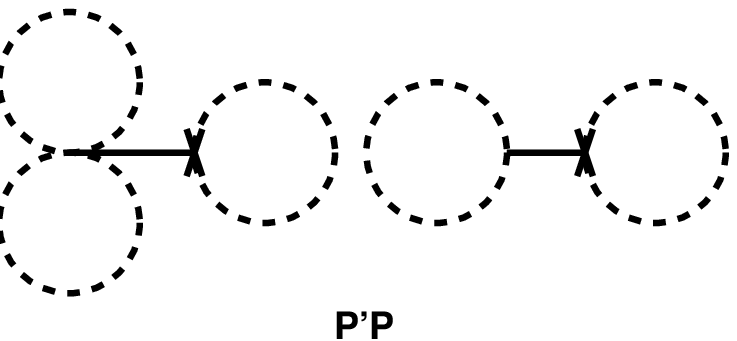}}
           }
\caption{
%*figCp3*Fig.C.3\  
Graphs for \ul{bridgeno}=0\ (\ disconnected )
}
\label{figCp3}
\end{figure}

\begin{figure}
\centerline{
{\epsfxsize=65pt  \epsfysize=95pt \epsfbox{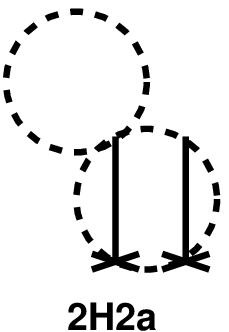}}\q
{\epsfxsize=65pt  \epsfysize=95pt \epsfbox{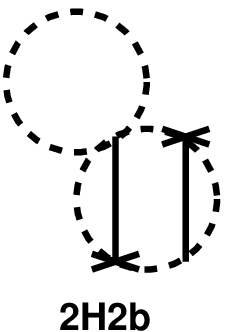}}\q
{\epsfxsize=86pt  \epsfysize=60pt \epsfbox{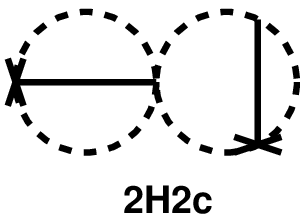}}
           }
\vspace{10pt}
\centerline{
{\epsfxsize=100pt  \epsfysize=95pt \epsfbox{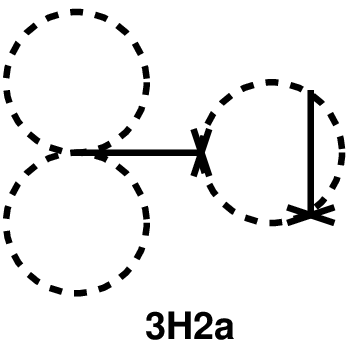}}\q
{\epsfxsize=100pt  \epsfysize=95pt \epsfbox{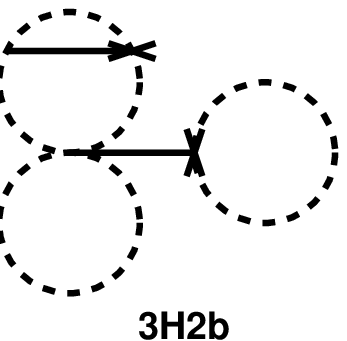}}\q
{\epsfxsize=107pt  \epsfysize=105pt \epsfbox{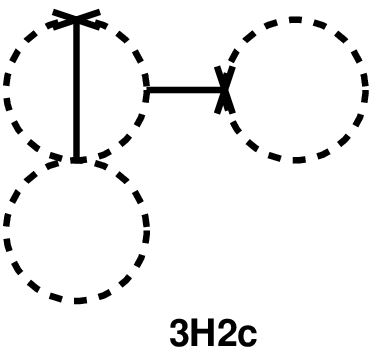}}
           }
\vspace{10pt}
\centerline{
{\epsfxsize=107pt  \epsfysize=106pt \epsfbox{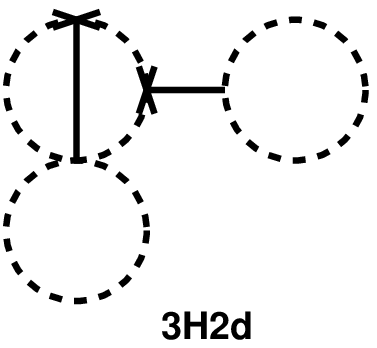}}\q
{\epsfxsize=147pt  \epsfysize=65pt \epsfbox{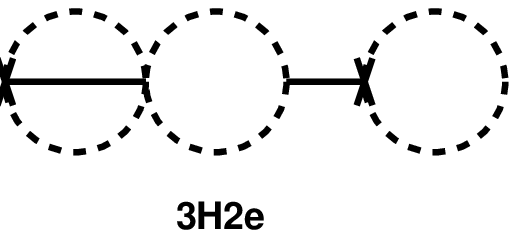}}\q
{\epsfxsize=145pt  \epsfysize=57pt \epsfbox{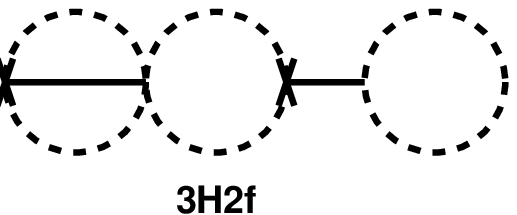}}
           }
\vspace{10pt}
\centerline{
{\epsfxsize=162pt  \epsfysize=95pt \epsfbox{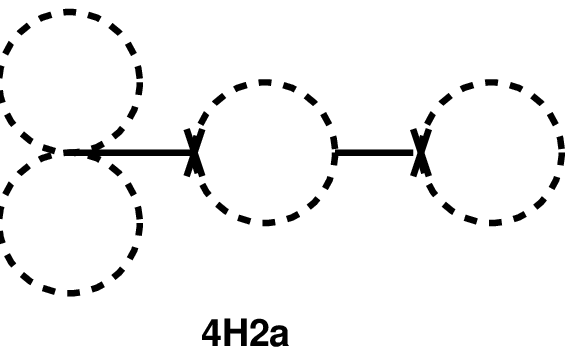}}\q
{\epsfxsize=163pt  \epsfysize=92pt \epsfbox{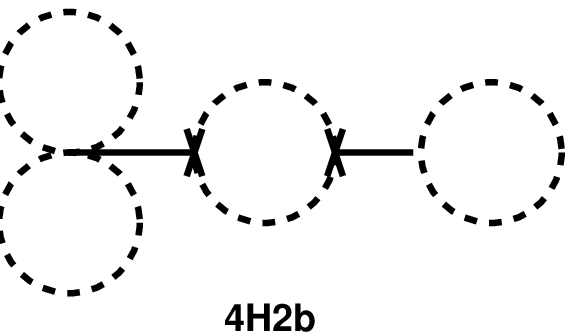}}
           }
\vspace{10pt}
\centerline{
{\epsfxsize=163pt  \epsfysize=102pt \epsfbox{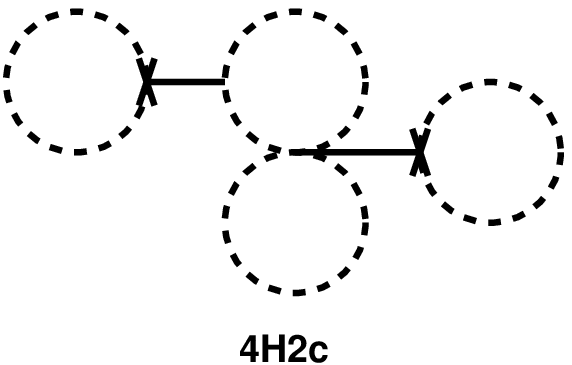}}\q
{\epsfxsize=163pt  \epsfysize=100pt \epsfbox{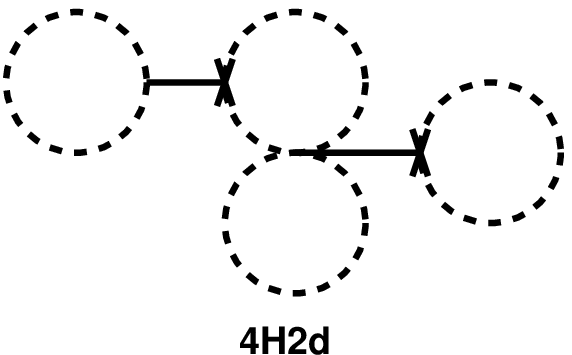}}
           }
\caption{
%*figCp4*Fig.C.4\  
Graphs for \ul{bridgeno}=2
}
\label{figCp4}
\end{figure}

\begin{figure}
\centerline{
{\epsfxsize=87pt  \epsfysize=63pt \epsfbox{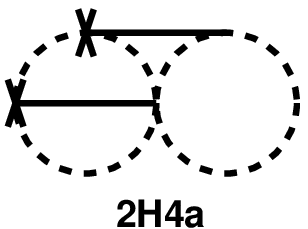}}\q
{\epsfxsize=87pt  \epsfysize=63pt \epsfbox{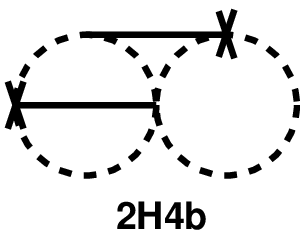}}\q
{\epsfxsize=87pt  \epsfysize=58pt \epsfbox{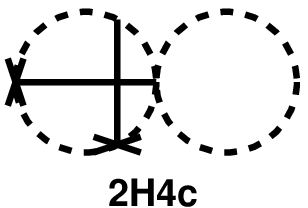}}
           }
\vspace{10pt}
\centerline{
{\epsfxsize=105pt  \epsfysize=95pt \epsfbox{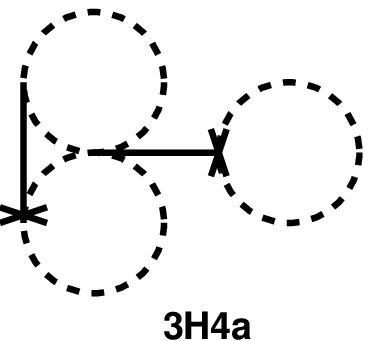}}\q
{\epsfxsize=96pt  \epsfysize=98pt \epsfbox{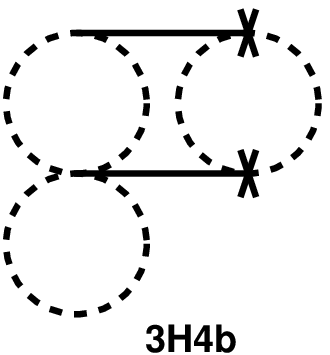}}\q
{\epsfxsize=95pt  \epsfysize=100pt \epsfbox{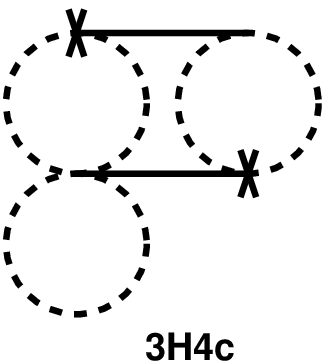}}
           }
\caption{
%*figCp5*Fig.C.5\  
Graphs for \ul{bridgeno}=4
}
\label{figCp5}
\end{figure}

The leading order of the weak-field expansion for $\na\na R\times R$-type
general invariants($T_1\sim T_4$), is given by a sum of
$\pl^4 h\cdot \pl^2 h$-invariants. 
%Besides Fig.2.1
%for the tensor $\pl_\m\pl_\n h_\ab$, 
In order to treat them graphically,
we introduce a graphical representation, 
in Fig.\ref{figCp1}, for a 6-tensor
$\pl_\m\pl_\n\pl_\la\pl_\si h_\ab$.
%%%%%%%%%%%%%%%%%%%%%%%  Fig.C.1  %%%%%%%%%%%%%%%%%%%%%%%%%%%%%%%%%%%%
%%%%%%%%%%%%%%%%%%%%%%%%%%%%%%%%%%%%%%%%%%%%%%%%%%%%%%%%%%%%%%%%%%%%%
There are two $\pl^4 h$-invariants, 
$P'\equiv \pl^2\pl^2 h_{\m\m}$ and $Q'\equiv \pl^2\pl_\m\pl_\n h_\mn$,
which are graphically shown in Fig.\ref{figCp2}.
%%%%%%%%%%%%%%%%%%%%%%%  Fig.C.2  %%%%%%%%%%%%%%%%%%%%%%%%%%%%%%%%%%%%
%%%%%%%%%%%%%%%%%%%%%%%%%%%%%%%%%%%%%%%%%%%%%%%%%%%%%%%%%%%%%%%%%%%%%
Let us consider$\pl^4 h\cdot \pl^2 h$-invariants and
list up all and independent ones.
For the classification, we must first
introduce a new index.
%%%%%%%%%%%%%%%%%%%%%%%%   Def    %%%%%%%%%%%%%%%%%%%%%%%%%%%%%%%%%%%%%%%
\begin{description}
\item[Def]\ 
Let us consider a general SO(n)-invariant of a binary type: 
$\pl^r h\cdot \pl^s h,\ r+s=\mbox{even}$.
(The case of $(r=4, s=2)$ is the present case.)
When we represent $(r+2)$-tensor $\pl^r h$ in a similar
way to Fig.\ref{figCp1} ($r=4$), the invariant 
$\pl^r h\cdot \pl^s h$ is
represented by a graph
with $(r+s+4)/2$ suffix-lines where each of them
connects two vertices in the graph. We define {\it bridge-lines}
as those suffix-lines which connect a vertex of one bond
with another vertex of the other bond. 
\end{description}
%%%%%%%%%%%%%%%%%%%%%%%%%%%%%%%%%%%%%%%%%%%%%%%%%%%%%%%%%%%%%%%%%%%%%%%%%%
%%%%%%%%%%%%%%%%%%%%%%%%   Def    %%%%%%%%%%%%%%%%%%%%%%%%%%%%%%%%%%%%%%%
\begin{description}
\item[Def]\ 
For a general SO(n)-invariant of a binary type: 
$\pl^r h\cdot \pl^s h,\ r+s=\mbox{even}$, we define
{\it bridge number} (\ul{bridgeno}) as the number of
bridge-lines of the graph.
\end{description}
%%%%%%%%%%%%%%%%%%%%%%%%%%%%%%%%%%%%%%%%%%%%%%%%%%%%%%%%%%%%%%%%%%%%%%%%%%
\ul{bridgeno} must be an even number
in this case because both $\pl^4 h$-tensor and $\pl^2 h$-tensor have
even number of suffixes. The classification is done by \ul{bridgeno}
and the number of suffix-loops, $\ul{l}$, as follows.
%%%%%%%%%%%%%%%%%%%%%%%%%%%%%  bridgeno=0 %%%%%%%%%%%%%%%%%%%%%%%%%%%%%%
\flushleft{(i) \ul{bridgeno}=0\ (\ disconnected ),Fig.\ref{figCp3}.}\nl
Q'Q($\ul{l}=3$);\ 
Q'P, P'Q ($\ul{l}=4$);\ 
P'P ($\ul{l}=5$)
%%%%%%%%%%%%%%%%%%%%%%%%%%%%%%%%%%%%%%%%%%%%%%%%%%%%%%%%%%%%%%%%%%%
%%%%%%%%%%%%%%%%%%%%%%%%%%%%%  bridgeno=2 %%%%%%%%%%%%%%%%%%%%%%%%%%%%%%
\flushleft{(ii) \ul{bridgeno}=2\ ,\ Fig.\ref{figCp4}.}\nl
2H2a, 2H2b, 2H2c ($\ul{l}=2$);\ 
3H2a, 3H2b, 3H2c, 3H2d, 3H2e, 3H2f ($\ul{l}=3$);\ 
4H2a, 4H2b, 4H2c, 4H2d ($\ul{l}=4$).
%%%%%%%%%%%%%%%%%%%%%%%%%%%%%%%%%%%%%%%%%%%%%%%%%%%%%%%%%%%%%%%%%%%
%%%%%%%%%%%%%%%%%%%%%%%%%%%%%  bridgeno=4 %%%%%%%%%%%%%%%%%%%%%%%%%%%%%%
\flushleft{(iii) \ul{bridgeno}=4\ ,\ Fig.\ref{figCp5}.}\nl
2H4a, 2H4b, 2H4c ($\ul{l}=2$);\ 
3H4a, 3H4b, 3H4c ($\ul{l}=3$).
%%%%%%%%%%%%%%%%%%%%%%%%%%%%%%%%%%%%%%%%%%%%%%%%%%%%%%%%%%%%%%%%%%%

%\vs {0.5}
%%%%%%%%%%%%%%%%%%%%%%%  Fig.C.3  %%%%%%%%%%%%%%%%%%%%%%%%%%%%%%%%%%%%
%%%%%%%%%%%%%%%%%%%%%%%%%%%%%%%%%%%%%%%%%%%%%%%%%%%%%%%%%%%%%%%%%%%%%

%%%%%%%%%%%%%%%%%%%%%%%  Fig.C.4  %%%%%%%%%%%%%%%%%%%%%%%%%%%%%%%%%%%%
%%%%%%%%%%%%%%%%%%%%%%%%%%%%%%%%%%%%%%%%%%%%%%%%%%%%%%%%%%%%%%%%%%%%%

%%%%%%%%%%%%%%%%%%%%%%%  Fig.C.5  %%%%%%%%%%%%%%%%%%%%%%%%%%%%%%%%%%%%
%%%%%%%%%%%%%%%%%%%%%%%%%%%%%%%%%%%%%%%%%%%%%%%%%%%%%%%%%%%%%%%%%%%%%

\vs {0.5}
\q In Table \ref{tab8}, the weak-field expansion of $T_1\sim T_4$ ,
the classification of $\pl^4 h\cdot \pl^2 h$-invariants  
and their weights are given. The total sum of weights is
$945=9\times 7\times 5\times 3\times 1$. 
We see $T_i$'s are independent each other.

\newpage
%%%%%%%%%%%%%%%%%%%%%%%%%%%%%%%%%%%%%%%%%%%%%%%%%%%%%%%%%%%%%%%%%%%%%%%%%%
%%%%%%%%%%%%%%%%%%%%%  Table AppC  %%%%%%%%%%%%%%%%%%%%%%%%%%%%%
%%%%%%%%%%%%%%%%%%%%%%%%%%%%%%%%%%%%%%%%%%%%%%%%%%%%%%%%%%%%%%%%%%%%%%%%%%
\begin{table}
\begin{tabular}{|c||c|c|c|c|c|c|c|}
\hline
bridge- & $\ul{l}$ & Graph  & Weight, & $T_1$ & $T_2$ & $T_3$ & $T_4$\\
no      &        & Name   & Total 945 &       &       &       &    \\
\hline\hline
        &   3    & Q'Q    & 24        & 1     & 0     & 0     & 0  \\
\cline{2-8}
0       &   4    & Q'P    & 12        & -1    & 0     & 0     & 0  \\
\cline{3-8}
        &        & P'Q    &  6        & -1    & 0     & 0     & 0   \\
\cline{2-8}
        &   5    & P'P    &  3        &  1    & 0     & 0     & 0    \\
\hline\hline
        &        & 2H2a   & 96        &  0    & $\half$& 0    & 0   \\
\cline{3-8}
        &   2    & 2H2b   & 96        &  0    & $\half$& 0    & 0   \\
\cline{3-8}
        &        & 2H2c   & 96        &  0    & 0      & 0    & 1   \\
\cline{2-8}
        &        & 3H2a   & 24        &  0    & $-\half$& 0   & 0 \\
\cline{3-8}
        &        & 3H2b   & 48        &  0    & $-\half$& 0   & -1 \\
\cline{3-8}
        &   3    & 3H2c   & 48        &  0    & $-\half$& 0   & 0  \\ 
\cline{3-8}
2       &        & 3H2d   & 48        &  0    & $-\half$& 0   & 0  \\
\cline{3-8}
        &        & 3H2e   & 24        &  0    & 0       & 0 &$-\half$\\
\cline{3-8}
        &        & 3H2f   & 24        &  0    & 0       & 0&$-\half$\\
\cline{2-8}
        &        & 4H2a   &  6        &  0    &$\fourth$& 0   & 0 \\
\cline{3-8}
        &   4    & 4H2b   &  6        &  0    &$\fourth$& 0   & 0 \\
\cline{3-8}
        &        & 4H2c   & 12        &  0    &$\fourth$& 0&$\half$\\
\cline{3-8}
        &        & 4H2d   & 12        &  0    &$\fourth$& 0&$\half$\\
\hline\hline
        &        & 2H4a   & 96        &  0    & 0       & 0   & 0 \\
\cline{3-8}
        &   2    & 2H4b   & 96        &  0    & 0       & 0   & 0 \\
\cline{3-8}
4       &        & 2H4c   & 96        &  0    & 0       & -2   & 0 \\
\cline{2-8}
        &        & 3H4a   & 24        &  0    & 0       & 0   & 0 \\
\cline{3-8}
        &   3    & 3H4b   & 24        &  0    & 0       & 1   & 0 \\
\cline{3-8}
        &        & 3H4c   & 24        &  0    & 0       & 1   & 0 \\
\hline
%\multicolumn{8}{c}{\q}                                                 \\
%\multicolumn{8}{c}{Table AppC\ \ Classification of
%$\pl^4 h\cdot \pl^2 h$-invariants, their weights and  }\\
%\multicolumn{8}{c}{weak-field expansion of 
%$\na\na R\times R$-type general invariants.
%                    }\\
\end{tabular}
\caption{ 
%*tab8*
Classification of
$\pl^4 h\cdot \pl^2 h$-invariants, their weights and
weak-field expansion of 
$\na\na R\times R$-type general invariants.}
\label{tab8}
\end{table}
%%%%%%%%%%%%%%%%%%%%%%%%%  END  of  Table AppC  %%%%%%%%%%%%%%%%%%%%%%%%%%%%%
%%%%%%%%%%%%%%%%%%%%%%%%%%%%%%%%%%%%%%%%%%%%%%%%%%%%%%%%%%%%%%%%%%%%%%%%%

\newpage
%%%%%%%%%%%%%%%%%%%%%%%%%%  App. D  %%%%%%%%%%%%%%%%%%%%%%%%%%%%%%%
%%%%%%%%%%%%%%%%%%%%%%%%%%%%%%%%%%%%%%%%%%%%%%%%%%%%%%%%%%%%%%%%%%%
\begin{flushleft}
{\Large\bf Appendix D.\ Classification of 
$(\pl\pl\pl h)^2$-Invariants
and Weak-Expansion of $\na R\times \na R$-terms}
\end{flushleft}

\begin{figure}
\centerline{
{\epsfxsize=117pt  \epsfysize=71pt \epsfbox{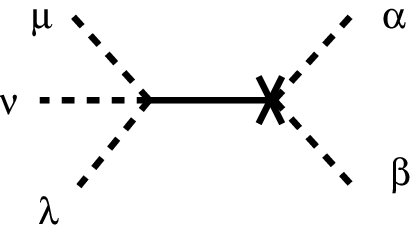}}
           }
\caption{
%*figDp1*Fig.D.1\ 
Graphs of 5-tensor $\pl_\m\pl_\n\pl_\la h_\ab$
        }
\label{figDp1}
\end{figure}

\begin{figure}
\centerline{
{\epsfxsize=87pt  \epsfysize=107pt \epsfbox{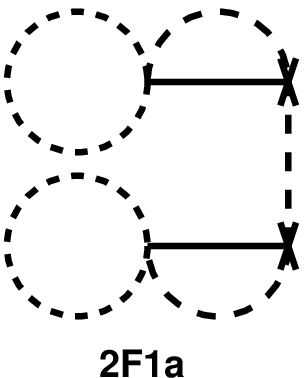}}\q
{\epsfxsize=45pt  \epsfysize=115pt \epsfbox{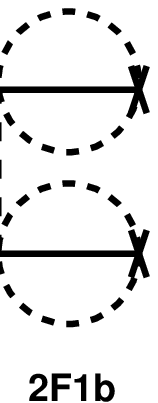}}\q
{\epsfxsize=87pt  \epsfysize=106pt \epsfbox{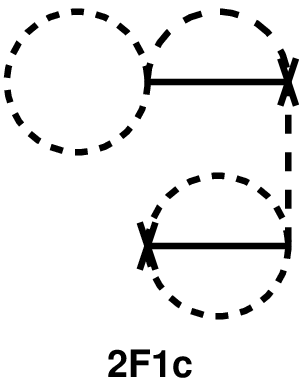}}
           }
\vspace{10pt}
\centerline{
{\epsfxsize=123pt  \epsfysize=103pt \epsfbox{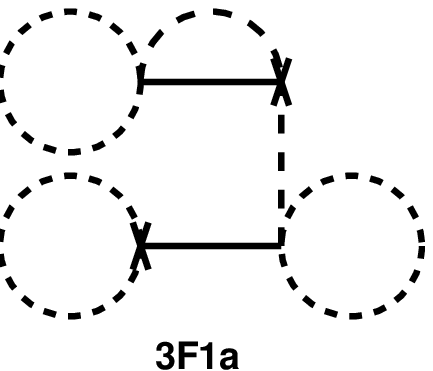}}\qq
{\epsfxsize=120pt  \epsfysize=105pt \epsfbox{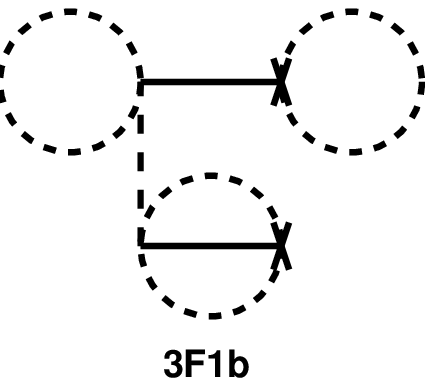}}
           }
\vspace{10pt}
\centerline{
{\epsfxsize=123pt  \epsfysize=105pt \epsfbox{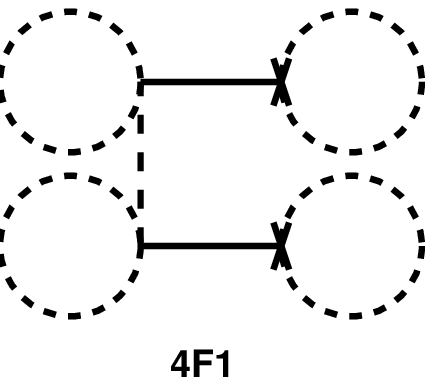}}
           }
\caption{
%*figDp2*Fig.D.2\  
Graphs for \ul{bridgeno}=1
}
\label{figDp2}
\end{figure}

\begin{figure}
\centerline{
{\epsfxsize=80pt  \epsfysize=108pt \epsfbox{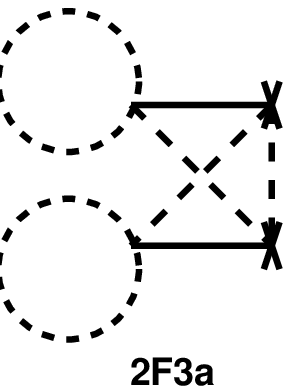}}\q
{\epsfxsize=100pt  \epsfysize=95pt \epsfbox{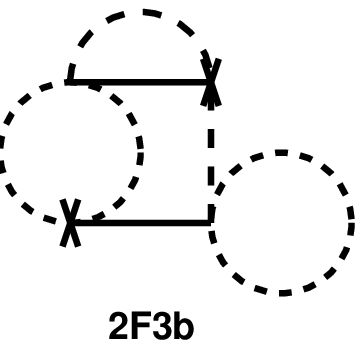}}\q
{\epsfxsize=83pt  \epsfysize=105pt \epsfbox{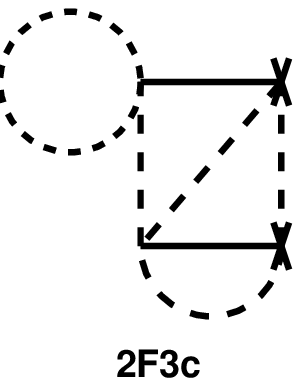}}
           }
\vspace{10pt}
\centerline{
{\epsfxsize=62pt  \epsfysize=97pt \epsfbox{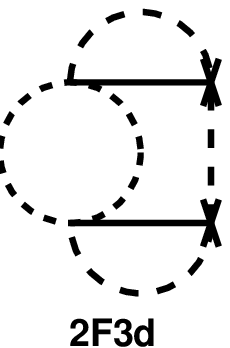}}\qq
{\epsfxsize=45pt  \epsfysize=97pt \epsfbox{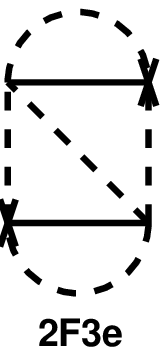}}
           }
\vspace{10pt}
\centerline{
{\epsfxsize=205pt  \epsfysize=60pt \epsfbox{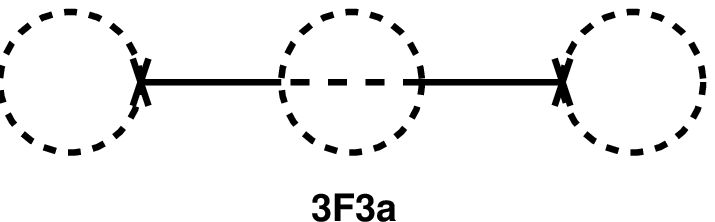}}\q
{\epsfxsize=102pt  \epsfysize=110pt \epsfbox{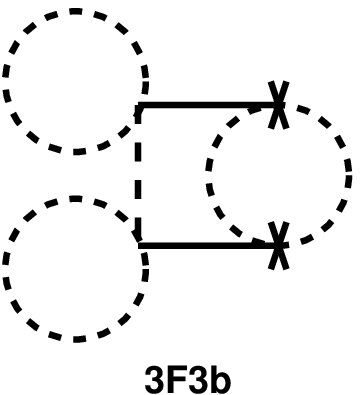}}
           }
\vspace{10pt}
\centerline{
{\epsfxsize=125pt  \epsfysize=110pt \epsfbox{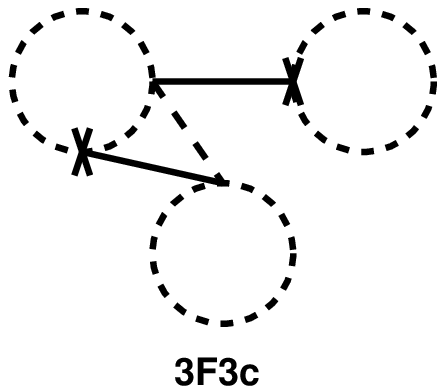}}\q
{\epsfxsize=120pt  \epsfysize=75pt \epsfbox{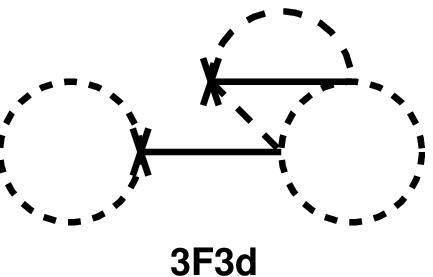}}
           }
\caption{
%*figDp3*Fig.D.3\  
Graphs for \ul{bridgeno}=3
}
\label{figDp3}
\end{figure}

\begin{figure}
\centerline{
{\epsfxsize=123pt  \epsfysize=56pt \epsfbox{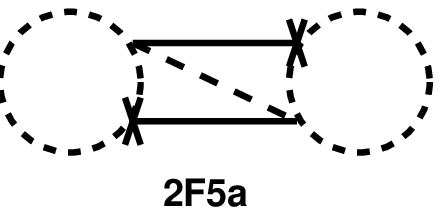}}\qq
{\epsfxsize=90pt  \epsfysize=55pt \epsfbox{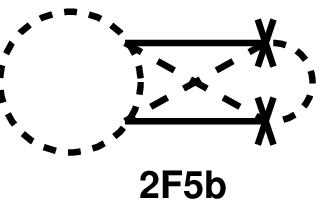}}
           }
\vspace{10pt}
\centerline{
{\epsfxsize=86pt  \epsfysize=65pt \epsfbox{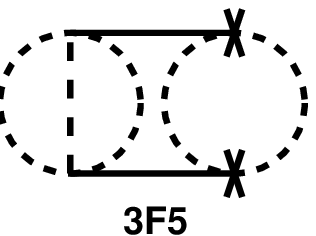}}
           }
\caption{
%*figDp4*Fig.D.4\  
Graphs for \ul{bridgeno}=5
}
\label{figDp4}
\end{figure}

The leading order of the weak-field expansion for $\na R\times \na R$-type
general invariants($O_1\sim O_4$), is given by a sum of
$(\pl\pl\pl h)^2$-invariants.
We introduce a graphical representation, in Fig.\ref{figDp1}, for a 5-tensor
$\pl_\m\pl_\n\pl_\la h_\ab$.
%%%%%%%%%%%%%%%%%%%%%%%  Fig.D.1  %%%%%%%%%%%%%%%%%%%%%%%%%%%%%%%%%%%%
%%%%%%%%%%%%%%%%%%%%%%%%%%%%%%%%%%%%%%%%%%%%%%%%%%%%%%%%%%%%%%%%%%%%%%%
We list here all and independent $(\pl\pl\pl h)^2$-invariants.
\ul{bridgeno} must be an odd number
in this case because the $\pl\pl\pl h$-tensor  has
an odd number of suffixes.
Especially there are no disconnected graphs.
We classify them  by \ul{bridgeno}
and  $\ul{l}$, as follows.
%%%%%%%%%%%%%%%%%%%%%%%%%%%%%  bridgeno=1 %%%%%%%%%%%%%%%%%%%%%%%%%%%%%%
\flushleft{(i) \ul{bridgeno}=1\ ,Fig.\ref{figDp2}.}\nl
2F1a, 2F1b, 2F1c ($\ul{l}=2$);\ 
3F1a, 3F1b ($\ul{l}=3$);\ 
4F1 ($\ul{l}=4$).
%%%%%%%%%%%%%%%%%%%%%%%%%%%%%%%%%%%%%%%%%%%%%%%%%%%%%%%%%%%%%%%%%%%
%%%%%%%%%%%%%%%%%%%%%%%%%%%%%  bridgeno=3 %%%%%%%%%%%%%%%%%%%%%%%%%%%%%%
\flushleft{(ii) \ul{bridgeno}=3\ ,\ Fig.\ref{figDp3}.}\nl
2F3a, 2F3b, 2F3c, 2F3d, 2F3e ($\ul{l}=2$);\ 
3F3a, 3F3b, 3F3c, 3F3d ($\ul{l}=3$).
%%%%%%%%%%%%%%%%%%%%%%%%%%%%%%%%%%%%%%%%%%%%%%%%%%%%%%%%%%%%%%%%%%%
%Graphs of Fig.D.3 are further classified by the vertex-types (dddd-vertex,
%h-vertex) of six ends of three 'bridge' suffix-lines.\nl
%a)\ (ddd)$^2$:\ F3a
%b)\ (ddh)$^2$:\ F3b,F3c
%c)\ (dhh)$^2$:\ F3d,F3e
%d)\ ddd$\times$ ddh:\ F3f
%e)\ ddd$\times$ dhh:\ F3g
%f)\ ddh$\times$ dhh:\ F3h,F3i
%%%%%%%%%%%%%%%%%%%%%%%%%%%%%  bridgeno=5 %%%%%%%%%%%%%%%%%%%%%%%%%%%%%%
\flushleft{(iii) \ul{bridgeno}=5\ ,\ Fig.\ref{figDp4}.}\nl
2F5a, 2F5b ($\ul{l}=2$);\ 
3F5 ($\ul{l}=3$).
%%%%%%%%%%%%%%%%%%%%%%%%%%%%%%%%%%%%%%%%%%%%%%%%%%%%%%%%%%%%%%%%%%%
%%%%%%%%%%%%%%%%%%%%%%%  Fig.D.2  %%%%%%%%%%%%%%%%%%%%%%%%%%%%%%%%%%%%
%%%%%%%%%%%%%%%%%%%%%%%%%%%%%%%%%%%%%%%%%%%%%%%%%%%%%%%%%%%%%%%%%%%%%

%%%%%%%%%%%%%%%%%%%%%%%  Fig.D.3  %%%%%%%%%%%%%%%%%%%%%%%%%%%%%%%%%%%%
%%%%%%%%%%%%%%%%%%%%%%%%%%%%%%%%%%%%%%%%%%%%%%%%%%%%%%%%%%%%%%%%%%%%%

%%%%%%%%%%%%%%%%%%%%%%%  Fig.D.4  %%%%%%%%%%%%%%%%%%%%%%%%%%%%%%%%%%%%
%%%%%%%%%%%%%%%%%%%%%%%%%%%%%%%%%%%%%%%%%%%%%%%%%%%%%%%%%%%%%%%%%%%%%

\vs {0.5}
\q In Table \ref{tab9}, the weak-field expansion of $O_1\sim O_4$ ,
the classification of $(\pl\pl\pl h)^2$-invariants  
and their weights are given. The total sum of weights is
$945=9\times 7\times 5\times 3\times 1$. 
We see $O_i$'s are independent each other.

\newpage
%%%%%%%%%%%%%%%%%%%%%%%%%%%%%%%%%%%%%%%%%%%%%%%%%%%%%%%%%%%%%%%%%%%%%%%%%%
%%%%%%%%%%%%%%%%%%%%%  Table AppD  %%%%%%%%%%%%%%%%%%%%%%%%%%%%%
%%%%%%%%%%%%%%%%%%%%%%%%%%%%%%%%%%%%%%%%%%%%%%%%%%%%%%%%%%%%%%%%%%%%%%%%%%
\begin{table}
\begin{tabular}{|c||c|c|c|c|c|c|c|}
\hline
bridge- & $\ul{l}$ & Graph  & Weight, & $O_1$ & $O_2$ & $O_3$ & $O_4$\\
no      &        & Name   & Total 945 &       &       &       &    \\
\hline\hline
        &        & 2F1a   & 36        & 0     & 0     & 0     & 0  \\
\cline{3-8}
        &   2    & 2F1b   & 36        & 1     & 0     & 0     & 0   \\
\cline{3-8}
1       &        & 2F1c   & 72        & 0     & 0     & 0     & 0   \\
\cline{2-8}
        &   3    & 3F1a   & 36        & 0     & 0     & 0     & 0   \\
\cline{3-8}
        &        & 3F1b   & 36        & -2     & 0     & 0     & 0   \\
\cline{2-8}
        &   4    & 4F1    &  9        & 1     & 0     & 0     & 0   \\
\hline\hline
        &        & 2F3a   & 36        & 0     & 0     & 0 &$\fourth$ \\
\cline{3-8}
        &        & 2F3b   & 72        & 0     & 0     & 0  &$-\half$ \\
\cline{3-8}
        &   2    & 2F3c   & 144       & 0     & -1    & 0  &$-\half$ \\
\cline{3-8}
        &        & 2F3d   & 72        & 0     &$\half$ & 0 & 
                                                        $\frac{3}{4}$ \\
\cline{3-8}
3       &        & 2F3e   & 144       & 0     &$\half$ & 0  &$\fourth$ \\
\cline{2-8}
        &        & 3F3a   &  6        & 0     &$\fourth$ & 0&$\fourth$ \\
\cline{3-8}
        &   3    & 3F3b   & 18       & 0       &$\fourth$ &  0  & 0 \\
\cline{3-8}
        &        & 3F3c   & 36       & 0       &$\half$&  0 &$\half$ \\
\cline{3-8}
        &        & 3F3d   & 72       & 0       & -1    &  0 &  -1   \\
\hline\hline
        &   2    & 2F5a   & 36        & 0     & 0     & 1 &   0  \\
\cline{3-8}
5       &        & 2F5b   & 72       & 0       & 0    &  -2 &  0   \\
\cline{2-8}
        &   3    & 3F5    &  12       & 0     &  0    & 1   &  0  \\
\hline
%\multicolumn{8}{c}{\q}                                                 \\
%\multicolumn{8}{c}{Table AppD\ \ Classification of
%$(\pl\pl\pl h)^3$-invariants, their weights and  }\\
%\multicolumn{8}{c}{weak-field expansion of 
%$\na R\times \na R$-type general invariants.
%                    }\\
\end{tabular}
\caption{ 
%*tab9*
Classification of
$(\pl\pl\pl h)^2$-invariants, their weights and 
weak-field expansion of 
$\na R\times \na R$-type general invariants.}
\label{tab9}
\end{table}
%%%%%%%%%%%%%%%%%%%%%%%%%  END  of  Table AppD  %%%%%%%%%%%%%%%%%%%%%%%%%%%%%
%%%%%%%%%%%%%%%%%%%%%%%%%%%%%%%%%%%%%%%%%%%%%%%%%%%%%%%%%%%%%%%%%%%%%%%%%

\newpage
%%%%%%%%%%%%%%%%%%%%%%%%%%  App. E  %%%%%%%%%%%%%%%%%%%%%%%%%%%%%%%
%%%%%%%%%%%%%%%%%%%%%%%%%%%%%%%%%%%%%%%%%%%%%%%%%%%%%%%%%%%%%%%%%%%
\begin{flushleft}
{\Large\bf Appendix E.\ 
Weak Field Expansion of General Invariants 
%($P_1\sim P_6,A_1,B_1
%,T_1\sim T_5,S
%$) 
}
\end{flushleft}

In this appendix we list the weak expansion of $RRR$-type general
invariants:$P_1\sim P_6,A_1,B_1$. We focus only in
$(\pl\pl h)^3$-terms among different types of expanded terms.
The classification of $(\pl\pl h)^3$-invariants is the main theme
of the text. The result is most fruitfully utilized in this
appendix. Especially the set of indices, which characterizes
every $(\pl\pl h)^3$-invariant by its graph topology,
 is exploited in the (computer)
calculation. The following results show the power of the present
approach. We see the 8 general invariants are locally independent
each other, furthermore they are
``orthogonal'' in the space of
$(\pl\pl h)^3$-invariants except in the ``directions'' of
$G3$ and $G13$\ ($A_1$ and $B_1$ only are mixed in those ``directions'')
\cite{foot10}
%\footnote{
%Among 90 terms listed in App.A, G3 and G13 only has the crossing
%number 3. As for the definition of the crossing number, see Subsec.5.2.
%         }
.

%\vs 2
%\newpage
%%%%%%%%%%%%%%%%%%%%%%%%%%%%%%%%%%%%%%%%%%%%%%%%%%%%%%%%%%%%%%%%%%%%%%%%%%
%%%%%%%%%%%%%%%%%%%%%  Table G1-G13  %%%%%%%%%%%%%%%%%%%%%%%%%%%%%
%%%%%%%%%%%%%%%%%%%%%%%%%%%%%%%%%%%%%%%%%%%%%%%%%%%%%%%%%%%%%%%%%%%%%%%%%%
\begin{table}
\begin{tabular}{|c||c|c|c|c|c|c|c|c|}
\hline
Graph  & $P_1$ & $P_2$ & $P_3$ & $P_4$ & $P_5$ & $P_6$ & $A_1$ & $B_1$ \\
\hline
  $G1$ &  $0$  & $0$   & $0$   & $-\fourth$  & $0$   & $0$   & $0$   & $0$ \\
\hline
  $G2$ &  $0$  & $0$   & $0$   & $0$  & $\fourth$   & $0$   & $0$   & $0$  \\
\hline
  $G3$ &  $0$  & $0$   & $0$   & $0$  & $0$   & $0$   & $-1$   & $-\fourth$  \\
\hline
  $G4$ &  $0$  & $0$   & $0$ & $-\frac{3}{4}$  & $0$ & $0$   & $0$   & $0$  \\
\hline
  $G5$ &  $0$  & $0$   & $0$   & $0$  & $0$   & $\half$   & $0$   & $0$   \\
\hline
  $G6$ &  $0$  & $0$   & $0$   & $0$  & $0$   & $\half$   & $0$   & $0$   \\
\hline
  $G7$ &  $0$  & $0$   & $0$   & $0$  & $0$   & $0$ & $0$ & $\frac{3}{2}$  \\
\hline
  $G8$ &  $0$  & $0$   & $0$   & $0$  & $\half$   & $0$   & $0$   & $0$   \\
\hline
  $G9$ &  $0$  & $0$   & $0$   & $0$  & $0$   & $\half$   & $0$   & $0$   \\
\hline
 $G10$ &  $0$  & $0$   & $0$   & $0$  & $\fourth$   & $0$   & $0$   & $0$  \\
\hline
 $G11$ &  $0$  & $0$   & $0$   & $0$  & $0$   & $\half$   & $0$   & $0$   \\
\hline
 $G12$ &  $0$  & $0$   & $0$   & $0$  & $0$   & $0$  & $0$ & $\frac{3}{2}$  \\
\hline
 $G13$ &  $0$  & $0$   & $0$   & $0$  & $0$   & $0$ & $-3$ & $-\frac{3}{4}$ \\
\hline
%\multicolumn{15}{c}{\q}                                                 \\
%\multicolumn{15}{c}{Table 4\ \ Weak-Expansion of 
%Invariants with (Mass)$^6$-Dim.:\  $(\pl\pl h)^3$-Part }\\
%\multicolumn{15}{c}{G1-G13(\ul{l}=1)           }\\
\end{tabular}
\caption{ 
%*tab10*
Weak-Expansion of 
Invariants with $M^6$-Dim.:\  $(\pl\pl h)^3$-Part,
G1-G13($\ul{l}$=1)           }
\label{tab10}
\end{table}
%%%%%%%%%%%%%%%%%%%%%%%%%  END  of  Table ** %%%%%%%%%%%%%%%%%%%%%%%%%%%%%
%%%%%%%%%%%%%%%%%%%%%%%%%%%%%%%%%%%%%%%%%%%%%%%%%%%%%%%%%%%%%%%%%%%%%%%%%
\newpage
%%%%%%%%%%%%%%%%%%%%%%%%%%%%%%%%%%%%%%%%%%%%%%%%%%%%%%%%%%%%%%%%%%%%%%%%%%
%%%%%%%%%%%%%%%%%%%%%  Table G14-G42 %%%%%%%%%%%%%%%%%%%%%%%%%%%%%
%%%%%%%%%%%%%%%%%%%%%%%%%%%%%%%%%%%%%%%%%%%%%%%%%%%%%%%%%%%%%%%%%%%%%%%%%%
\begin{table}
\begin{tabular}{|c||c|c|c|c|c|c|c|c|}
\hline
Graph  & $P_1$ & $P_2$ & $P_3$ & $P_4$ & $P_5$ & $P_6$ & $A_1$ & $B_1$ \\
\hline
 $G14$ &  $0$  & $0$   & $0$   & $0$  & $0$   & $0$   & $1$   & $0$   \\
\hline
 $G15$ &  $0$  & $0$   & $0$   & $0$  & $0$   & $0$   & $3$   & $0$   \\
\hline
 $G16$ &  $0$  & $0$   & $0$   & $0$  & $-1$   & $0$   & $0$   & $0$   \\
\hline
 $G17$ &  $0$  & $0$   & $0$   & $0$  & $0$   & $-\half$   & $0$   & $0$  \\
\hline
 $G18$ &  $0$  & $0$   & $0$   & $0$  & $0$   & $0$ & $0$ & $-\frac{3}{4}$ \\
\hline
 $G19$ &  $0$  & $0$   & $0$   & $0$  & $0$   & $-\half$   & $0$   & $0$   \\
\hline
 $G20$ &  $0$  & $0$   & $0$   & $0$  & $0$   & $0$ & $0$ & $-\frac{3}{4}$  \\
\hline
 $G21$ &  $0$  & $0$   & $0$   & $0$  & $0$   & $-\half$   & $0$   & $0$   \\
\hline
 $G22$ &  $0$  & $0$   & $0$   & $0$  & $0$   & $0$ & $0$ & $-\frac{3}{2}$  \\
\hline
 $G23$ &  $0$  & $-\half$   & $0$   & $0$  & $0$   & $0$   & $0$   & $0$   \\
\hline
 $G24$ &  $0$  & $0$   & $2$   & $0$  & $0$   & $0$   & $0$   & $0$   \\
\hline
 $G25$ &  $0$  & $0$   & $0$   & $0$  & $0$   & $-\half$   & $0$   & $0$  \\
\hline
 $G26$ &  $0$  & $-\half$   & $0$   & $0$  & $0$   & $0$   & $0$   & $0$  \\
\hline
 $G27$ &  $0$  & $0$   & $0$   & $\frac{3}{8}$ & $0$ & $0$   & $0$   & $0$ \\
\hline
 $G28$ &  $0$  & $0$   & $0$   & $0$  & $0$   & $-\fourth$   & $0$   & $0$  \\
\hline
 $G29$ &  $0$  & $0$   & $0$   & $0$  & $-\half$   & $0$   & $0$   & $0$   \\
\hline
 $G30$ &  $0$  & $0$   & $0$   & $0$  & $0$   & $-\half$   & $0$   & $0$   \\
\hline
 $G31$ &  $0$  & $0$   & $0$   & $0$  & $-\half$   & $0$   & $0$   & $0$   \\
\hline
 $G32$ &  $0$  & $0$   & $0$   & $\frac{3}{4}$ & $0$ & $0$   & $0$   & $0$  \\
\hline
 $G33$ &  $0$  & $0$   & $0$   & $\frac{3}{8}$ & $0$ & $0$   & $0$   & $0$  \\
\hline
 $G34$ &  $0$  & $0$   & $0$   & $0$  & $0$   & $-\fourth$   & $0$   & $0$  \\
\hline
 $G35$ &  $0$  & $0$   & $0$   & $\frac{3}{8}$ & $0$ & $0$   & $0$   & $0$  \\
\hline
 $G36$ &  $0$  & $0$   & $0$   & $0$  & $0$   & $-\fourth$   & $0$   & $0$  \\
\hline
 $G37$ &  $0$  & $0$   & $0$   & $0$  & $-\half$   & $0$   & $0$   & $0$   \\
\hline
 $G38$ &  $0$  & $0$   & $0$   & $0$  & $0$   & $-\half$   & $0$   & $0$  \\
\hline
 $G39$ &  $0$  & $0$   & $0$   & $\frac{3}{8}$ & $0$ & $0$   & $0$   & $0$ \\
\hline
 $G40$ &  $0$  & $0$   & $0$   & $0$  & $0$   & $-\fourth$   & $0$   & $0$  \\
\hline
 $G41$ &  $0$  & $0$   & $0$   & $0$  & $-\half$   & $0$   & $0$   & $0$   \\
\hline
 $G42$ &  $0$  & $0$   & $0$   & $\frac{3}{4}$ & $0$ & $0$   & $0$   & $0$  \\
\hline
%\multicolumn{15}{c}{\q}                                                 \\
%\multicolumn{15}{c}{Table 4\ \ Weak-Expansion of 
%Invariants with (Mass)$^6$-Dim.:\  $(\pl\pl h)^3$-Part }\\
%\multicolumn{15}{c}{ G14-G42(\ul{l}=2)           }\\
\end{tabular}
\caption{ 
%*tab11*
Weak-Expansion of 
Invariants with $M^6$-Dim.:\  $(\pl\pl h)^3$-Part, G14-G42($\ul{l}$=2)           }
\label{tab11}
\end{table}
%%%%%%%%%%%%%%%%%%%%%%%%%  END  of  Table ** %%%%%%%%%%%%%%%%%%%%%%%%%%%%%
%%%%%%%%%%%%%%%%%%%%%%%%%%%%%%%%%%%%%%%%%%%%%%%%%%%%%%%%%%%%%%%%%%%%%%%%%

\newpage
%%%%%%%%%%%%%%%%%%%%%%%%%%%%%%%%%%%%%%%%%%%%%%%%%%%%%%%%%%%%%%%%%%%%%%%%%%
%%%%%%%%%%%%%%%%%%%%%  Table G43-G69 %%%%%%%%%%%%%%%%%%%%%%%%%%%%%%%%%%%%%%%%
%%%%%%%%%%%%%%%%%%%%%%%%%%%%%%%%%%%%%%%%%%%%%%%%%%%%%%%%%%%%%%%%%%%%%%%%%%
\begin{table}
\begin{tabular}{|c||c|c|c|c|c|c|c|c|}
\hline
Graph  & $P_1$ & $P_2$ & $P_3$ & $P_4$ & $P_5$ & $P_6$ & $A_1$ & $B_1$ \\
\hline
 $G43$ &  $0$  & $0$   & $0$   & $0$  & $\fourth$   & $0$   & $0$   & $0$ \\
\hline
 $G44$ &  $0$  & $0$   & $0$   & $-\frac{3}{4}$ & $0$ & $0$   & $0$   & $0$ \\
\hline
 $G45$ &  $0$  & $0$   & $0$   & $0$  & $0$   & $\fourth$   & $0$   & $0$  \\
\hline
 $G46$ &  $0$  & $0$   & $0$   & $0$  & $0$   & $\fourth$   & $0$   & $0$  \\
\hline
 $G47$ &  $0$  & $0$   & $0$   & $0$  & $0$   & $\fourth$   & $0$   & $0$  \\
\hline
 $G48$ &  $0$  & $0$   & $0$   & $0$  & $\half$   & $0$   & $0$   & $0$   \\
\hline
 $G49$ &  $0$  & $0$   & $0$   & $0$  & $0$   & $0$   & $0$ & $\frac{3}{4}$ \\
\hline
 $G50$ &  $0$  & $0$   & $-1$   & $0$  & $0$   & $0$   & $0$   & $0$   \\
\hline
 $G51$ &  $0$  & $0$   & $0$   & $-\frac{3}{4}$ & $0$ & $0$   & $0$   & $0$\\
\hline
 $G52$ &  $0$  & $0$   & $0$   & $0$  & $\half$   & $0$   & $0$   & $0$   \\
\hline
 $G53$ &  $0$  & $\half$   & $0$   & $0$  & $0$   & $0$   & $0$   & $0$   \\
\hline
 $G54$ &  $0$  & $0$   & $-2$   & $0$  & $0$   & $0$   & $0$   & $0$   \\
\hline
 $G55$ &  $0$  & $0$   & $0$   & $-\frac{3}{4}$ & $0$ & $0$   & $0$   & $0$ \\
\hline
 $G56$ &  $0$  & $\half$   & $0$   & $0$  & $0$   & $0$   & $0$   & $0$   \\
\hline
 $G57$ &  $0$  & $0$   & $0$   & $0$  & $0$   & $\fourth$   & $0$   & $0$  \\
\hline
 $G58$ &  $0$  & $0$   & $0$   & $0$  & $\half$   & $0$   & $0$   & $0$   \\
\hline
 $G59$ &  $0$  & $0$   & $0$   & $0$  & $0$   & $\half$   & $0$   & $0$   \\
\hline
 $G60$ &  $0$  & $0$   & $0$   & $0$  & $\half$   & $0$   & $0$   & $0$   \\
\hline
 $G61$ &  $0$  & $1$   & $0$   & $0$  & $0$   & $0$   & $0$   & $0$   \\
\hline
 $G62$ &  $0$  & $0$   & $0$   & $-\frac{3}{4}$ & $0$ & $0$   & $0$   & $0$\\
\hline
 $G63$ &  $0$  & $0$   & $0$   & $0$  & $\fourth$   & $0$   & $0$   & $0$ \\
\hline
 $G64$ &  $0$  & $0$   & $0$   & $0$  & $0$   & $\half$   & $0$   & $0$  \\
\hline
 $G65$ &  $0$  & $0$   & $0$   & $0$  & $\half$   & $0$   & $0$   & $0$  \\
\hline
 $G66$ &  $0$  & $1$   & $0$   & $0$  & $0$   & $0$   & $0$   & $0$   \\
\hline
 $G67$ &  $0$  & $0$   & $0$   & $0$  & $0$   & $0$   & $0$   & $\fourth$  \\
\hline
 $G68$ &  $0$  & $0$   & $-1$   & $0$  & $0$   & $0$   & $0$   & $0$   \\
\hline
 $G69$ &  $-1$  & $0$   & $0$   & $0$  & $0$   & $0$   & $0$   & $0$   \\
\hline
%\multicolumn{15}{c}{\q}                                                 \\
%\multicolumn{15}{c}{Table 4\ \ Weak-Expansion of 
%Invariants with (Mass)$^6$-Dim.:\  $(\pl\pl h)^3$-Part }\\
%\multicolumn{15}{c}{G43-G69(\ul{l}=3)           }\\
\end{tabular}
\caption{ 
%*tab12*
Weak-Expansion of 
Invariants with $M^6$-Dim.:\  $(\pl\pl h)^3$-Part
,G43-G69($\ul{l}$=3)           }
\label{tab12}
\end{table}
%%%%%%%%%%%%%%%%%%%%%%%%%  END  of  Table G43-G69 %%%%%%%%%%%%%%%%%%%%%%%%%%%
%%%%%%%%%%%%%%%%%%%%%%%%%%%%%%%%%%%%%%%%%%%%%%%%%%%%%%%%%%%%%%%%%%%%%%%%%
\newpage
%%%%%%%%%%%%%%%%%%%%%%%%%%%%%%%%%%%%%%%%%%%%%%%%%%%%%%%%%%%%%%%%%%%%%%%%%%
%%%%%%%%%%%%%%%%%%%%%  Table G70-G90 %%%%%%%%%%%%%%%%%%%%%%%%%%%%%%%%%%%%%%%%
%%%%%%%%%%%%%%%%%%%%%%%%%%%%%%%%%%%%%%%%%%%%%%%%%%%%%%%%%%%%%%%%%%%%%%%%%%
\begin{table}
\begin{tabular}{|c||c|c|c|c|c|c|c|c|}
\hline
Graph  & $P_1$ & $P_2$ & $P_3$ & $P_4$ & $P_5$ & $P_6$ & $A_1$ & $B_1$ \\
\hline
 $G70$ &  $0$  & $0$   & $0$   & $\frac{1}{8}$ & $0$ & $0$   & $0$   & $0$ \\
\hline
 $G71$ &  $0$  & $0$   & $0$   & $0$  & $-\fourth$   & $0$   & $0$   & $0$ \\
\hline
 $G72$ &  $0$  & $0$   & $1$   & $0$  & $0$   & $0$   & $0$   & $0$   \\
\hline
 $G73$ &  $0$  & $0$   & $0$   & $\frac{3}{8}$ & $0$ & $0$   & $0$   & $0$ \\
\hline
 $G74$ &  $0$  & $-1$   & $0$   & $0$  & $0$   & $0$   & $0$   & $0$   \\
\hline
 $G75$ &  $0$  & $0$   & $0$   & $0$  & $-\fourth$   & $0$   & $0$   & $0$ \\
\hline
 $G76$ &  $0$  & $-\fourth$   & $0$   & $0$  & $0$   & $0$   & $0$   & $0$ \\
\hline
 $G77$ &  $0$  & $0$   & $0$   & $\frac{1}{8}$  & $0$   & $0$   & $0$   & $0$\\
\hline
 $G78$ &  $0$  & $0$   & $0$   & $0$  & $-\fourth$   & $0$   & $0$   & $0$  \\
\hline
 $G79$ &  $0$  & $-\fourth$   & $0$   & $0$  & $0$   & $0$   & $0$   & $0$  \\
\hline
 $G80$ &  $0$  & $0$   & $0$   & $\frac{3}{8}$  & $0$   & $0$   & $0$   & $0$\\
\hline
 $G81$ &  $0$  & $-1$   & $0$   & $0$  & $0$   & $0$   & $0$   & $0$   \\
\hline
 $G82$ &  $0$  & $0$   & $0$   & $0$  & $-\fourth$   & $0$   & $0$   & $0$ \\
\hline
 $G83$ &  $0$  & $-\half$   & $0$   & $0$  & $0$   & $0$   & $0$   & $0$  \\
\hline
 $G84$ &  $0$  & $0$   & $1$   & $0$  & $0$   & $0$   & $0$   & $0$   \\
\hline
 $G85$ &  $3$  & $0$   & $0$   & $0$  & $0$   & $0$   & $0$   & $0$   \\
\hline
\hline
 $G86$ &  $0$  & $\fourth$   & $0$   & $0$  & $0$   & $0$   & $0$   & $0$  \\
\hline
 $G87$ &  $0$  & $\fourth$   & $0$   & $0$  & $0$   & $0$   & $0$   & $0$  \\
\hline
 $G88$ &  $-3$  & $0$   & $0$   & $0$  & $0$   & $0$   & $0$   & $0$   \\
\hline
 $G89$ &  $0$  & $\half$   & $0$   & $0$  & $0$   & $0$   & $0$   & $0$  \\
\hline
\hline
 $G90$ &  $1$  & $0$   & $0$   & $0$  & $0$   & $0$   & $0$   & $0$   \\
\hline
%\multicolumn{15}{c}{\q}                                                 \\
%\multicolumn{15}{c}{Table 4\ \ Weak-Expansion of 
%Invariants with (Mass)$^6$-Dim.:\  $(\pl\pl h)^3$-Part }\\
%\multicolumn{15}{c}{G70-G85(\ul{l}=4),G86-G89(\ul{l}=5),G90(\ul{l}=6)     }\\
\end{tabular}
\caption{ 
%*tab13*
Weak-Expansion of 
Invariants with $M^6$-Dim.:\  $(\pl\pl h)^3$-Part
,G70-G85($\ul{l}$=4),G86-G89($\ul{l}$=5),G90($\ul{l}$=6) }
\label{tab13}
\end{table}
%%%%%%%%%%%%%%%%%%%%%%%%%  END  of  Table G70-G90 %%%%%%%%%%%%%%%%%%%%%%%%%%%%%
%%%%%%%%%%%%%%%%%%%%%%%%%%%%%%%%%%%%%%%%%%%%%%%%%%%%%%%%%%%%%%%%%%%%%%%%%

\newpage
%%%%%%%%%%%%%%%%%%%%%%%%%%  App. F  %%%%%%%%%%%%%%%%%%%%%%%%%%%%%%%
%%%%%%%%%%%%%%%%%%%%%%%%%%%%%%%%%%%%%%%%%%%%%%%%%%%%%%%%%%%%%%%%%%%
\begin{flushleft}
{\Large\bf Appendix F.\ 
Graphs of General Invariants with $M^6$ Dimension
}
\end{flushleft}

In this section we graphically list all independent 
(in n-dim space) 
general invariants with $M^6$ dimension.
They are classified in the following ways.

\begin{figure}
     \centerline{
{\epsfxsize=220pt  \epsfysize=50pt \epsfbox{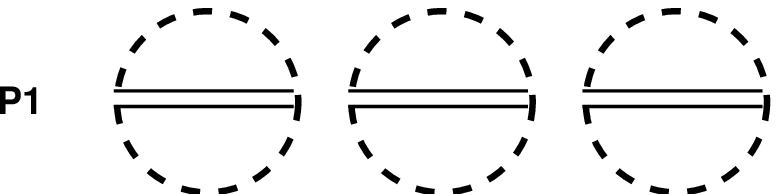}}
\q
{\epsfxsize=150pt  \epsfysize=50pt \epsfbox{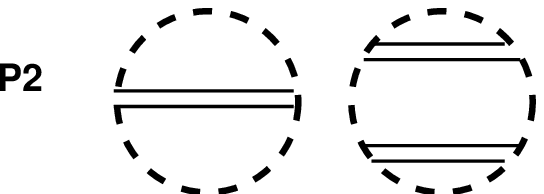}}
                 }
\vspace{10pt}
     \centerline{
{\epsfxsize=135pt  \epsfysize=135pt \epsfbox{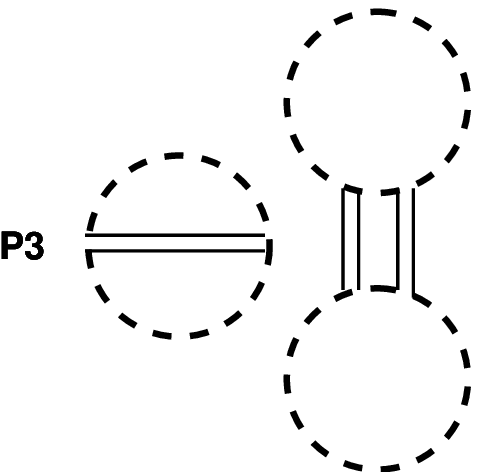}}
\q
{\epsfxsize=90pt  \epsfysize=72pt \epsfbox{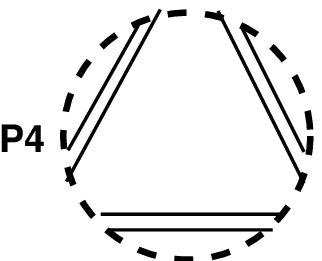}}
%                 }
%\vspace{10pt}
%     \centerline{
\q
{\epsfxsize=92pt  \epsfysize=72pt \epsfbox{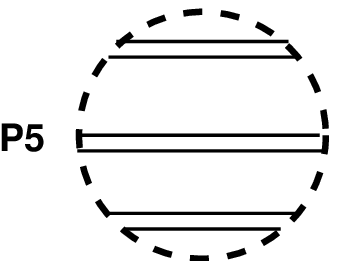}}
\q
{\epsfxsize=70pt  \epsfysize=130pt \epsfbox{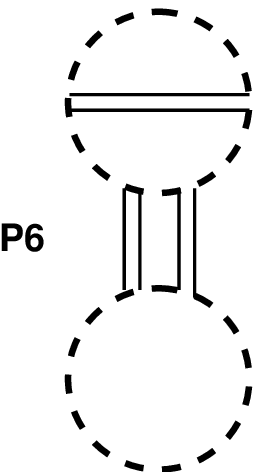}}
                 }
\caption{
%*figP1t6* Fig.P1t6\ 
Graphs for 
$P_1=RRR\com\ \ P_2=RR_\mn R^\mn, P_3=RR_{\mn\ls}R^{\mn\ls},
P_4=R_\mn R^{\n\la}R_\la^{~\mu}, P_5=R_{\mn\ls}R^{\mu\la}R^{\nu\si}$ and
$ P_6=R_{\mn\ls}R_\tau^{~\nu\ls}R^{\mu\tau}$.
        }
\label{figP1t6}
\end{figure}

\flushleft{(i)\ Fig.\ref{figP1t6}:\ $RRR$-type (on-shell vanishing)}

\begin{figure}
     \centerline{
{\epsfxsize=130pt  \epsfysize=105pt \epsfbox{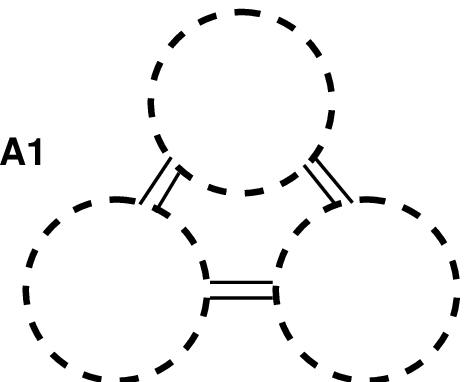}}
\qqq
{\epsfxsize=72pt  \epsfysize=132pt \epsfbox{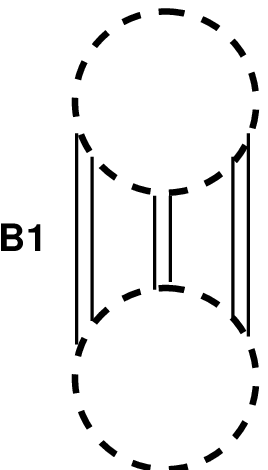}}
                 }
\caption{
%*figAB*Fig.AB\ 
Graphs for
$A_1=R_{\mn\ls}R^{\si\la}_{~~~\tau\om}R^{\om\tau\n\m}$ and 
$B_1=R_{\mn\tau\si}R^{\n~~~\tau}_{~\la\om}R^{\la\mu\si\om}$.
        }
\label{figAB}
\end{figure}

\flushleft{(ii)\ Fig.\ref{figAB}:\ $RRR$-type (on-shell non-vanishing)}

\begin{figure}
     \centerline{
{\epsfxsize=150pt  \epsfysize=53pt \epsfbox{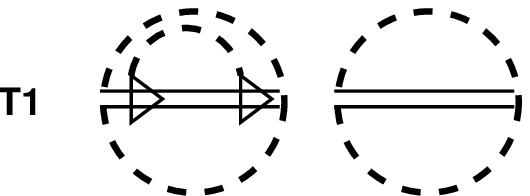}}
\qqq
{\epsfxsize=95pt  \epsfysize=70pt \epsfbox{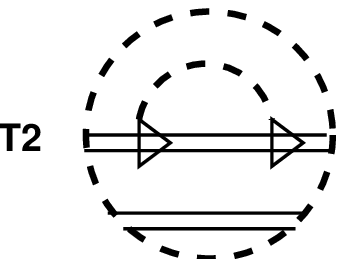}}
                 }
\vspace{10pt}
     \centerline{
{\epsfxsize=153pt  \epsfysize=55pt \epsfbox{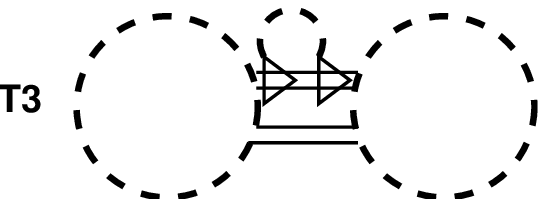}}
\qqq
{\epsfxsize=90pt  \epsfysize=70pt \epsfbox{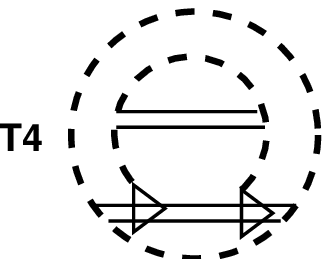}}
                 }
\caption{
%*figT1t4*Fig.T1t4\ 
Graphs for 
$T_1=\na^2R\cdot R, T_2=\na^2R_\ls\cdot R^\ls, 
T_3=\na^2R_{\la\rho\si\tau}\cdot R^{\la\rho\si\tau}$, and
$T_4=\na^\m\na^\n R\cdot R_\mn$. 
        }
\label{figT1t4}
\end{figure}

\flushleft{(iii)\ Fig.\ref{figT1t4}:\ $\na\na R \times R$-type}

\begin{figure}
     \centerline{
{\epsfxsize=155pt  \epsfysize=55pt \epsfbox{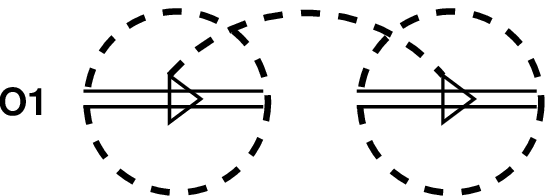}}
\qqq
{\epsfxsize=100pt  \epsfysize=72pt \epsfbox{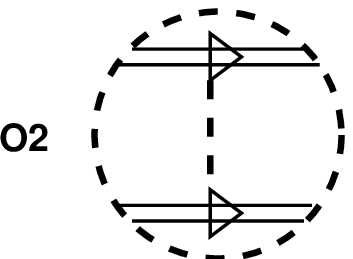}}
                 }
\vspace{10pt}
     \centerline{
{\epsfxsize=153pt  \epsfysize=53pt \epsfbox{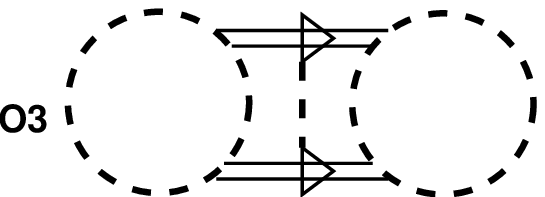}}
\qqq
{\epsfxsize=87pt  \epsfysize=72pt \epsfbox{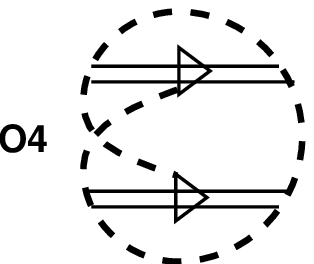}}
                 }
\caption{
%*figO1t4*Fig.O1t4\ 
Graphs for 
$O_1=\na^\mu R\cdot \na_\mu R, 
O_2=\na^\mu R_\ls\cdot \na_\mu R^\ls, 
O_3=\na^\mu R^{\la\rho\si\tau}\cdot \na_\mu R_{\la\rho\si\tau}$, and 
$O_4=\na^\mu R_{\la\n}\cdot \na^\n R^\la_{~\m}$. 
        }
\label{figO1t4}
\end{figure}

\flushleft{(iv)\ Fig.\ref{figO1t4}:\ $\na R \times \na R$-type}

\begin{figure}
     \centerline{
{\epsfxsize=92pt  \epsfysize=70pt \epsfbox{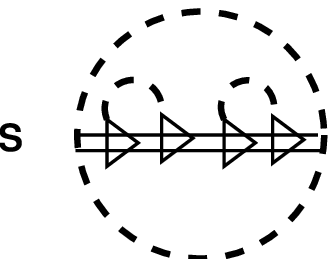}}
                 }
\caption{
%*figS*Fig.S\ 
Graphs for
$S=\na^2\na^2R$.
        }
\label{figS}
\end{figure}

\flushleft{(v)\ Fig.\ref{figS}:\ $\na^2 \na^2 R $-type}

\vs 2

%%%%%%%%%%%%%%%%%%%%%%%% Fig.P1t6 %%%%%%%%%%%%%%%%%%%%%%%%%%%%%%%%%
%%%%%%%%%%%%%%%%%%%%%%%%%%%%%%%%%%%%%%%%%%%%%%%%%%%%%%%%%%%%%%%%%%%%%

%%%%%%%%%%%%%%%%%%%%%%%% Fig.AB %%%%%%%%%%%%%%%%%%%%%%%%%%%%%%%%%
%%%%%%%%%%%%%%%%%%%%%%%%%%%%%%%%%%%%%%%%%%%%%%%%%%%%%%%%%%%%%%%%%%%%%

%%%%%%%%%%%%%%%%%%%%%%%% Fig.T1t4 %%%%%%%%%%%%%%%%%%%%%%%%%%%%%%%%%
%%%%%%%%%%%%%%%%%%%%%%%%%%%%%%%%%%%%%%%%%%%%%%%%%%%%%%%%%%%%%%%%%%%%%

%%%%%%%%%%%%%%%%%%%%%%%% Fig.O1t4 %%%%%%%%%%%%%%%%%%%%%%%%%%%%%%%%%
%%%%%%%%%%%%%%%%%%%%%%%%%%%%%%%%%%%%%%%%%%%%%%%%%%%%%%%%%%%%%%%%%%%%%

%%%%%%%%%%%%%%%%%%%%%%%% Fig.S %%%%%%%%%%%%%%%%%%%%%%%%%%%%%%%%%
%%%%%%%%%%%%%%%%%%%%%%%%%%%%%%%%%%%%%%%%%%%%%%%%%%%%%%%%%%%%%%%%%%%%%

\newpage
%%%%%%%%%%%%%%%%%%%%%%%%%%  App. G  %%%%%%%%%%%%%%%%%%%%%%%%%%%%%%%
%%%%%%%%%%%%%%%%%%%%%%%%%%%%%%%%%%%%%%%%%%%%%%%%%%%%%%%%%%%%%%%%%%%
\begin{flushleft}
{\Large\bf Appendix G.\ 
Graphical Definitions of Totally Anti-symmetrized Quantities
}
\end{flushleft}

\begin{figure}
     \centerline{
{\epsfxsize=345pt  \epsfysize=50pt \epsfbox{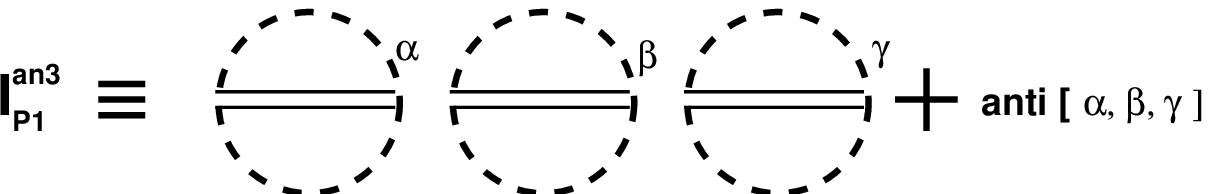}}
                 }
\vspace{10pt}
     \centerline{
{\epsfxsize=290pt  \epsfysize=105pt \epsfbox{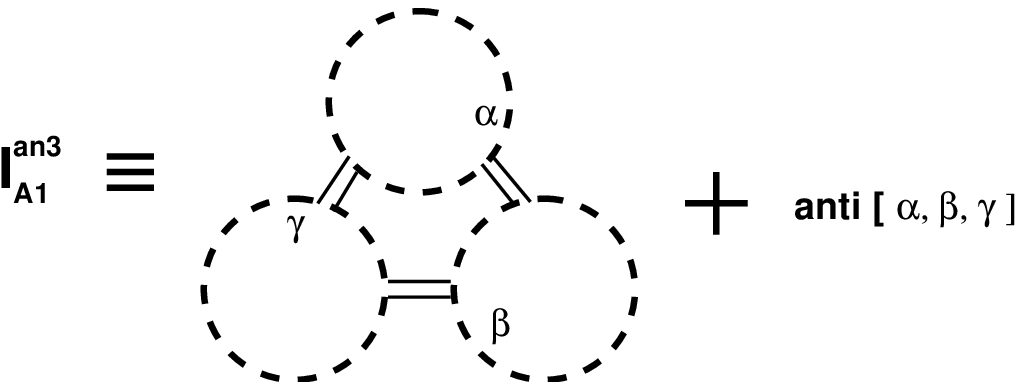}}
                 }
\vspace{10pt}
     \centerline{
{\epsfxsize=290pt  \epsfysize=130pt \epsfbox{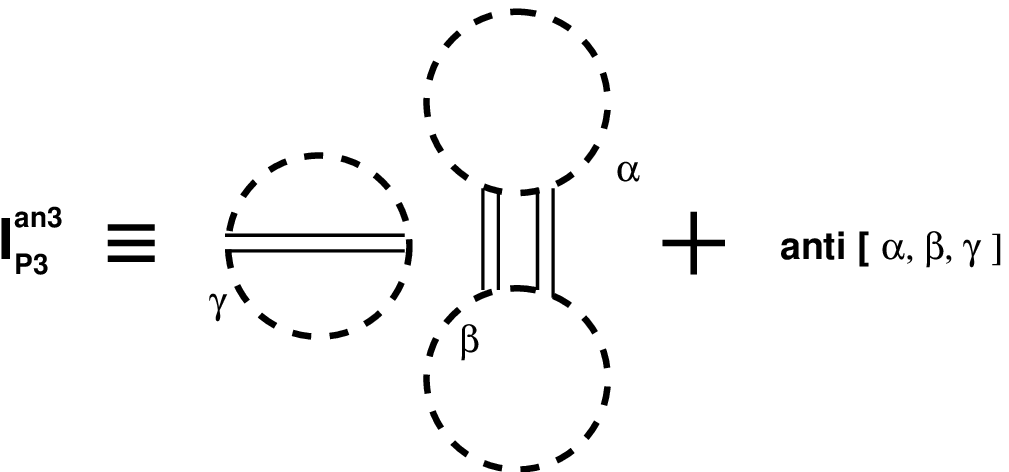}}
                 }
\vspace{10pt}
     \centerline{
{\epsfxsize=225pt  \epsfysize=130pt \epsfbox{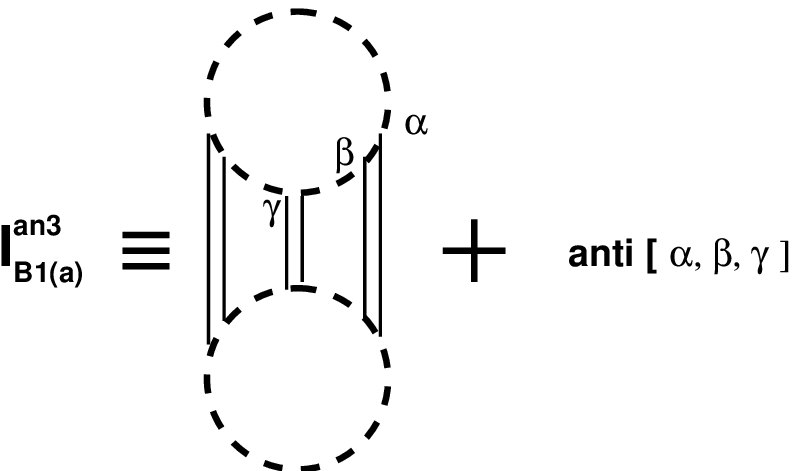}}
                 }
\vspace{10pt}
     \centerline{
{\epsfxsize=230pt  \epsfysize=130pt \epsfbox{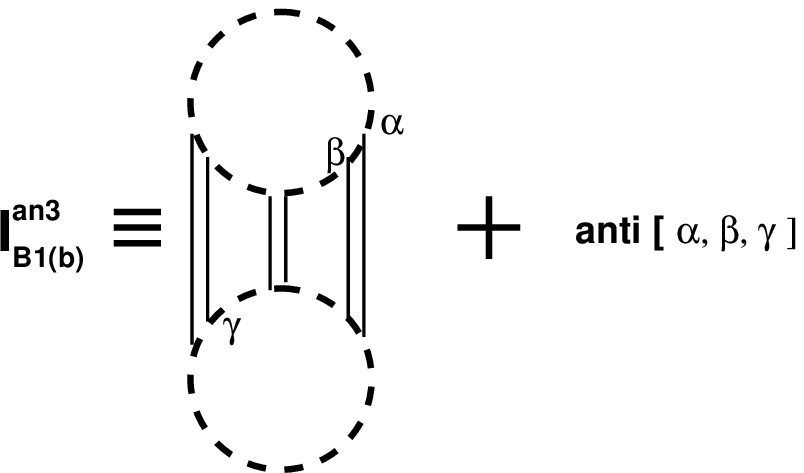}}
                 }
     \centerline{
{\epsfxsize=260pt  \epsfysize=85pt \epsfbox{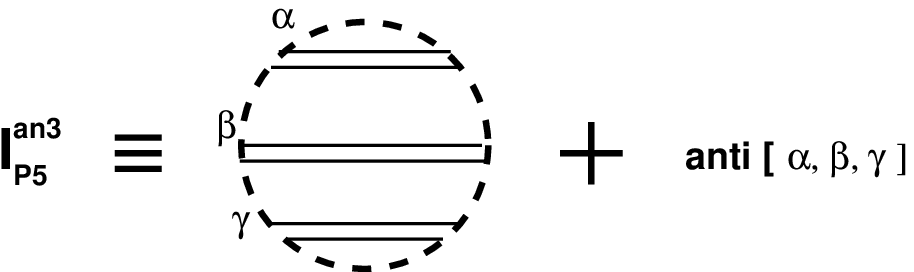}}
                 }
\caption{
%*fig9p3a*Fig.9.3a\ 
Graphical definition for $I^{an3}_{P_1},I^{an3}_{A_1},
I^{an3}_{P_3},I^{an3}_{B_1(a)},I^{an3}_{B_1(b)}$
\ and $I^{an3}_{P_5}$.
        }
\label{fig9p3a}
\end{figure}

\begin{figure}
     \centerline{
{\epsfxsize=355pt  \epsfysize=50pt \epsfbox{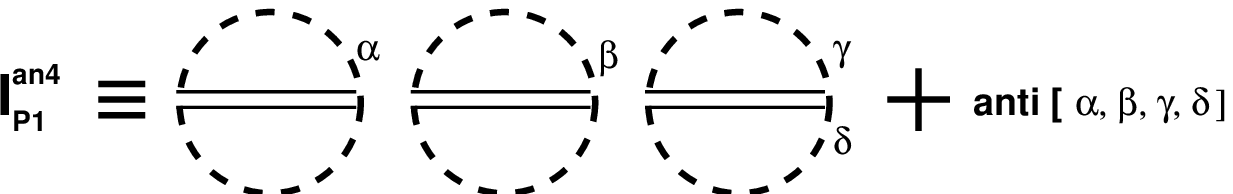}}
                 }
\vspace{10pt}
     \centerline{
{\epsfxsize=295pt  \epsfysize=105pt \epsfbox{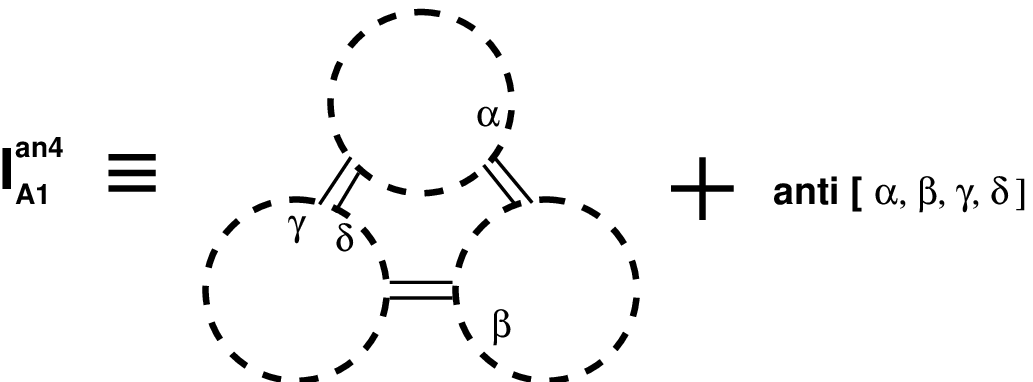}}
                 }
\vspace{10pt}
     \centerline{
{\epsfxsize=310pt  \epsfysize=130pt \epsfbox{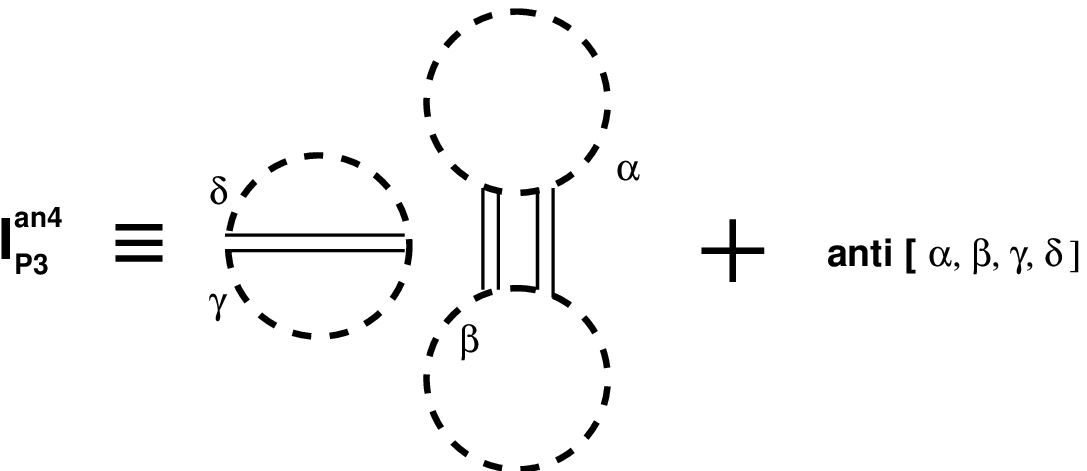}}
                 }
\vspace{10pt}
     \centerline{
{\epsfxsize=235pt  \epsfysize=130pt \epsfbox{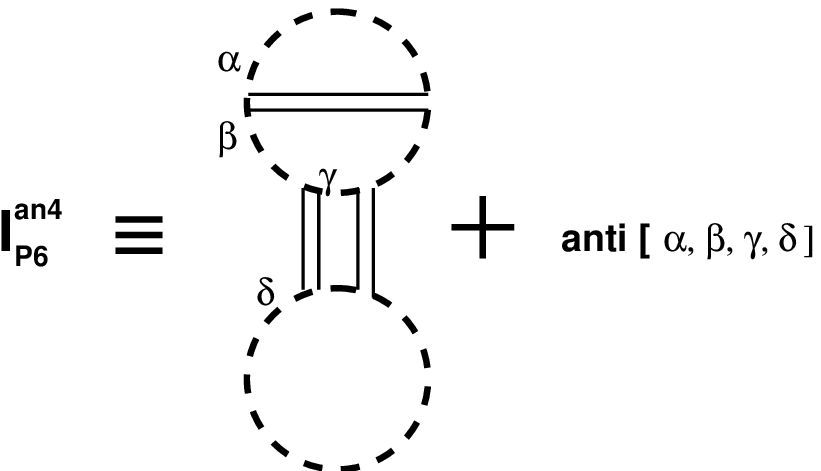}}
                 }
\caption{
%*fig9p3b*Fig.9.3b\ 
Graphical definition for $
I^{an4}_{P_1},I^{an4}_{A_1},I^{an4}_{P_3}$
\ and $I^{an4}_{P_6}$.
        }
\label{fig9p3b}
\end{figure}

\begin{figure}
     \centerline{
{\epsfxsize=350pt  \epsfysize=55pt \epsfbox{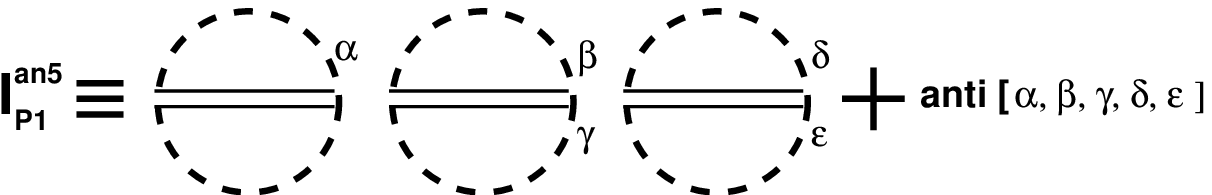}}
                 }
\vspace{10pt}
     \centerline{
{\epsfxsize=310pt  \epsfysize=105pt \epsfbox{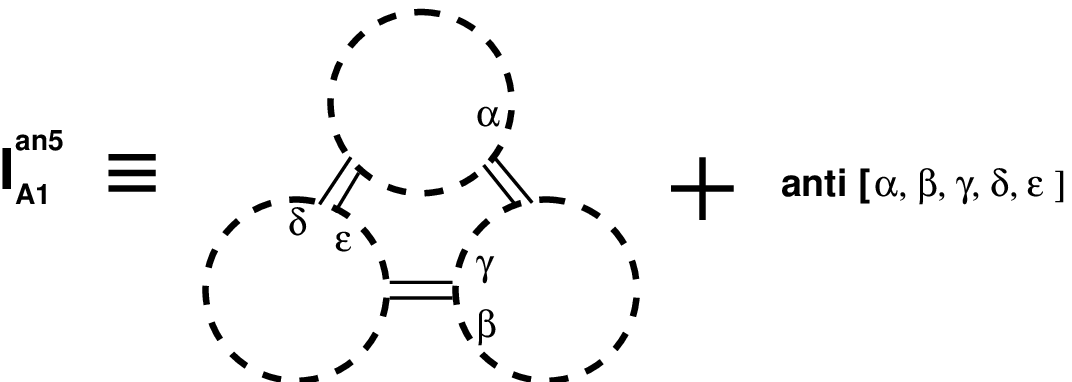}}
                 }
\vspace{10pt}
     \centerline{
{\epsfxsize=350pt  \epsfysize=55pt \epsfbox{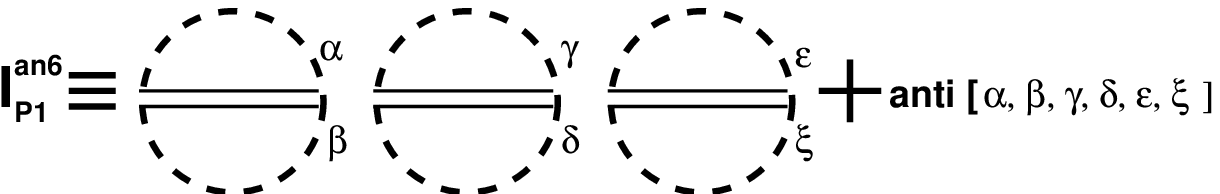}}
                 }
\caption{
%*fig9p3c*Fig.9.3c\ 
Graphical definition for $I^{an5}_{P_1},I^{an5}_{A_1}$
\ and $I^{an6}_{P_1}$.
        }
\label{fig9p3c}
\end{figure}

\begin{figure}
     \centerline{
{\epsfxsize=315pt  \epsfysize=55pt \epsfbox{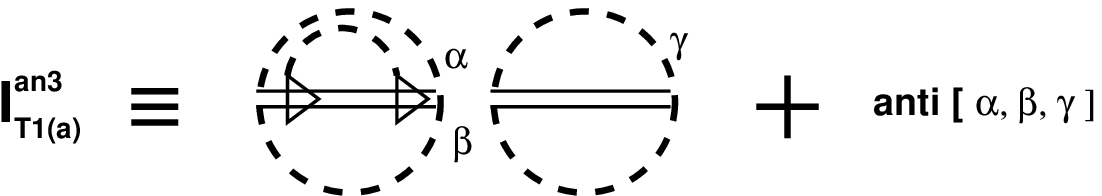}}
                 }
\vspace{10pt}
     \centerline{
{\epsfxsize=315pt  \epsfysize=55pt \epsfbox{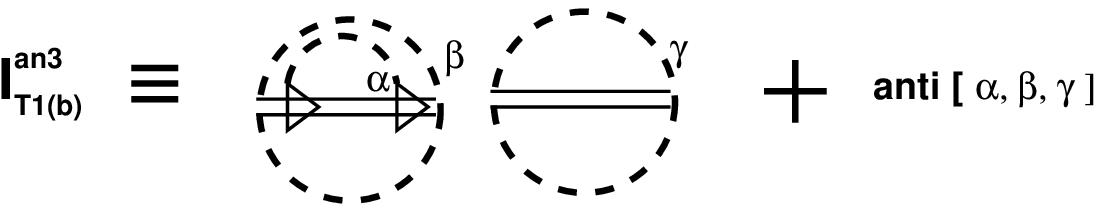}}
                 }
\vspace{10pt}
     \centerline{
{\epsfxsize=315pt  \epsfysize=55pt \epsfbox{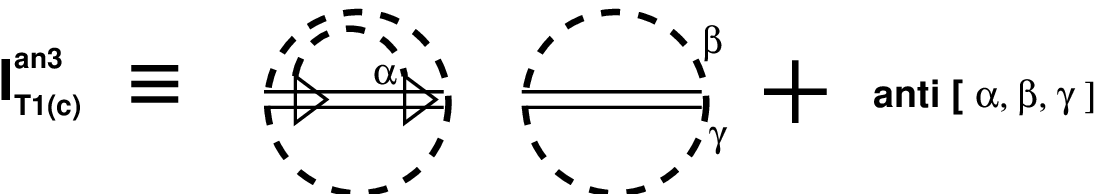}}
                 }
\vspace{10pt}
     \centerline{
{\epsfxsize=315pt  \epsfysize=55pt \epsfbox{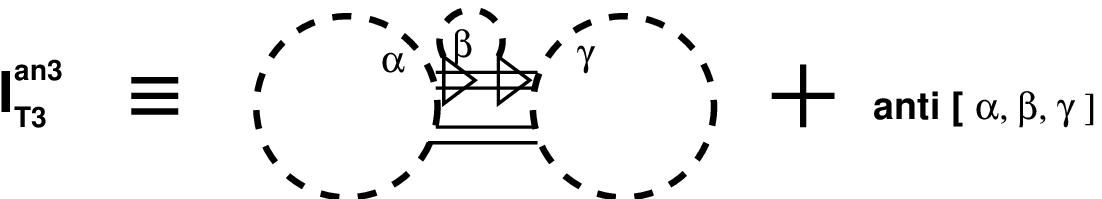}}
                 }
\vspace{10pt}
     \centerline{
{\epsfxsize=310pt  \epsfysize=50pt \epsfbox{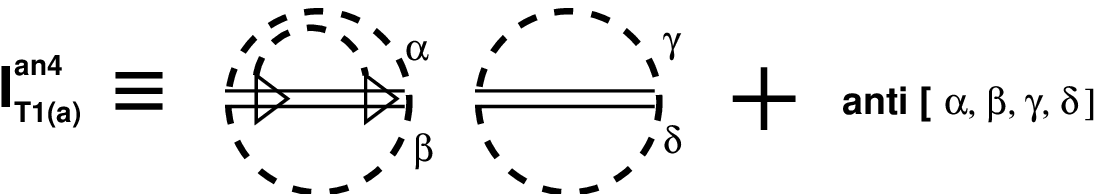}}
                 }
\vspace{10pt}
     \centerline{
{\epsfxsize=315pt  \epsfysize=55pt \epsfbox{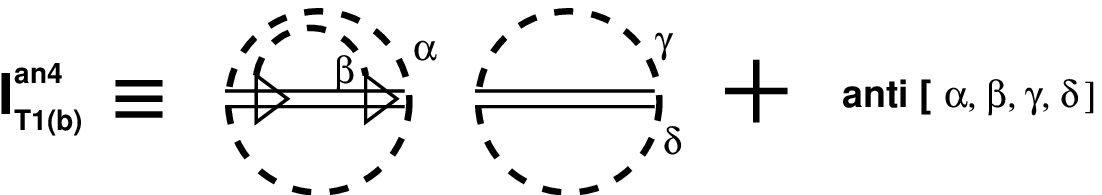}}
                 }
\caption{
%*fig9p4*Fig.9.4\ 
Graphical definition for $I^{an3}_{T_1(a)},I^{an3}_{T_1(b)},
I^{an3}_{T_1(c)},I^{an3}_{T_3},I^{an4}_{T_1(a)}$
\ and $I^{an4}_{T_1(b)}$.
        }
\label{fig9p4}
\end{figure}

\begin{figure}
     \centerline{
{\epsfxsize=313pt  \epsfysize=52pt \epsfbox{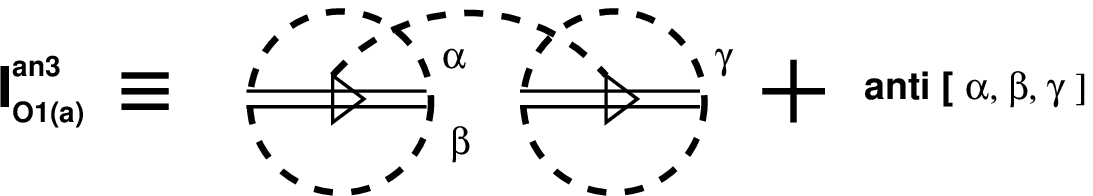}}
                 }
\vspace{10pt}
     \centerline{
{\epsfxsize=305pt  \epsfysize=55pt \epsfbox{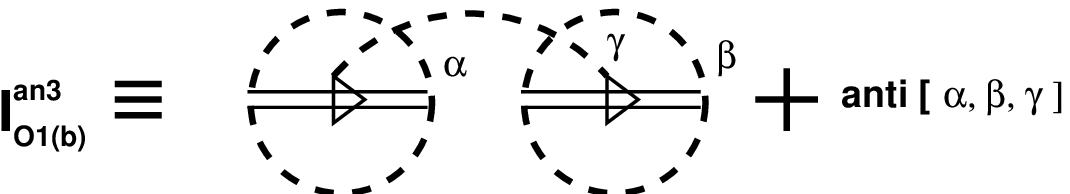}}
                 }
\vspace{10pt}
     \centerline{
{\epsfxsize=320pt  \epsfysize=55pt \epsfbox{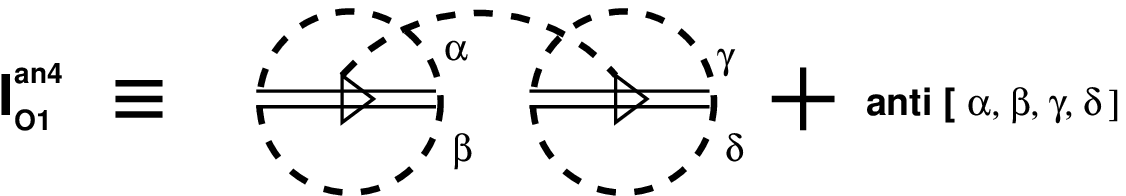}}
                 }
\caption{
%*fig9p5*Fig.9.5\ 
Graphical definition for $I^{an3}_{O_1(a)},I^{an3}_{O_1(b)}$
\ and $I^{an4}_{O_1}$.
        }
\label{fig9p5}
\end{figure}

All independent non-vanishing
totally anti-symmetrized quantities with the dimension $M^6$
are graphically defined in this appendix.
They are used, in Sec.IX of the text, to derive special relations, 
between general invariants, valid only in each dimension. 
The anti-symmetrized quantities are grouped, in the following, 
by the type of
a starting general invariant: Fig.\ref{fig9p3a}-\ref{fig9p3c}
($R\times R\times R$-type),
Fig.\ref{fig9p4} ($\na\na R\times R$-type) 
and Fig.\ref{fig9p5}($\na R\times \na R$-type).

\vs 2
%%%%%%%%%%%%%%%%%%%%%%%% Fig.9.3a %%%%%%%%%%%%%%%%%%%%%%%%%%%%%%%%%
%%%%%%%%%%%%%%%%%%%%%%%%%%%%%%%%%%%%%%%%%%%%%%%%%%%%%%%%%%%%%%%%%%%%%

%%%%%%%%%%%%%%%%%%%%%%%% Fig.9.3b %%%%%%%%%%%%%%%%%%%%%%%%%%%%%%%%%
%%%%%%%%%%%%%%%%%%%%%%%%%%%%%%%%%%%%%%%%%%%%%%%%%%%%%%%%%%%%%%%%%%%%%

%%%%%%%%%%%%%%%%%%%%%%%% Fig.9.3c %%%%%%%%%%%%%%%%%%%%%%%%%%%%%%%%%
%%%%%%%%%%%%%%%%%%%%%%%%%%%%%%%%%%%%%%%%%%%%%%%%%%%%%%%%%%%%%%%%%%%%%

%%%%%%%%%%%%%%%%%%%%%%%% Fig.9.4 %%%%%%%%%%%%%%%%%%%%%%%%%%%%%%%%%
%%%%%%%%%%%%%%%%%%%%%%%%%%%%%%%%%%%%%%%%%%%%%%%%%%%%%%%%%%%%%%%%%%%%%

%%%%%%%%%%%%%%%%%%%%%%%% Fig.9.5 %%%%%%%%%%%%%%%%%%%%%%%%%%%%%%%%%
%%%%%%%%%%%%%%%%%%%%%%%%%%%%%%%%%%%%%%%%%%%%%%%%%%%%%%%%%%%%%%%%%%%%%

%%%%%%%%%%%%%%%%%%%%%%%%%%%%%
\newpage
\begin{flushleft}
{\Large\bf Figure Captions}
\end{flushleft}
\begin{itemize}
\item
Fig.\ref{fig1}\ 
4-tensor $\pl_\m\pl_\n h_\ab$ 
\item
Fig.\ref{fig2}\ 
Graphical representation of 
$A1=\pl_\si\pl_\la h_\mn\cdot\pl_\si\pl_\n h_{\m\la}$. 
\item
Fig.\ref{fig3}\ 
Bondless diagram for $A1$ of Fig.\ref{fig2}.
dd-vertices are explicitly represented by small circles.
\item
Fig.\ref{fig4}\ 
Graphical representation of 
$PQ
=\pl^2 h_{\la\la}\cdot\pl_\m\pl_\n h_{\mn}$.
\item
Fig.\ref{fig5}\ 
Bondless diagrams and values of ($v$, $w$, \ul{vcn}).
\item
Fig.\ref{fig6}\ 
Reduction procedure of identifying two vertex-types:
dd-vertex and h-vertex.
\item
Fig.\ref{f7p1}\ 
Classification of 
$(\pl\pl h)^3$-graphs by \ul{bcn}[\ ], $\ul{l}=1$.
\item
Fig.\ref{f7p2}\ 
Classification of $(\pl\pl h)^3$-graphs 
by \ul{bcn}[\ ], $\ul{l}=2$.
\item
Fig.\ref{f7p3}\ 
Classification of $(\pl\pl h)^3$-graphs by \ul{bcn}[\ ], $\ul{l}=3$.
\item
Fig.\ref{f7p4}\ 
Classification of $(\pl\pl h)^3$-graphs by \ul{bcn}[\ ], $\ul{l}=4$.
\item
Fig.\ref{fig8}\ 
Reduction of Graphs.
\item
Fig.\ref{fig9}\ 
Reduced Graphs by the procedure Fig.8.
\item
Fig.\ref{fig10}\ 
G51:\ $3F_a\Th$.
\item
Fig.\ref{fig11}\ 
Three Graphs with the same $\ul{l}$,\ul{vcn} and \ul{bcn}. \ul{Vorder}
discriminate them.
\item
Fig.\ref{fig12}\ 
Two Graphs (G28,G30) with the same $\ul{l}$,\ul{vcn}[\ ] 
and \ul{bcn}[\ ]. \ul{Vorder} discriminate them.
\item
Fig.\ref{fig13}\ 
Two Graphs with the same $\ul{l}$,\ul{vcn}[\ ] and \ul{bcn}[\ ]. 
\ul{ddverno}[\ ] and \ul{hverno}[\ ] discriminate them.
\item
Fig.\ref{fig14}\ 
Two Graphs with the same $\ul{l}$,\ul{vcn} and \ul{bcn}. \ul{crossno}[\ ]
discriminates them.
\item
Fig.\ref{fig15p1}\ 
Vertices and propagators of (\ref{weight.3}).
\item
Fig.\ref{fig15p2}\ 
Vertices and propagators of (\ref{weight.4}).
\item
Fig.\ref{fig15p3}\ 
The vertex  of (\ref{weight.5}).
\item
Fig.\ref{fig8p1}\ 
Graphical representations for
\ (a)\ $R_{\mn\ls}$,\ (b)\ $\na_\al R_{\mn\ls}$\  
and\ (c)\ $\na_\al\na_\be R_{\mn\ls}$.
\item
Fig.\ref{fig8p2}\ 
Graphical representation for the Riemann scalar $R$.
\item
Fig.\ref{fig8p3}\ 
Graphical representation for
\ (a)\ $\na^2R$, (b)\ $R^2$, (c)\ $R_\mn R^\mn$,  
and (d)\ $R_{\mn\ls}R^{\mn\ls}$.
\item
Fig.\ref{fig9p1}\ 
Graphical representation for $I^{an2}_R\equiv R^{\ab}_{~\be\al}
-R^\ab_{~\ab}$. 
In the figure, anti[$\al,\be$] means anti-symmetrization
w.r.t. $\al$ and $\be$.
The second-line figure demonstrates the present
notation used in the following.
\item
Fig.\ref{fig9p2}\ 
Graphical definition for $I^{an3}_{RR}$\ and $I^{an4}_{RR}$.
In the figure, anti[$\al,\be,\ga$]
and anti[$\al,\be,\ga,\del$] mean totally anti-symmetrization
w.r.t. ($\al,\be,\ga$) and ($\al,\be,\ga,\del$) respectively.
\item
Fig.\ref{G1-13}\ 
(i) $\ul{l}=1$\ (\ 13(con)+0(discon)=13 terms
G1-13.
\item
Fig.\ref{G14-42a}\ 
(ii) $\ul{l}=2$\ (\ 26(con)+3(discon)=29\ terms \ G14-42 \ No.1
\item
Fig.\ref{G14-42b}\ 
(ii) $\ul{l}=2$\ (\ 26(con)+3(discon)=29\ terms \ G14-42 \ No.2
\item
Fig.\ref{G43-69a}\ 
(iii) $\ul{l}=3$\ (\ 19(con)+8(discon)=27 terms\ G43-69\ No.1
\item
Fig.\ref{G43-69b}\ 
(iii) $\ul{l}=3$\ (\ 19(con)+8(discon)=27 terms\ G43-69\ No.2
\item
Fig.\ref{G70-85a}\ 
(iv) $\ul{l}=4$\ (\ 8(con)+8(discon)=16 terms\ G70-85, No.1
\item
Fig.\ref{G70-85b}\ 
(iv) $\ul{l}=4$\ (\ 8(con)+8(discon)=16 terms\ G70-85, No.2
\item
Fig.\ref{G86-89}\ 
(v) $\ul{l}=5$\ (\ 0(con)+4(discon)=4 terms,\ G86-89
\item
Fig.\ref{G90}\ 
(vi) $\ul{l}=6$\ (\ 0(con)+1(discon)=1 term,\ G90
\item
Fig.\ref{figCp1}\ 
Graphs of 6-tensor $\pl_\m\pl_\n\pl_\la\pl_\si h_\ab$
\item
Fig.\ref{figCp2}\ 
Graphs for
$P'\equiv \pl^2\pl^2 h_{\m\m}$ and $Q'\equiv \pl^2\pl_\m\pl_\n h_\mn$.
\item
Fig.\ref{figCp3}\ 
Graphs for \ul{bridgeno}=0\ (\ disconnected )
\item
Fig.\ref{figCp4}\ 
Graphs for \ul{bridgeno}=2
\item
Fig.\ref{figCp5}\ 
Graphs for \ul{bridgeno}=4
\item
Fig.\ref{figDp1}\ 
Graphs of 5-tensor $\pl_\m\pl_\n\pl_\la h_\ab$
\item
Fig.\ref{figDp2}\ 
Graphs for \ul{bridgeno}=1
\item
Fig.\ref{figDp3}\ 
Graphs for \ul{bridgeno}=3
\item
Fig.\ref{figDp4}\ 
Graphs for \ul{bridgeno}=5
\item
Fig.\ref{figP1t6}\ 
Graphs for 
$P_1=RRR\com\ \ P_2=RR_\mn R^\mn, P_3=RR_{\mn\ls}R^{\mn\ls},
P_4=R_\mn R^{\n\la}R_\la^{~\mu}, P_5=R_{\mn\ls}R^{\mu\la}R^{\nu\si}$ and
$ P_6=R_{\mn\ls}R_\tau^{~\nu\ls}R^{\mu\tau}$.
\item
Fig.\ref{figAB}\ 
Graphs for
$A_1=R_{\mn\ls}R^{\si\la}_{~~~\tau\om}R^{\om\tau\n\m}$ and 
$B_1=R_{\mn\tau\si}R^{\n~~~\tau}_{~\la\om}R^{\la\mu\si\om}$.
\item
Fig.\ref{figT1t4}\ 
Graphs for 
$T_1=\na^2R\cdot R, T_2=\na^2R_\ls\cdot R^\ls, 
T_3=\na^2R_{\la\rho\si\tau}\cdot R^{\la\rho\si\tau}$, and
$T_4=\na^\m\na^\n R\cdot R_\mn$. 
\item
Fig.\ref{figO1t4}\ 
Graphs for 
$O_1=\na^\mu R\cdot \na_\mu R, 
O_2=\na^\mu R_\ls\cdot \na_\mu R^\ls, 
O_3=\na^\mu R^{\la\rho\si\tau}\cdot \na_\mu R_{\la\rho\si\tau}$, and 
$O_4=\na^\mu R_{\la\n}\cdot \na^\n R^\la_{~\m}$. 
\item
Fig.\ref{figS}\ 
Graphs for
$S=\na^2\na^2R$.
\item
Fig.\ref{fig9p3a}\ 
Graphical definition for $I^{an3}_{P_1},I^{an3}_{A_1},
I^{an3}_{P_3},I^{an3}_{B_1(a)},I^{an3}_{B_1(b)}$
\ and $I^{an3}_{P_5}$.
\item
Fig.\ref{fig9p3b}\ 
Graphical definition for $
I^{an4}_{P_1},I^{an4}_{A_1},I^{an4}_{P_3}$
\ and $I^{an4}_{P_6}$.
\item
Fig.\ref{fig9p3c}\ 
Graphical definition for $I^{an5}_{P_1},I^{an5}_{A_1}$
\ and $I^{an6}_{P_1}$.
\item
Fig.\ref{fig9p4}\ 
Graphical definition for $I^{an3}_{T_1(a)},I^{an3}_{T_1(b)},
I^{an3}_{T_1(c)},I^{an3}_{T_3},I^{an4}_{T_1(a)}$
\ and $I^{an4}_{T_1(b)}$.
\item
Fig.\ref{fig9p5}\ 
Graphical definition for $I^{an3}_{O_1(a)},I^{an3}_{O_1(b)}$
\ and $I^{an4}_{O_1}$.
\end{itemize}


\begin{thebibliography}{99}
\bibitem{foot1}
{
When we say ``dimension'', there are two different meanings in this
paper. One is the dimension of space, which we call the space-dimension
when clear separation is necessary. The other is the mass dimension
of an operator. For example, the Riemann tensor has an $M^2$ dimension
in any space-dimension. We call it the mass-dimension when necessary.
}
\bibitem{foot2}
{
The relations come from
``Type 2'' symmetry in Sec.2 of \cite{SI}.
Note that the statement does not apply to those relations which
come from ``Type 3'' symmetry. It
is the symmetry depending on each dimension.
         }
\bibitem{foot3}
{
\ul{bcn}[\ ] imply an general element of an
array:\ul{bcn}[0],$\cdots$,\ul{bcn}[$\ul{l}$]. 
The same thing is for \ul{vcn}[\ ].
All quantities of indices
are underlined in the following.
}
\bibitem{foot4}
{
The introduced indices here are the same as 
those explained in Lemma of Sec.II:\ 
\ul{ddverno}[$i$]=$w_i$,\ul{hverno}[$i$]=$v_i$. 
}
\bibitem{foot5}
{
As a general tendency, we see, from the result of App.B, 
the weight of a graph decreases as its suffix-loop no (\ul{$l$})
increases. 
         }
\bibitem{foot6}
{
In contrast with \cite{SI},
we take here double solid lines to express the 'bond' of $R_{\mn\ls}$
in order to avoid the confusion with $\pl_\m\pl_\n h_\ab$ of 
Fig.\ref{fig1}.
         }
\bibitem{foot7}
{
Note that in \cite{SI}, total derivative terms are neglected, whereas
they are not neglected in the present case.
          }
\bibitem{foot8}
{
\{$P_1-P_6$\},$A_1$,$B_1$ and \{$O_1-O_2$\} are the same notation 
as that in \cite{SI}.
$O_3$ and $O_4$ in (\ref{ind.3}) correspond to $Q_2$ and $Q_{10}$ in \cite{SI}
respectively.(There is a typographical mistake in the figure caption
of Fig.15 of \cite{SI} where a minus sign is missing in the literal
(not graphical) definition of $P_5$. The one given in (\ref{ind.3})
of the present text is the right one.)
          }
\bibitem{foot9}
{
The space-dimension is taken to be 2 in order to
obtain the mass-dimensions of the couplings, (\ref{conc.3}), 
without
introducing any additional mass parameters. Note that
the present purpose is the graph classification.
There the important thing is the topological structure
of graphs. 
The space-dimension of the field theory is irrelevant. 
}
\bibitem{foot10}
{
Among 90 terms listed in App.A, G3 and G13 only has the crossing
number 3. As for the definition of the crossing number, see Subsec.V.ii.
         }
\end{thebibliography}
\end{document}